\documentclass[12pt]{article}
\usepackage[utf8]{inputenc}
\usepackage[top=1in, bottom=1in, left=0.75in, right=0.75in]{geometry}
\usepackage{setspace}
\usepackage{natbib}
\usepackage{amsmath}
\usepackage{amssymb}
\usepackage[dvipsnames]{xcolor}
\usepackage{graphicx}
\usepackage{amsthm}
\usepackage{mathtools}
\usepackage{algpseudocode}
\usepackage{algorithm}
\usepackage{appendix}
\usepackage{authblk}
\usepackage{subfigure}
\usepackage{bbm}
\usepackage{enumitem}
\RequirePackage[colorlinks,citecolor=blue,linkcolor=blue,urlcolor=blue,pagebackref]{hyperref}
\newtheorem{theorem}{Theorem}[section]

\newtheorem{lemma}[theorem]{Lemma}
\newtheorem{proposition}[theorem]{Proposition}
\newtheorem{remark}{Remark}
\newtheorem{condition}{Condition}

\newcommand{\calT}{\mathcal{T}}

\newcommand{\vectorize}[1]{\textnormal{vec}(#1)}
\newcommand{\argmax}{\arg\max}
\DeclareMathOperator*{\argmin}{arg\,min}
\newcommand{\ind}[1]{\mathbbm{1}\left\{#1\right\}}
\newcommand{\Var}{\textnormal{Var}}
\newcommand{\Cov}{\textnormal{Cov}}
\newcommand{\EE}{\mathbb{E}}

\title{Sampling-based federated inference for M-estimators with non-smooth objective functions}
\author[1]{Xiudi Li}
\author[2]{Lu Tian}
\author[3]{Tianxi Cai}
\affil[1]{Division of Biostatistics, University of California, Berkeley}
\affil[2]{Department of Biomedical Data Science, Stanford University}
\affil[3]{Department of Biostatistics, Harvard T.H. Chan School of Public Health}
\date{}

\begin{document}

\maketitle
\begin{abstract}
We propose a novel sampling-based federated learning framework for statistical inference on M-estimators with non-smooth objective functions, which frequently arise in modern statistical applications such as quantile regression and AUC maximization. Classical inference methods for such estimators are often computationally intensive or require nonparametric estimation of nuisance quantities. Our approach circumvents these challenges by leveraging Markov Chain Monte Carlo (MCMC) sampling and a second-stage perturbation scheme to efficiently estimate both the parameter of interest and its variance. In the presence of multiple sites with data-sharing constraints, we introduce an adaptive strategy to borrow information from potentially heterogeneous source sites without transferring individual-level data. This strategy selects source sites based on a dissimilarity measure and constructs an optimally weighted estimator using a lasso regularization. The resulting estimator has an oracle property, i.e., it achieves the optimal asymptotical efficiency by borrowing information from eligible sites while guarding against negative transfer. We establish consistency and asymptotic normality of our proposed estimators and validate the method through extensive simulations and a real-data application on type 2 diabetes. Our results demonstrate substantial gains in inference precision and underscore the importance of inclusive, data-adaptive analysis frameworks in federated learning settings.
\end{abstract}

\section{Introduction}

$M$-estimators refer to a general class of estimators obtained by optimizing an objective function which encompasses various frequently used estimators including the classical maximum likelihood estimators. On the one hand, asymptotic normality of $M$-estimators can be established via standard argument under general smoothness assumptions on the objective function such as differentiability and Lipschitz continuity \citep[see, for example,][]{van2000asymptotic}. On the other hand, many useful estimators are obtained by maximizing a non-smooth loss function, and establishing their asymptotic normality is more challenging and requires often case-by-case analysis. In either case, under asymptotic normality, we can construct a Wald type CI centered around the consistent estimator of the parameter of interest, provided that a consistent estimate of the asymptotic variance is available.

When the objective function is non-smooth, the closed-form expression of the asymptotic variance of the corresponding $M$-estimator often involves certain nuisance functions. To obtain a consistent variance estimator based on the closed-form expression requires consistent estimation of the nuisance functions involved, which typically calls for nonparametric function estimation. For example, the variance of sample median from $n$ independent and identically distributed (i.i.d.) observations, which can be viewed as a $M$-estimator minimizing a $L_1$ loss function, is $(4nf_0)^{-1},$ where $f_0$ is the density of the underlying distribution at the true median. Clearly, a direct estimator of this variance requires nonparametric estimation of the density function. In another example, the asymptotic variance of the estimator that maximizes the area under the empirical receiver operating characteristic curve (AUC) \citep{pepe2006combining}, which is a special case of the maximum rank correlation estimator \citep{han1987non}, also involves the density function of the linear combination of predictors as well as derivatives of certain conditional mean function. Consistently estimating these functions are difficult, and the results may be sensitive to choices of tuning parameters such as the bandwidth used in local kernel smoothing.

Bootstrap is another approach for direct variance estimation without the need for nonparametric function estimation. The ordinary bootstrap involves repeated sampling from observed data and optimizing the objective functions based on these resamples. When the objective function is non-smooth, standard gradient-based optimization method may not be applicable, and repeatedly solving these optimization problems can be computationally burdensome. Additionally, in a multi-site study with data-sharing constraints, one has to run the bootstrap procedure in each site before aggregating the results, which further adds to the logistic complexity. 

In this work, we develop a novel sampling-based framework for statistical inference for $M$-estimators with non-smooth objective functions. Specifically, our framework enables integrative analysis of multi-site data, where the inference in a target site can be augmented by leveraging additional data from eligible source sites. Our specific contributions are two folds:
\begin{enumerate}
    \item develop a computationally efficient inferential method based on Markov Chain Monte Carlo (MCMC) sampling for $M$-estimators associated with a non-smooth objective function.
    \item extend our framework to integrative analysis of multi-site data and select eligible source sites to improve statistical inference in a target site. Our method accounts for the potential heterogeneity between the source and target data and requires only one round of communication without sharing individual-level data.
\end{enumerate}

This paper is organized as follows. In the remainder of this section, we review relevant literature and introduce some useful notations. In Section~\ref{sec:method}, we introduce our MCMC sampling-based inferential framework in a single site study and extend it to integrative analysis of multi-site data. In Section~\ref{sec:theory}, we establish theoretical guarantees for our proposed method. In Section~\ref{sec:example}, we illustrate our framework with two concrete examples, quantile regression and AUC maximization. In Section~\ref{sec:experiments}, we demonstrate the performance of our method in simulation studies and apply it to study the association between type 2 diabetes and various genetic markers. Section~\ref{sec:discussion} concludes with some further discussion.

\subsection{Relevant literature}
A classical sampling-based approach to constructing confidence intervals (CIs) is the nonparametric bootstrap \citep{efron1992bootstrap,tibshirani1993introduction}, which is valid for general Hadamard differentiable functionals of the data generating distribution. Since then, different extensions and variations of the bootstrap have been developed \citep[see, for example, ][for a review]{horowitz2019bootstrap, chernick2011bootstrap}. In the context of $M$-estimation, the nonparametric bootstrap involves repeatedly resampling data points from the original sample and optimizing the objective function defined using the resampled data. Instead of resampling, \citet{jin2001simple} proposed to repeatedly perturb the objective function using a set of random weights and solve the perturbed optimization problem. Compared to these methods, our procedure is computationally efficient in that it does not require solving any optimization problems and instead relies on a single run of MCMC and repeated evaluations of the objective functions at a given set of parameter values. 

When the objective function is smooth, one can solve the estimating equation defined by its first-order condition. This is related to the general strategy of $Z$-estimation. \citet{hu2000estimating} developed a procedure for hypothesis testing and CI construction by resampling the individual terms in the estimating equation. \citet{parzen1994resampling} and \citet{goldwasser2004statistical} studied inferential procedure for $Z$-estimators that took advantage of the pivotal property of the estimating functions and involved repeatedly solving perturbed equations. For non-smooth estimating equations, obtaining a good solution may be difficult. \citet{tian2004statistical} tackled this problem using the importance sampling technique and proposed an inferential method without the need to solve any equations. \citet{zeng2008efficient} estimated the asymptotic slope of the estimating function by evaluating the estimating function at perturbed values of the parameter, and their inferential procedure also did not involve repeatedly solving the estimating equations. Here, our focus is on $M$-estimators, and we note that the corresponding estimating equation may not exist when the objective function is highly non-smooth, for example, in the AUC maximization problem. 

There have been a variety of methods using MCMC for estimation and statistical inference. The most prominent example is in Bayesian statistics where MCMC are often used to sample realizations from the posterior distribution to construct point estimate and Bayesian credible interval \citep[see, for example, ][for a review]{li2021review}. \citet{chernozhukov2003mcmc} extended the use of MCMC and developed the class of Laplace type estimators for M-estimation. It is worth pointing out that their quasi-posterior interval is a valid CI, only when the generalized information equality holds, which is not the case with many non-smooth objective functions, where the construction of a valid CI still requires an estimate of the variance of the ``score function". More recently, \citet{giordano2023bayesian} proposed the infinitesimal jackknife to estimate the frequentist covariance of Bayesian estimators based on a single MCMC run assuming appropriate smoothness of the ``likelihood'' functions. In contrast, this work focuses on $M$-estimators with non-smooth objective functions and develops computationally efficient methods to estimate its standard error, which is extended to multi-site transfer learning setting using only a single MCMC run. 

Our work is also related to the growing body of literature on estimation and statistical inference in federated and transfer learning. Many previous works have focused on the problem of estimation in transfer learning \citep[e.g.,][]{li2022transfer,gu2022robust,xiong2023distributionally} and federated learning \citep[e.g.,][]{lee2017communication, maity2022meta,cai2022individual}. Other works studied statistical inference using multi-source data in various contexts, often aiming at addressing the potential heterogeneity in the data generating distributions in the source populations and/or data-sharing constraints. For example, \citet{xiong2023federated} studied federated inference of an average treatment effect using multi-source data; \citet{han2021federated} proposed a framework to efficiently infer about the treatment effect in a target population leveraging data from multiple heterogeneous source populations; \citet{liu2021integrative} developed a federated approach to testing the overall effect of a covariate in high-dimensional regression models; \citet{tian2023transfer} studied estimation and inference in transfer learning of sparse high-dimensional generalized linear models (GLM), and it's worth noting that their method requires individual-level data from all sites. For inference about more general statistical parameters, \citet{wang2023robust} considered the problem of combining summary statistics from multiple data sources that is robust to potentially biased sources, while \citet{guo2023robust} developed a general framework for federated inference of a prevailing parameter under the majority rule assumption that utilizes site-specific point estimates along with their estimated standard errors. Here, we focus on developing efficient inference of a general $M$-estimator in a target population in the presence of heterogeneous source populations and data-sharing constraints and a computationally efficient method to construct CIs when direct estimation of the asymptotic variance is difficult.

\subsection{Notation and problem set-up}
\paragraph{Notations.} For a generic vector $v \in \mathbb{R}^d$, we use $\|v\|_p$ to denote its $l_p$-norm defined as $(\sum_{i=1}^d |v_i|^p)^{1/p}$ for $p>0$ and write $\|\cdot\|_2$ as $\|\cdot\|$ for simplicity. For a generic matrix $V \in \mathbb{R}^{m \times n}$, we use $\|V\|$ to denote its operator norm and $\|V\|_1$ to denote its element-wise $l_1$ norm, that is, $\sum_{i=1}^m\sum_{j=1}^n|V_{ij}|$. We use $I_d$ to denote the $d$-dimensional identity matrix and use $\ind{\cdot}$ to denote the indicator function.  

\paragraph{Problem set-up.} Suppose that we want to infer about a parameter $\theta_\mathcal{T} \in \mathbb{R}^d$ in a target population defined as $\theta_\mathcal{T} = \argmin_\theta \mathbb{E}_\mathcal{T}[M_{n,\mathcal{T}}(\theta)]$, where $M_{n,\mathcal{T}}(\theta)$ is a potentially non-smooth objective function defined using an i.i.d. sample from the target population and $\mathbb{E}_\mathcal{T}$ denotes the expectation under the distribution in this target population. Moreover, suppose we have access to data from $K$ source populations. We can similarly define the source-population-specific parameter $\theta_k = \argmin_\theta \mathbb{E}_k[M_{n,k}(\theta)]$ for $k = 1, \ldots, K$. Here, $M_{n,k}(\theta)$ is the objective function in the $k$-th source and $\mathbb{E}_k$ denotes the expectation under the data distribution in the $k$-th source. 

We assume that the source populations can be heterogeneous, and some of them may be biased in the sense that $\theta_k \neq \theta_\mathcal{T}$. Let $\mathcal{K}$ denote the set of eligible sources defined as $\{1\leq k \leq K : \theta_k = \theta_\mathcal{T}\}$, which is the set of source populations from which we wish to borrow information to improve the inference about $\theta_\mathcal{T}$. However, we do not know $\mathcal{K}$ nor its cardinality \textit{a priori}.

\section{Federated sampling-based inference}\label{sec:method}

In this section, we introduce our sampling-based inferential framework. We first discuss statistical inference in the target site based on MCMC and perturbations of the target objective function, which leads to a target-only estimate $\widehat{\theta}_\mathcal{T}$. 
When data are available from other source sites, borrowing information from similar source sites can greatly improve efficiency in statistical inference for parameters in the target site. In this section, we propose an information borrowing technique based on transferring the scores from a source site to the target site.  Specifically, we will combine scores from relevant source sites with a target-site estimate $\widehat{\theta}_\mathcal{T}.$ The rationale is that $\widehat{\theta}_\mathcal{T}$ from the target site should approximately satisfy the first order condition of the loss function for source site $k \in \mathcal{K},$  i.e., $S_{n, k}(\widehat{\theta}_\mathcal{T})$ follows a mean zero normal distribution, where $S_{n,k}(\theta)$ is the ``score function'' in the $k$th source site. Therefore, 
$$\widehat{\theta}_{\mathcal{T}}+\sum_{k \in \mathcal{K}} \Lambda_k S_{n,k}(\widehat{\theta}_{\mathcal{T}}),$$
is also a valid regular estimator for $\theta_\calT$, the parameter of interest, for any set of weight matrices $\{\Lambda_k\mid k\in \mathcal{K}\}.$  It is possible to select weight matrices $\Lambda_k$ to minimize the variance of the resulting estimator of $\theta_\mathcal{T}.$  In particular, the resulting estimator is expected to be more efficient than $\widehat{\theta}_{\mathcal{T}}$, the estimator based on the target site data only.  To construct the optimal combination, one needs to know the variance of $\widehat{\theta}_\mathcal{T}$, the variance of $S_{n,k}(\widehat{\theta}_\mathcal{T})$ and their covariance. In the next subsections, we will discuss their estimation.

\subsection{A sampling-based inference in target site}\label{subsec:target}
Let $n_{\mathcal{T}}$ denote the sample size in the target site. Given an i.i.d. sample $\{X_{\mathcal{T},1},\ldots,X_{\mathcal{T},n_{\mathcal{T}}}\}$, we define the following, possibly non-smooth, objective function,
\begin{equation}\label{eq: target objective}
    M_{n,\mathcal{T}}(\theta) = \binom{n_\mathcal{T}}{D}^{-1} \sum_{1\leq i_1 < i_2 < \cdots < i_D \leq n_\mathcal{T}} h\left(X_{\mathcal{T},i_1},\ldots,X_{\mathcal{T},i_D};\theta\right),
\end{equation}
where $h(\cdot;\theta)$ is the kernel of a $D$-degree $U$-process symmetric in the $X$'s, indexed by parameter $\theta \in \mathbb{R}^d$. The $M$-estimator for a target site can be defined as 
\begin{equation}
    \tilde\theta_\mathcal{T} = \argmin_\theta M_{n,\mathcal{T}}(\theta).
\label{eq:targetM} \end{equation}

Instead of directly solving the optimization problem (\ref{eq:targetM}), which can be computationally demanding, we propose to use a sampling-based Monte-Carlo method. To this end, define the following density function
\begin{equation}
    L_\mathcal{T}(\theta) = C \exp\left\{-n_\mathcal{T}M_{n,\mathcal{T}}(\theta) + n_\mathcal{T}M_{n,\mathcal{T}}(\tilde\theta_\mathcal{T}) 
    \right\}\ind{\|\theta 
    \| \leq R},
\end{equation}
where $C$ is a normalizing constant such that $L_\mathcal{T}(\theta)$ integrates to 1 and $R$ is the radius of a ball centered at zero outside of which the targeted density is 0. 
Let $\{\widehat\theta_\mathcal{T}^{*,1},\ldots,\widehat\theta_\mathcal{T}^{*,B}\}$ be $B$ i.i.d. samples from the density $L_\mathcal{T}(\theta)$, and define their sample mean as 
\begin{equation}\label{eq:target-only estimator}
    \widehat\theta_\mathcal{T} = \frac{1}{B}\sum_{b=1}^B \widehat\theta_{\mathcal{T}}^{*,b},
\end{equation}
and the inverse of sample covariance (scaled) as 
$$\widehat A_\mathcal{T} = \frac{1}{n_\mathcal{T}}\left\{\frac{1}{B}\sum_{j=1}^B (\widehat\theta_{\mathcal{T}}^{*,j} - \widehat\theta_\mathcal{T})(\widehat\theta_{\mathcal{T}}^{*,j} - \widehat\theta_\mathcal{T})^\top \right\}^{-1}.$$ 
We will use $\widehat\theta_\mathcal{T}$ as our estimator in the target site. Heuristically, if the objective function is approximately quadratic in the sense that
\begin{equation*}
    M_{n,\mathcal{T}}(\theta) \approx M_{n,\mathcal{T}}(\tilde\theta_\mathcal{T}) + \frac{1}{2}\left(\theta - \tilde\theta_\mathcal{T}\right)^\top A_{\mathcal{T}}\left(\theta - \tilde\theta_\mathcal{T}\right)
\end{equation*}
for a positive definite matrix $A_\mathcal{T}$, then 
$$ L_\mathcal{T}(\theta) \propto \exp\{-n_\mathcal{T}M_{n,\mathcal{T}}(\theta) + n_\mathcal{T}M_{n,\mathcal{T}}(\tilde\theta_\mathcal{T}) \} \approx \exp\left\{-\frac{n_\mathcal{T}}{2}(\theta - \tilde\theta_\mathcal{T})^\top A_{\mathcal{T}}(\theta - \tilde\theta_\mathcal{T})\right\},$$ and the density $L_\mathcal{T}(\theta)$ is approximately a Gaussian density with mean $\tilde\theta_\mathcal{T}$ and covariance matrix $(n_\mathcal{T}A_\mathcal{T})^{-1}.$ Thus, we expect that $\widehat\theta_\mathcal{T}$ and $\widehat A_\mathcal{T}$ are good approximations of $\tilde\theta_\mathcal{T}$ and $A_\mathcal{T},$ respectively. These heuristics will be made rigorous in Section~\ref{sec:theory}. 

As $n_\mathcal{T}M_{n,\mathcal{T}}(\tilde\theta_\mathcal{T})$ in the definition of $L_\mathcal{T}(\theta)$ is independent of $\theta$ and can be absorbed into the normalizing constant $C,$ in practice we can generate $\{\widehat\theta_{\mathcal{T}}^{*,b}\}_{b=1}^B$ according to a density proportional to $\exp\{-n_\mathcal{T}M_{n,\mathcal{T}}(\theta)\}\ind{\|\theta \| \leq R}$ using Markov chain Monte Carlo (MCMC).  Other sampling techniques such as rejection sampling can also be used, when appropriate.

The MCMC sampling provides a means to compute the point estimator $\widehat\theta_\mathcal{T}$ and the matrix $\widehat{A}_\mathcal{T}$ which is a building block in the asymptotic variance of $\widehat\theta_\mathcal{T}$. To calculate the variance of $\widehat\theta_\mathcal{T}$, we also need the variance of the ``score function", which is the sample analogue of the first derivative of the expected objective function at the true parameter (see Section~\ref{sec:theory}). To obtain this variance, we use a second layer of resampling. Specifically, condition on the target sample, we define a perturbed objective function as follows,
\begin{equation}\label{target perturbation}
    M^\dag_{n,\mathcal{T}}(\theta) = \binom{n_\mathcal{T}}{D}^{-1} \sum_{1\leq i_1 < i_2 < \cdots < i_D \leq n_{\mathcal{T}}} W_{\mathcal{T},i_1}W_{\mathcal{T},i_2}\cdots W_{\mathcal{T},i_D} \cdot h\left(x_{\mathcal{T},i_1},\ldots,x_{\mathcal{T},i_D};\theta\right),
\end{equation}
where $\{W_{\mathcal{T},1},\ldots,W_{\mathcal{T},n_\mathcal{T}}\}$ is a set of positive random weights induced by the nonparametric bootstrap, i.e., $(W_{\mathcal{T}, 1}, \cdots, W_{\mathcal{T}, n_\mathcal{T}})$ follows a multinomial distribution $\mbox{MN}\left(n_\mathcal{T}, (n_\mathcal{T}^{-1}, \cdots, n_\mathcal{T}^{-1})\right),$
and $\{x_{\mathcal{T},1},\ldots,x_{\mathcal{T},n_{\mathcal{T}}}\}$ are observed data from the target site. 

\begin{remark}\label{alternative perturbation}
 The choice of weights in \eqref{target perturbation} can be more flexible. For example, any random weights, which are independent of the target site data and satisfy Assumption A in \citet{han2022multiplier}, can be used. Alternatively, we can introduce the perturbations in the same way as in \citet{jin2001simple}. Specifically, let $\{W_{\mathcal{T},1},\ldots,W_{\mathcal{T},n_\mathcal{T}}\}$ be $n_\mathcal{T}$ i.i.d. samples of a nonnegative, known random variable $W$ with mean $D^{-1}$  and variance 1, independent of the observed data from the target site. Then, the perturbed objective function $M^\dag_{n,\mathcal{T}}(\theta)$ can be defined as
\begin{equation*}
    M^\dag_{n,\mathcal{T}}(\theta) = \binom{n_\mathcal{T}}{D}^{-1} \sum_{1\leq i_1 < i_2 < \cdots < i_D \leq n_{\mathcal{T}}}\left(W_{\mathcal{T},i_1}+ \cdots +W_{\mathcal{T},i_D}\right)h\left(x_{\mathcal{T},i_1},\ldots,x_{\mathcal{T},i_D};\theta\right).
\end{equation*}
\end{remark}

For any given $\theta$, define $V_{\mathcal{T}}(\theta) = \textnormal{Var}_W[M^\dag_{n,\mathcal{T}}(\theta)-M^\dag_{n,\mathcal{T}}(\widehat\theta_\mathcal{T})]$, where the variance is taken with respect to the weight $W$'s condition on the target site sample. Suppose that $M^\dag_{n,\mathcal{T}}(\theta)$ is approximately quadratic near $\widehat\theta_\mathcal{T}$ in the sense that
\begin{equation}\label{target boot quadratic}
    M^\dag_{n,\mathcal{T}}(\theta) - M^\dag_{n,\mathcal{T}}(\widehat\theta_{\mathcal{T}}) \approx S^\dag_{n,\mathcal{T}}(\widehat\theta_\mathcal{T})(\theta - \widehat\theta_\mathcal{T}) + \frac{1}{2}(\theta - \widehat\theta_\mathcal{T})^\top A_{\mathcal{T}} (\theta - \widehat\theta_\mathcal{T})
\end{equation}
for $\theta$ close to $\widehat\theta_\mathcal{T}$, where $S^\dag_{n,\mathcal{T}}(\cdot)$ can be regarded as a perturbation of the ``score function". Then, 
$$V_\mathcal{T}(\theta)\approx (\theta - \widehat\theta_{\mathcal{T}})^\top \Sigma_{S,\mathcal{T}}^\dag(\theta - \widehat\theta_{\mathcal{T}})/n_\mathcal{T},$$ 
where 
$$\Sigma_{S,\mathcal{T}}^\dag = \textnormal{Var}_W[\sqrt{n_\mathcal{T}}S^\dag_{n,\mathcal{T}}(\widehat\theta_\mathcal{T})].$$ 
Similar to the ordinary bootstrap, the inner variance $\Sigma_{S,\mathcal{T}}^\dag$ is expected to be a consistent estimator of the variance of the ``score function" $$\Sigma_{S,\mathcal{T}} = \textnormal{Var}[\sqrt{n_\mathcal{T}}S_{n,\mathcal{T}}(\theta_\mathcal{T})],$$ suggesting that 
$$ V_{\mathcal{T}}(\theta)\approx n_\mathcal{T}^{-1}(\theta - \widehat\theta_{\mathcal{T}})^\top \Sigma_{S,\mathcal{T}}(\theta - \widehat\theta_{\mathcal{T}}).$$
This motivates us to regress $V_\mathcal{T}(\theta)$ against the vector obtained by vectorizing $(\theta - \widehat\theta_\mathcal{T})(\theta - \widehat\theta_\mathcal{T})^\top$ to obtain an estimate of the variance of the score. Note that, in practice, for any given $\theta$, $V_{\mathcal{T}}(\theta)$ can be well approximated by the empirical variance of $M^\dag_{n,\mathcal{T}}(\theta)-M^\dag_{n,\mathcal{T}}(\widehat\theta_\mathcal{T})$ over a large number of weight vectors $W,$ and the computational demand is light, since there is no need for repeated optimization. 

In order to regress $V_{{\cal T}}(\theta)$ against vectorized $(\theta - \widehat\theta_\mathcal{T})(\theta - \widehat\theta_\mathcal{T})^\top$,  we consider $\theta \in \{\widehat\theta_{\mathcal{T}}^{*,1},\ldots,\widehat\theta_{\mathcal{T}}^{*,B}\}$, the collection of parameter values we draw from the density $L_\mathcal{T}(\theta)$. To remove potential influential points in the regression, we further restrict the selected $\theta$ within a ball centered at $\widehat\theta_\mathcal{T}$ with a radius on the order of $n^{-1/2}_\mathcal{T},$ that is, $\|\widehat\theta_{\mathcal{T}}^{*,j} - \widehat\theta_\mathcal{T}\| \leq C_1 n_\mathcal{T}^{-1/2}$ for some constant $C_1.$ Let $\mathcal{S}$ be a subset of $\{1,\ldots,B\}$ of cardinality $B_1$ indexing the selected elements in $\{\widehat\theta_{\mathcal{T}}^{*,1},\ldots,\widehat\theta_{\mathcal{T}}^{*,B}\}$.
Furthermore, let $\widehat\Theta^{*,j}_\mathcal{T}$ denote the vector obtained by vectorizing the upper triangular part (including the diagonal) of $(\widehat\theta_{\mathcal{T}}^{*,j} - \widehat\theta_\mathcal{T})(\widehat\theta_{\mathcal{T}}^{*,j} - \widehat\theta_\mathcal{T})^\top$. Let $\widehat\gamma$ denote the ordinary least squares (OLS) coefficient obtained by regressing $V_\mathcal{T}(\widehat\theta_{\mathcal{T}}^{*,j})$ against $\widehat\Theta^{*,j}_\mathcal{T}$ with $j \in \mathcal{S}$, and let $\widehat\gamma_{u,v}$ denote the regression coefficient corresponding to the $(u,v)$-th entry of the matrix $(\widehat\theta_{\mathcal{T}}^{*,j} - \widehat\theta_\mathcal{T})(\widehat\theta_{\mathcal{T}}^{*,j} - \widehat\theta_\mathcal{T})^\top$ for $u \leq v$. We then define an estimate of $\Sigma_{S,\mathcal{T}}$, denoted $\widehat\Sigma_{S,\mathcal{T}}$, with the entries defined as follows
\begin{equation}\label{target score var}
    (\widehat\Sigma_{S,\mathcal{T}})_{u,v} = 
    \begin{cases}
    n_\mathcal{T}\times \widehat\gamma_{u,u}, & \text{if } 1\leq u=v \leq d\\
    n_\mathcal{T}/2 \times \widehat\gamma_{u,v}, &\text{if } 1 \leq u < v \leq d \\
    n_\mathcal{T}/2 \times \widehat\gamma_{v,u}, &\text{if } 1 \leq v < u \leq d
    \end{cases}
\end{equation}

We now have all the necessary ingredients for statistical inference using only the target site data: the variance of the point estimator $\widehat\theta_\mathcal{T}$ can be estimated as 
$$n_\mathcal{T}^{-1}\widehat{A}_\mathcal{T}^{-1}\widehat\Sigma_{S,\mathcal{T}}\widehat{A}_\mathcal{T}^{-1},$$ and we can construct a corresponding Wald-type CI.  In summary, the final output of this step of analysis includes a set of estimates and selected samples: 
$$ \left(\widehat{\theta}_\mathcal{T},  \widehat{A}_\mathcal{T},  \widehat{\Sigma}_{S,\mathcal{T}}\right) ~~\mbox{and}~~ \left\{ \widehat{\theta}_{\mathcal{T}}^{*,j}\mid j\in \mathcal{S} \right\}$$

\subsection{Estimation of the score and associated variance in source sites}
 In this section, we present a sampling-based procedure to estimate the score vector (first derivative of the expected objective function) and variance of the score vector at $\widehat{\theta}_{\mathcal{T}}.$  Let $\{X_{k,1},\ldots,X_{k,n_k}\}$ denote the $n_k$ i.i.d. sample in the $k$-th source site for $k \in \{1,\ldots,K\}$. We define the objective function in the $k$-th source site as
\begin{equation}
    M_{n,k}(\theta) = \binom{n_k}{D}^{-1} \sum_{1\leq i_1 < i_2 < \cdots < i_D \leq n_k} h\left(X_{k,i_1},\ldots,X_{k,i_D};\theta\right).
\end{equation}
The target site first broadcasts its point estimator $\widehat\theta_\mathcal{T}$ and the MCMC sample $\{\widehat\theta_{\mathcal{T}}^{*,j}: j \in \mathcal{S}\}$ to the source sites, and each source site evaluates its objective function at these parameter values, that is, $M_{n,k}(\widehat\theta_\mathcal{T})$ and $M_{n,k}(\widehat\theta_{\mathcal{T}}^{*,j})$ for $j \in \mathcal{S}$. 
Suppose that the source site objective function is approximately quadratic near the true parameter value $\theta_\mathcal{T}$ as well as its estimator $\widehat{\theta}_{\cal T}.$ We would expect the following quadratic approximation:
\begin{equation}\label{eq: source quadratic approx}
    M_{n,k}(\widehat\theta_{\mathcal{T}}^{*,j}) - M_{n,k}(\widehat\theta_{\mathcal{T}}) \approx S_{n,k}(\widehat\theta_\mathcal{T})(\widehat\theta_{\mathcal{T}}^{*,j} - \widehat\theta_{\mathcal{T}}) + \frac{1}{2}(\widehat\theta_{\mathcal{T}}^{*,j} - \widehat\theta_{\mathcal{T}})^\top A_k(\widehat\theta_\mathcal{T}) (\widehat\theta_{\mathcal{T}}^{*,j} - \widehat\theta_{\mathcal{T}}),
\end{equation}
where $S_{n,k}(\cdot)$ is the ``score function" in the source site and $A_k(\cdot)$ is the second derivative of the expected objective function. This suggests we can approximate $S_{n,k}(\widehat\theta_\mathcal{T})$ and $A_k(\widehat\theta_\mathcal{T})$ via regression. Specifically, let $\widehat\delta_{\mathcal{T}}^{*,j} = \widehat\theta_{\mathcal{T}}^{*,j}-\widehat\theta_{\mathcal{T}}$. We then regress the difference $M_{n,k}(\widehat\theta_{\mathcal{T}}^{*,j}) - M_{n,k}(\widehat\theta_{\mathcal{T}})$ against the covariates $\widehat\delta_{\mathcal{T}}^{*,j}$ and $ \widehat\Theta_{\mathcal{T}}^{*,j}$ for $j \in \mathcal{S}$. Denote the OLS estimates of the coefficient for $\widehat\delta_{\mathcal{T}}^{*,j}$ and $\widehat\Theta_{\mathcal{T}}^{*,j}$
as $\widehat\alpha^k$ and $\widehat\beta^k$, respectively. We approximate $S_{n,k}(\widehat\theta_\mathcal{T})$ with $\widehat S_{n,k}(\widehat\theta_\mathcal{T}) = \widehat\alpha^k$ and estimate $A_k(\widehat\theta_\mathcal{T})$ with $\widehat A_k$ defined as follows
\begin{equation}\label{second derivative source}
    (\widehat A_k)_{u,v} = 
    \begin{cases}
    2\widehat\beta^k_{u,u}, & \text{if } 1\leq u=v \leq d\\
    \widehat\beta^k_{u,v}, &\text{if } 1 \leq u < v \leq d \\
    \widehat\beta^k_{v,u}, &\text{if} 1 \leq v < u \leq d,
    \end{cases}
\end{equation}
where $\widehat\beta^k_{u,v}$ is the entry in $\widehat\beta^k$ corresponding to the $(u,v)$-th entry of $(\widehat\theta_{\mathcal{T}}^{*,j} - \widehat\theta_{\mathcal{T}})(\widehat\theta_{\mathcal{T}}^{*,j} - \widehat\theta_{\mathcal{T}})^\top$ for $u\leq v$. Note that here we use the MCMC samples generated in the target site, and no additional MCMC is needed in the source site.


Similar to Section~\ref{subsec:target}, we need to estimate the variance of $S_{n,k}$ for CI construction, and we again use a sampling-based approach by perturbing the source objective function. Define a perturbation $M_{n,k}^\dag(\theta)$ of the observed source-site objective function
\begin{align}
    M^\dag_{n,k}(\theta) &= \binom{n_k}{D}^{-1} \sum_{1\leq i_1 < i_2 < \cdots < i_D \leq n_k} W_{k,i_1}W_{k,i_2} \cdots W_{k,i_D} \cdot h\left(x_{k,i_1},\ldots,x_{k,i_D};\theta\right), \label{source perturbation bootstrap} 
\end{align}
where $x_{k,i}$ is the observed value of $X_{k,i}$ and $W_{k,i}$'s are random weights induced by the nonparametric bootstrap, i.e., $(W_{k,1}, \cdots, W_{k, n_k})\sim \mbox{MN}\left(n_k, (n_k^{-1}, \cdots, n_k^{-1})\right).$ 
For any given $\theta$, define 
\begin{align*}
V_k(\theta) = \textnormal{Var}_W[M^\dag_{n,k}(\theta)-M^\dag_{n,k}(\widehat\theta_\mathcal{T})]
\end{align*}
where the expectation and variance are again with respect to the random weights $W_k$'s condition on the observed samples in the source and target sites. If a quadratic approximation similar to \eqref{target boot quadratic} holds in the source site in the sense that
\begin{equation}\label{source boot quadratic}
    M^\dag_{n,k}(\theta) - M^\dag_{n,k}(\widehat\theta_{\mathcal{T}}) \approx S^\dag_{n,k}(\widehat\theta_\mathcal{T})(\theta - \widehat\theta_\mathcal{T}) + \frac{1}{2}(\theta - \widehat\theta_\mathcal{T})^\top A_k(\widehat\theta_{\mathcal{T}}) (\theta - \widehat\theta_\mathcal{T})
\end{equation}
for $\theta$ close to $\widehat\theta_{\mathcal{T}}$, 
\begin{align*}
V_k(\theta)&\approx n_k^{-1}(\theta-\widehat\theta_\mathcal{T})^\top\Sigma_{S,k}^\dag (\theta-\widehat\theta_\mathcal{T}),
\end{align*}
where 
$$ \Sigma_{S,k}^\dag = \textnormal{Var}_W[\sqrt{n_k}S^\dag_{n,k}(\widehat\theta_\mathcal{T})],$$
a consistent estimator of 
$$\Sigma_{S,k} = \textnormal{Var}[\sqrt{n_k}S_{n,k}(\theta_\mathcal{T})],$$ if $\{W_{k,1}, \cdots W_{k, n_k}\}$ were induced from nonparametric bootstrap.
We can again estimate $\Sigma_{S,k}^\dagger$ via a regression. Specifically,  we regress $V_k(\widehat\theta_{\mathcal{T}}^{*,j})$ against the vector $\widehat\Theta_\mathcal{T}^{*,j}$ defined in Section~\ref{subsec:target} for $j \in \mathcal{S}$, and let $\widehat\gamma^k$ denote the estimated regression coefficient. We can then construct an estimator $\widehat\Sigma_{S,k}$ similar to \eqref{target score var}, but with $\widehat\gamma$ and $n_\mathcal{T}$ replaced by $\widehat\gamma^k$ and $n_k$, respectively. 

In summary, the approximations to the score value $S_{n, k}$ and its variance $\Sigma_{S,k}$ can be obtained by evaluating $M_{n,k}(\cdot)$ and $M_{n, k}^\dagger(\cdot)$ at chosen values. The final output of this step of analysis is 
$$(\widehat{S}_{n,k}, \widehat{A}_k, \widehat{\Sigma}_{S,k}).$$ 
There is no need to find the minimizer of $M_{n, k}(\theta)$ or $M_{n, k}^\dagger(\theta)$ in this process, and thus the computation is expected to be fast. 


\begin{remark}\label{non-convex A}
\normalfont We assume that the quadratic approximations in \eqref{eq: source quadratic approx} and \eqref{source boot quadratic} hold in all source sites and this assumption needs to be verified in specific application. For source site $k$ where $\theta_k \neq \theta_\mathcal{T}$, the expected objective function may not be convex near $\theta_\mathcal{T}$ and consequently both $A_k(\widehat\theta_\mathcal{T})$ and its estimator $\widehat A_k$ may not necessarily be positive-definite. As a by-product of this observation, we may check the positive-definiteness of the matrix $\widehat A_k$  to detect ineligible sites, which may introduce bias in transfer (see Section~\ref{subsec: adaptive combination} for more details.)   
\end{remark}

\subsection{Adaptive combination}\label{subsec: adaptive combination}
To prevent ``biased'' transfer, it is important to identify and exclude source sites that have very different parameter value from the target site. To this end, we develop a dissimilarity measure between a source site and the target site resembling the classical score test statistic. We then combine $\widehat\theta_\mathcal{T}$ and $\widehat S_{n,k}$ in a data adaptive manner via $l_1$ regularization, where larger penalty is assigned to sites with larger dissimilarity measures.

With the true site-specific parameter $\theta_k$, the expectation of the score $S_{n,k}(\theta_k)$ is 0. For an eligible source site  $k \in \mathcal{K}$, $\theta_k=\theta_{\mathcal{T}},$ and thus $\widehat\theta_\mathcal{T}$ is centered at $\theta_k$. Consequently, $S_{n,k}(\widehat\theta_\mathcal{T})$ should be centered at 0 as well. This motivates us to define the following dissimilarity measure:
\begin{equation}\label{test statistic}
    T_k = n_k\widehat S_{n,k}(\widehat\theta_\mathcal{T})^\top \left\{\widehat\Sigma_{S,k} + n_kn_\mathcal{T}^{-1}\widehat A_k \widehat A_\mathcal{T}^{-1}\widehat\Sigma_{S,\mathcal{T}}\widehat A_\mathcal{T}^{-1}\widehat A_k\right\}^{-1} \widehat S_{n,k}(\widehat\theta_\mathcal{T})
\end{equation}
for source site $k,$ where $\widehat A_k$ is positive definite. Compared to the classical score test statistic, there is an additional term $\widehat A_k \widehat A_\mathcal{T}^{-1}\widehat\Sigma_{S,\mathcal{T}}\widehat A_\mathcal{T}^{-1}\widehat A_k$ in the variance part. This term accounts for the additional variation in  $\widehat\theta_\mathcal{T}$, which is independent of the variation in the source data. 
A small value of $T_k$ can be viewed as evidence that $k \in {\cal K}.$  Lastly, when $\widehat A_k$ is not positive definite, we set $T_k = +\infty,$ as one would expect to observe a positive definite $\widehat{A}_k$ for $k\in {\cal K}.$ As we will show in Section~\ref{sec:theory}, 
for $k \notin \mathcal{K}$, $T_k$ will converge in probability to infinity as sample size increases.

Next, we construct a linear combination of the scores in the form 
$$\sum_{k=1}^K \Lambda_k \widehat S_{n,k}(\widehat\theta_\mathcal{T})$$ 
to augment $\widehat{\theta}_\mathcal{T}$ in estimating $\theta_{\mathcal{T}},$ where $\Lambda_k \in \mathbb{R}^{d\times d}$ is the weight matrix associated with the $k$-th score. Now, we discuss data-adaptive selection of the weights $\Lambda_k$'s. Specifically, we define the following matrices: 
$$\widehat A_1^K = \left(\begin{array}{c}\widehat A_1\\ \vdots \\ \widehat A_K \end{array}\right) \in \mathbb{R}^{dK\times d}, $$ a concatenation of the matrices $\widehat A_k$'s; 
$$\widehat\Sigma_1^K = n_{\mathcal{T}}\left(\begin{array}{cccc} n_1^{-1}\widehat\Sigma_{S,1} & 0 & \cdots & 0 \\
0 & n_2^{-1}\widehat\Sigma_{S,2} & \cdots & 0\\
0 & 0 &  \cdots & 0\\
0 & 0 & \cdots & n_K^{-1}\widehat\Sigma_{S,K}
\end{array}\right)
\in \mathbb{R}^{dK\times dK},$$ 
a block diagonal matrix with diagonal block $\widehat\Sigma_{S,k}$ scaled by respective sample size. For some large integer $Q$ and $1\leq q \leq Q$, we sample i.i.d. vectors following the normal distribution
\begin{equation*}\label{eq: large normal vector}
    \begin{pmatrix}
        \widehat\theta_\mathcal{T}^{(q)} \\
        \widehat S_{n,1}^{(q)} \\
        \vdots \\
        \widehat S_{n,K}^{(q)}
    \end{pmatrix}
    \stackrel{i.i.d}{\sim}
    N\left(
    \boldsymbol{0}, \widehat\Omega\right), \textnormal{where} \quad \widehat\Omega = 
    \begin{pmatrix}
        \widehat A_\mathcal{T}^{-1}\widehat\Sigma_{S,\mathcal{T}}\widehat A_\mathcal{T}^{-1} & \widehat A_\mathcal{T}^{-1}\widehat\Sigma_{S,\mathcal{T}}\widehat A_\mathcal{T}^{-1} (\widehat A_1^K)^\top \\
        \widehat A_1^K \widehat A_\mathcal{T}^{-1}\widehat\Sigma_{S,\mathcal{T}}\widehat A_\mathcal{T}^{-1} & \widehat\Sigma_1^K + \widehat A_1^K \widehat A_\mathcal{T}^{-1}\widehat\Sigma_{S,\mathcal{T}}\widehat A_\mathcal{T}^{-1}(\widehat A_1^K)^\top 
    \end{pmatrix}.
\end{equation*}
We select the weights by solving a penalized regression problem
\begin{equation}\label{eq: adaptive lasso}
    (\widehat\Lambda_1,\ldots,\widehat\Lambda_K) = \argmin_{\Lambda_k \in \mathbb{R}^{d\times d}, 1\leq k \leq K} \left\{\frac{1}{Q}\sum_{q=1}^Q \left\|\widehat\theta_\mathcal{T}^{(q)} - \sum_{k=1}^K \Lambda_k \widehat S_{n,k}^{(q)}\right\|_2^2 + \lambda\sum_{k=1}^K \frac{1}{p_k}\left\|\Lambda_k\right\|_1 \right\},
\end{equation}
where $p_k = P(\chi^2_d > T_k)$ for $1 \leq k \leq K$, and $\lambda$ is a tuning parameter. The value $p_k$ is analogous to the p-value in a score test, and hence a larger dissimilarity measure $T_k$ leads to a smaller value of $p_k$ and a larger penalty factor $p_k^{-1}$. In \eqref{eq: adaptive lasso}, we compute the mean squared error for predicting the entire vector $\widehat\theta_\mathcal{T}^{(q)}$.  In practice, the optimization in (\ref{eq: adaptive lasso}) can be solved component-wise with different lasso penalty parameters. 
The $\ell_1$-penalty can also be replaced by group-lasso penalty to facilitate the exclusion of ineligible source sites, with all components of the corresponding $\Lambda_k$ being zeros.  Given the weights $\widehat\Lambda_k$'s, we construct the combined estimator $\widehat\theta_C$ for $\theta_\mathcal{T}$ as 
\begin{equation}\label{eq: combined estimate}
    \widehat\theta_C = \widehat\theta_\mathcal{T} - \sum_{k=1}^K \widehat\Lambda_k \widehat S_{n,k},
\end{equation}
and estimate its variance as 
\begin{equation}\label{eq: combined estimate variance}
    \widehat V_C = (\widehat\Lambda_1^K)^\top \widehat\Omega \widehat\Lambda_1^K, \quad \textnormal{where} \quad \widehat\Lambda_1^K = (\textnormal{I}_{p}, -\widehat\Lambda_1,\ldots, -\widehat\Lambda_K)^\top,
\end{equation} 
where $\textnormal{I}_{p}$ is a $p$ by $p$ identity matrix. A Wald-type CI can then be constructed for $\theta_\mathcal{T}$ or a transformation thereof using the delta method. We show in Section~\ref{sec:theory} that the matrix $\widehat\Lambda_k$ will converge to $\boldsymbol{0}$ asymptotically for $k \notin \mathcal{K}$, i.e., ineligible sites would be excluded in estimating $\theta_{\cal T}.$ On the other hand, for $k \in \mathcal{K}$, the matrix $\widehat\Lambda_k$ will converge to the optimal weights in the sense that it minimizes the variance of the combined estimator. The final output of the analysis is 
$$(\widehat{\theta}_C,  \widehat{V}_C).$$

We summarize the full workflow of our proposed inferential procedure in Algorithm~\ref{alg:workflow}.

\begin{algorithm}[ht]
\singlespacing
\caption{Federated sampling-based inference for M-estimators}\label{alg:workflow}
\hspace*{\algorithmicindent} \textbf{Input:} target site data $\{X_{\mathcal{T},1},\ldots,X_{\mathcal{T},n_\mathcal{T}}\}$ and objective function $M_{n,\mathcal{T}}(\cdot)$, source site data $\{X_{k,1},\ldots,X_{k,n_k}\}$ and objective function $M_{n,k}(\cdot)$ for $1\leq k \leq K$, \\
\hspace*{\algorithmicindent} \textbf{Output:} combined estimate $\widehat\theta_C$ for $\theta_\mathcal{T} = \argmin_\theta \EE_\mathcal{T}[M_{n,\mathcal{T}}(\theta)]$ and estimated variance of $\widehat\theta_C$
\begin{algorithmic}[1]
\For {target site}
    \State sample $\{\widehat\theta_{\mathcal{T}}^{*,j}\}_{j=1}^B$ from density $L_\mathcal{T}(\theta) \propto \exp\left\{-n_\mathcal{T}M_{n,\mathcal{T}}(\theta)\right\}\ind{\|\theta\| \leq R}$
    \State compute the sample mean and variance of $\{\widehat\theta_{\mathcal{T}}^{*,j}\}_{j=1}^B$ to obtain $\widehat\theta_\mathcal{T}$ and $\widehat A_\mathcal{T}$
    \State estimate the variance of the score function $\widehat\Sigma_{S,\mathcal{T}}$ via \eqref{target score var} by perturbing the target objective function as in \eqref{target perturbation} and regressing $V_{\mathcal{T}}(\widehat\theta_{\mathcal{T}}^{*,j})$ against $\widehat\Theta_{\mathcal{T}}^{*,j}$ 
    \State broadcast $\{\widehat\theta_{\mathcal{T}}^{*,j}: j \in \mathcal{S}\}$, $\widehat\theta_{\mathcal{T}}$, $\widehat A_\mathcal{T}$ and $\widehat\Sigma_{S,\mathcal{T}}$ to source sites
\EndFor
\For {source site $k$, $k \in \{1,\ldots,K\}$}
    \State obtain the score function and second derivative matrix $\widehat S_{n,k}(\widehat\theta_\mathcal{T})$ and $\widehat A_k$ via \eqref{second derivative source} by regressing $M_{n,k}(\widehat\theta_{\mathcal{T}}^{*,j})-M_{n,k}(\widehat\theta_{\mathcal{T}})$ against $(\widehat\delta_{\mathcal{T}}^{*,j},\widehat\Theta_{\mathcal{T}}^{*,j})$
    \State estimate the variance of the score function $\widehat\Sigma_{S,k}$ by perturbing the source objective function as in \eqref{source perturbation bootstrap} and regressing $V_k(\widehat\theta_{\mathcal{T}}^{*,j})$ against $\widehat\Theta_\mathcal{T}^{*,j}$
    \State send $\widehat S_{n,k}(\widehat\theta_\mathcal{T})$, $\widehat A_k$ and $\widehat\Sigma_{S,k}$ to target site
\EndFor
\For {target site}
    \State construct the dissimilarity measure $T_k$ for $1\leq k \leq K$ as in \eqref{test statistic}
    \State compute the weight matrix $\widehat\Lambda_k$ via penalized regression in \eqref{eq: adaptive lasso}
\EndFor \\
\Return combined estimate $\widehat\theta_C$ as in \eqref{eq: combined estimate} and its estimated variance $\widehat V_C$ in \eqref{eq: combined estimate variance}
\end{algorithmic}
\end{algorithm}

\begin{remark}\label{alternative approach to combine}
\normalfont An alternative approach is to obtain source-site-specific estimators for $\theta_k$ and its variance-covariance matrix via the proposal described in Section~\ref{subsec:target}, and then 
\begin{enumerate}
    \item select eligible source sites through hypothesis testing and combine the corresponding estimates using inverse variance weighting, or
    \item directly combine the source- and target-estimates through heterogeneity-aware approaches such as \citet{guo2023robust}.
\end{enumerate} 
 The major downside of this approach is the potential numerical difficulty in computing $\widehat\theta_k$ and its variance in a source site. 
\end{remark}

\section{Theoretical Results}\label{sec:theory}


\subsection{Theoretical properties of the target site estimator}\label{subsec: theory target}

We now formalize the heuristic arguments in Section~\ref{sec:method} and present theoretical guarantees for our proposed procedure. We first introduce some more notations. For a generic distribution $P$ and a generic function $f$ of $(X_1,\ldots,X_D)$ indexed by parameter $\theta$, we define $Pf(\theta) = \EE_P[f(X_1,\ldots,X_D;\theta)]$. We use $P_\mathcal{T}$ to denote the distribution of $X_{i,\mathcal{T}}$ in the target site and $P_k$ to denote the distribution of $X_{i,k}$ in the source site. Suppose that there exists a ``score'' function $s_{\mathcal{T}}(\cdot;\theta)$ with $P_\mathcal{T}\|s_\mathcal{T}(\theta)\|^2 = \mathbb{E}_{P_\mathcal{T}}\|s_\mathcal{T}(X_{\mathcal{T},1},\ldots,X_{\mathcal{T},D}; \theta)\|^2 < \infty$ such that 
\begin{equation} \label{eq: target score}
    P_\mathcal{T}s_\mathcal{T}(\theta) = \frac{\partial\{P_\mathcal{T}h(\theta)\}}{\partial \theta},
\end{equation}
and $P_\mathcal{T}s_\mathcal{T}(\theta_\mathcal{T}) = 0$, where $h$ is the kernel of the U-statistic in \eqref{eq: target objective}. When $h$ is smooth in $\theta$ for almost every $(X_1,\ldots,X_D)$, we can typically take $s_\mathcal{T}$ as the gradient of $h$ with respect to $\theta$. However, even if $h$ is not differentiable everywhere, it is still possible to construct a ``score'' function $s_\mathcal{T}$ to satisfy (\ref{eq: target score}) as in Section 4.  Suppose that $P_\mathcal{T}s_\mathcal{T}(\theta)$ is differentiable with respect to $\theta$ and define the matrix $A_\mathcal{T}(\theta)$ as
\begin{equation}
    A_\mathcal{T}(\theta) = \frac{\partial P_\mathcal{T}s_\mathcal{T}(\theta)}{\partial\theta}=\frac{\partial^2 P_\mathcal{T}h(\theta)}{\partial\theta^2}.
\end{equation}
For simplicity, we write $A_\mathcal{T}(\theta_\mathcal{T})$ as $A_\mathcal{T}$. Let $S_{n,\mathcal{T}}(\theta)$ be the $U$-statistic with kernel $s_\mathcal{T}$, that is,
\begin{equation*}
    S_{n,\mathcal{T}}(\theta) = \binom{n_\mathcal{T}}{D}^{-1} \sum_{1\leq i_1 < i_2 < \cdots < i_D \leq n_\mathcal{T}} s_\mathcal{T}(X_{\mathcal{T},i_1},\ldots,X_{\mathcal{T},i_D};\theta).
\end{equation*}
The functions $s_{\mathcal{T}}$ and $S_{n,\mathcal{T}}$ are analogous to the score function in maximum likelihood estimation. In Section~\ref{sec:example} we will show that $s_\mathcal{T}$ and $A_\mathcal{T}$ exist in various examples even if the kernel $h$ is non-smooth. Similar to \eqref{target perturbation}, we also define perturbation of $S_{n,\mathcal{T}}$ as follows
\begin{equation*}
    S^\dagger_{n,\mathcal{T}}(\theta) = \binom{n_\mathcal{T}}{D}^{-1} \sum_{1\leq i_1 < i_2 < \cdots < i_D \leq n_\mathcal{T}} W_{\mathcal{T},i_1} W_{\mathcal{T},i_2} \cdots W_{\mathcal{T},i_D} \cdot s_\mathcal{T}(X_{\mathcal{T},i_1},\ldots,X_{\mathcal{T},i_D};\theta),
\end{equation*}
with weights $W_\mathcal{T}$ defined below \eqref{target perturbation}.

First, we show that under suitable conditions, the target site estimator $\widehat\theta_\mathcal{T}$ constructed using the MCMC samples is asymptotically normal, and its variance can be approximated by 
$n_\mathcal{T}^{-1}\widehat{A}_\mathcal{T}^{-1}\widehat\Sigma_{S,\mathcal{T}}\widehat{A}_\mathcal{T}^{-1}.$

\begin{condition}[local quadratic approximation in target site]\label{cond:target quad}
Let $$R_n(\theta_1,\theta_2) = M_{n,\mathcal{T}}(\theta_1) - M_{n,\mathcal{T}}(\theta_2) - S_{n,\mathcal{T}}(\theta_2)(\theta_1 - \theta_2) - \frac{1}{2}(\theta_1 - \theta_2)^\top A_\mathcal{T}(\theta_1 - \theta_2).$$ 
Let $\{\delta_{n_\mathcal{T}}\}$ be some deterministic sequence such that (a) $\delta_{n_\mathcal{T}} \rightarrow 0$ as $n_\mathcal{T} \rightarrow \infty$, and (b) $n_\mathcal{T}^{1/2}\delta_{n_\mathcal{T}} \rightarrow \infty$ as $n_\mathcal{T} \rightarrow \infty$. 
For any $\epsilon > 0$, there exists $\delta > 0$ such that
\begin{equation}
    \lim_{n_\mathcal{T} \rightarrow \infty} P\left(\sup_{ \|\theta_1 - \theta_2\|_2 \leq \delta, \ \|\theta_2 - \theta_\mathcal{T}\|_2 \leq \delta_{n_\mathcal{T}} } \frac{|R_n(\theta_1, \theta_2)|}{\|\theta_1 - \theta_2\|_2^2 + n_\mathcal{T}^{-1}} > \epsilon \right) = 0.
\end{equation}
\end{condition}

\begin{condition}[identifiability, tail behavior of the target objective]\label{cond:target tail}
For any $\delta>0$, there exists $\epsilon>0$, such that
\begin{equation}
    \lim_{n_\mathcal{T}\rightarrow \infty} P\left(\inf_{\|\theta-\tilde\theta_\mathcal{T}\|_2\geq \delta} \left\{M_{n,\mathcal{T}}(\theta) - M_{n,\mathcal{T}}(\tilde\theta_\mathcal{T})\right\} \geq \epsilon \right) = 1,
\end{equation}
where $\tilde\theta_\mathcal{T}$ is the minimizer of $M_{n,\mathcal{T}}(\theta)$.
\end{condition}

\begin{condition}[local quadratic approximation of perturbed objective function]\label{cond:target perturb quad} Let 
$$R_n^\dag(\theta_1,\theta_2) = M_{n,\mathcal{T}}^\dag(\theta_1) - M_{n,\mathcal{T}}^\dag(\theta_2) - S_{n,\mathcal{T}}^\dag(\theta_2)(\theta_1 - \theta_2) - \frac{1}{2}(\theta_1 - \theta_2)^\top A_\mathcal{T}(\theta_1 - \theta_2).$$ 
Let $\{\delta_{n_\mathcal{T}}\}$ be some deterministic sequence such that (a) $\delta_{n_\mathcal{T}} \rightarrow 0$ as $n_\mathcal{T} \rightarrow \infty$, and (b) $n_\mathcal{T}^{1/2}\delta_{n_\mathcal{T}} \rightarrow \infty$ as $n_\mathcal{T} \rightarrow \infty$. For any $\epsilon > 0$,
    \begin{equation*}
    \lim_{n_\mathcal{T} \rightarrow \infty} P\left(\sup_{\|\theta_1 - \theta_2\|_2 \leq \delta_{n_\mathcal{T}}, \ \|\theta_2 - \theta_\mathcal{T}\|_2 \leq \delta_{n_\mathcal{T}}} \frac{\mathbb{E}_W[R_n^\dagger(\theta_1, \theta_2)^2]}{\{\|\theta_1 - \theta_2\|_2^2 + n_\mathcal{T}^{-1}\}^2} > \epsilon \right) = 0.
    \end{equation*}
\end{condition}

\noindent Condition~\ref{cond:target quad} assumes that the target site objective function is well-approximated by a quadratic function near the true parameter value $\theta_\mathcal{T}$. A similar but stronger condition was assumed in \citet{jin2001simple}, where almost sure convergence was required. A similar condition was also assumed in \citet{chernozhukov2003mcmc} (see Assumption 4 and proof of Proposition 1 therein.) Condition~\ref{cond:target quad} can be shown to hold for many objective functions of practical interest, and some concrete examples will be given in Section~\ref{sec:example}. Condition~\ref{cond:target tail} assumes that with probability approaching 1, for any parameter value $\theta$ that is bounded away from $\tilde\theta_\mathcal{T}$, the minimizer of the objective function $M_{n,\mathcal{T}}(\theta)$, $M_{n,\mathcal{T}}(\theta)$ is also bounded away from $M_{n,\mathcal{T}}(\tilde\theta_\mathcal{T})$. This essentially requires that the $M$-estimator $\tilde\theta_\mathcal{T}$ is an isolated minimizer of the objective function. A similar condition was used in \citet{chernozhukov2003mcmc}, which is closely related to standard identifiability condition in $M$-estimation \citep[see, for example,][]{van2000asymptotic}. Condition \ref{cond:target perturb quad} is similar to Condition~\ref{cond:target quad} in that it assumes that the perturbed objective function is approximately quadratic. Specifically, it requires that the second moment of the remainder in this quadratic approximation with respect to the weights $W$ is negligible.  

Under these conditions, the following theorem establishes the asymptotic properties of $\widehat\theta_\mathcal{T}$.

\begin{theorem}[asymptotic normality of target site estimator]\label{thm:normality target estimator} 
Suppose that Conditions~\ref{cond:target quad} and \ref{cond:target tail} hold and that $A_\mathcal{T}$ is positive definite. In addition, suppose that the $M$-estimator $\tilde{\theta}_\mathcal{T}$ is asymptotically Gaussian such that 
$$\sqrt{n_\mathcal{T}}(\tilde{\theta}_\mathcal{T} - \theta_\mathcal{T}) \xrightarrow[]{d} N(0, A_\mathcal{T}^{-1}\Sigma_\mathcal{T}A_\mathcal{T}^{-1})$$
as $n_\mathcal{T} \rightarrow \infty$, where $ \Sigma_\mathcal{T}=\lim_{n_\mathcal{T}\rightarrow \infty}\textnormal{Var}[\sqrt{n_\mathcal{T}}S_{n,\mathcal{T}}(\theta_\mathcal{T})].$ Then, the target site resampling estimator $\widehat\theta_\mathcal{T}$ is also asymptotically normal
\begin{equation}
    \sqrt{n_\mathcal{T}}(\widehat{\theta}_\mathcal{T} - \theta_\mathcal{T}) \xrightarrow[]{d} N\left(0, A_\mathcal{T}^{-1}\Sigma_\mathcal{T} A_\mathcal{T}^{-1}\right),
\end{equation}
as $n_\mathcal{T} \rightarrow \infty.$ Moreover, if Condition~\ref{cond:target perturb quad} also holds, then $$\widehat{A}_\mathcal{T}^{-1}\widehat\Sigma_{S,\mathcal{T}}\widehat{A}_\mathcal{T}^{-1} =  A_\mathcal{T}^{-1}\Sigma_\mathcal{T}A_\mathcal{T}^{-1}+o_p(1),$$ 
that is, the proposed variance estimator is consistent.
\end{theorem}

There has been a rich body of literature that establishes the asymptotic normality of $M$-estimators, even in cases where the objective functions may not be smooth \citep[see, for example,][]{sherman1993limiting}. Theorem~\ref{thm:normality target estimator} states that, under conditions, the MCMC-based estimator $\widehat\theta_\mathcal{T}$ is asymptotically equivalent to the $M$-estimator $\tilde\theta_\mathcal{T}$. Furthermore, the MCMC step combined with the additional layer of perturbation provides an effective way to consistently estimate its asymptotic variance. We can then readily perform statistical inference using data from the target site alone.

\subsection{Behavior of the dissimilarity measure $T_k$}
Similar to Section~\ref{subsec: theory target}, suppose that for all $k$ in $\{1,\ldots,K\}$ there exists a function $s_k(\cdot;\theta)$ with $P_k\|s_k(\theta)\|^2 = \EE_{P_k}\|s_k(X_{k,1},\ldots,X_{k,D}; \theta)\|^2 < \infty$ such that 
\begin{equation}
    P_ks_k(\theta) = \frac{\partial P_kh(\theta)}{\partial \theta},
\end{equation}
and $P_ks_k(\theta_k) = 0$. Suppose that $P_ks_k(\theta)$ is differentiable with respect to $\theta$ and define the matrix $A_k(\theta)$ as
\begin{equation}
    A_k(\theta) = \frac{\partial P_ks_k(\theta)}{\partial\theta}.
\end{equation}
Let 
$$S_{n,k}(\theta) = \binom{n_k}{D}^{-1} \sum_{1\leq i_1 < i_2 < \cdots < i_D \leq n_k} s_k(X_{k,i_1},\ldots,X_{k,i_D};\theta),$$ and 
$$S_{n,k}^\dag(\theta) = \binom{n_k}{D}^{-1} \sum_{1\leq i_1 < i_2 < \cdots < i_D \leq n_k} W_{k,i_1} W_{k,i_2} \cdots W_{k,i_D} s_k(X_{k,i_1},\ldots,X_{k,i_D};\theta).$$

In this subsection, we establish the asymptotic property of the dissimilarity measure $T_k$ defined in \eqref{test statistic}. We first introduce the relevant conditions. 

\begin{condition}[local quadratic approximation in source site]\label{cond:source quad}
Let 
$$R_{n,k}(\theta_1,\theta_2) = M_{n,k}(\theta_1) - M_{n,k}(\theta_2) - S_{n,k}(\theta_2)(\theta_1 - \theta_2) - \frac{1}{2}(\theta_1 - \theta_2)^\top A_k(\theta_\mathcal{T})(\theta_1 - \theta_2).$$
Let $\{\delta_{n_k}\}$ be some deterministic sequence such that (a) $\delta_{n_k} \rightarrow 0$ as $n_k \rightarrow \infty$, and (b) $n_k^{1/2}\delta_{n_k} \rightarrow \infty$ as $n_k \rightarrow \infty$. For any $k \in \{1,\ldots,K\}$ and $\epsilon > 0$,
\begin{equation}
    \lim_{n_k \rightarrow \infty} P\left(\sup_{\|\theta_1 - \theta_2\|_2 \leq \delta_{n_k}, \ \|\theta_2 - \theta_\mathcal{T}\|_2 \leq \delta_{n_k}} \frac{|R_{n,k}(\theta_1, \theta_2)|}{\|\theta_1 - \theta_2\|_2^2 + n_k^{-1}} > \epsilon \right) = 0.
\end{equation}
\end{condition}

\begin{condition}[local quadratic approximation of perturbed objective function in source site]\label{cond:source perturb quad} Let 
$$R_{n,k}^\dag(\theta_1,\theta_2) = M_{n,k}^\dag(\theta_1) - M_{n,k}^\dag(\theta_2) - S_{n,k}^\dag(\theta_2)(\theta_1 - \theta_2) - \frac{1}{2}(\theta_1 - \theta_2)^\top A_k(\theta_\mathcal{T})(\theta_1 - \theta_2).$$ 
Let $\{\delta_{n_k}\}$ be some deterministic sequence such that (a) $\delta_{n_k} \rightarrow 0$ as $n_k \rightarrow \infty$, and (b) $n_k^{1/2}\delta_{n_k} \rightarrow \infty$ as $n_k \rightarrow \infty$. For any $k \in \{1,\ldots,K\}$ and $\epsilon > 0$
    \begin{equation*}
    \lim_{n_k \rightarrow \infty} P\left(\sup_{\|\theta_1 - \theta_2\|_2 \leq \delta_{n_k}, \ \|\theta_2 - \theta_\mathcal{T}\|_2 \leq \delta_{n_k}} \frac{\mathbb{E}_W[\{R_{n,k}^\dagger(\theta_1, \theta_2)\}^2]}{\{\|\theta_1 - \theta_2\|_2^2 + n_k^{-1}\}^2} > \epsilon \right) = 0.
    \end{equation*}
\end{condition}

\begin{condition}[continuity of second derivative matrix]\label{cond:continuity of A}
For all $k \in \{1,\ldots,K\}$, $A_k(\theta)$ is continuous at $\theta_\mathcal{T}$ as a function of $\theta$ and positive definite for $k \in \mathcal{K}.$
\end{condition}

\begin{condition}[stochastic equi-continuity]\label{cond:empirical process} 
For $1 \leq k \leq K$, 
$$S_{n,k}(\widehat\theta_\mathcal{T}) - S_{n,k}(\theta_\mathcal{T}) - P_ks_k(\widehat\theta_\mathcal{T}) + P_ks_k(\theta_\mathcal{T}) = o_P(n_k^{-1/2}).$$    
\end{condition}

\begin{condition}[sample size ratio]\label{cond:sample size ratio}
Let $\rho_k = \lim_{n_{\mathcal{T}} \rightarrow \infty} n_k/n_\mathcal{T}$. There exist finite constants $0<m_1<m_2<+\infty$ such that $m_1 < \rho_k < m_2$ for all $1 \leq k \leq K$.
\end{condition}

\begin{condition}[identifiability of ineligible sites]\label{cond: spurious root}
    Either $P_ks_k(\theta_\calT) \neq 0$ or $A_k(\theta_\calT)$ is not positive definite for any  $k \notin \mathcal{K}.$ 
\end{condition}

\noindent Conditions~\ref{cond:source quad} and \ref{cond:source perturb quad} assume that the source site objective function and the corresponding perturbation are approximately quadratic locally near $\theta_\mathcal{T}$. This pair of conditions is similar to Conditions~\ref{cond:target quad} and \ref{cond:target perturb quad}. Note that for eligible sites, this is equivalent to a local quadratic approximation near $\theta_k$ as $\theta_k = \theta_\mathcal{T}$. These conditions can be established in various examples of interest (see Section~\ref{sec:example} for concrete examples.) Condition~\ref{cond:continuity of A} assumes that the second order derivative matrix is continuous in the parameter $\theta$, which holds when the expectation of the objective function is sufficiently smooth. Condition~\ref{cond:empirical process} controls the remainder term in the asymptotic expansion of $S_{n,k}(\widehat\theta_\mathcal{T})$. When $D=1$, $S_{n,k}$ is a sample average, and Condition~\ref{cond:empirical process} can be established using appropriate results of empirical process such as Lemma 19.24 in \cite{van2000asymptotic}. 
When $D>1$, similar results hold for U-processes under suitable regularity conditions (see, for example, \cite{sherman1993limiting}.) Condition~\ref{cond:sample size ratio} assumes that the sample sizes in the source and target sites approach infinity at the same rate. Condition~\ref{cond: spurious root} assumes that $\theta_\mathcal{T}$ can not be a local minimizer of the objective function in an ineligible site, as otherwise we cannot distinguish $\theta_\calT$ from $\theta_k$ based on the values of the score function alone. \citet{freedman2007can} discusses similar issues in terms of consistency of the score test in likelihood-based inference. 

\begin{theorem}[asymptotic distribution of dissimilarity measure $T_k$]\label{thm: test consistency}
Suppose that Conditions~\ref{cond:target quad} to \ref{cond: spurious root} hold. Then, as $n_k, n_\mathcal{T} \rightarrow \infty$, 
$$T_k \xrightarrow[]{p} \begin{cases} +\infty, \mbox{ if } k \notin \mathcal{K}\\  \chi^2_d, \mbox{ if } k \in \mathcal{K}\end{cases}.$$
\end{theorem}

Theorem~\ref{thm: test consistency} implies that under appropriate conditions, the dissimilarity measure stays bounded in probability when $\theta_k = \theta_\mathcal{T}$ but diverges to $+\infty$ when $\theta_k \neq \theta_\mathcal{T}$. As a consequence, the penalty factors used for constructing $\Lambda_k$ in (\ref{eq: adaptive lasso}) are much larger for ineligible sites compared to those for eligible sites, which helps prevent negative transfer.

\subsection{Oracle property and efficiency of the combined estimator}
Recall that given the estimated scores $\widehat S_{n,k}$, we use the linear combination \eqref{eq: combined estimate} as the combined estimator where the coefficients $\widehat\Lambda_k$ are obtained via the penalized regression \eqref{eq: adaptive lasso}. The following lemma characterizes the properties of these estimated coefficients $\widehat\Lambda_k$. Before stating the result, we first define an oracle coefficient $\Lambda_k^{opt}$ for $k \in \mathcal{K}$
\begin{equation}
    \Lambda_k^{opt} = \left(A_\mathcal{T}\Sigma_{S,\mathcal{T}}^{-1}A_\mathcal{T} + \sum_{k \in \mathcal{K}} \rho_k A_k\Sigma_{S,k}^{-1}A_k\right)^{-1} \rho_k A_k\Sigma_{S,k}^{-1}.
\end{equation}

\begin{lemma}\label{adaptive lasso solution}
Suppose that Conditions~\ref{cond:target quad} to \ref{cond: spurious root} hold. If $\lambda = n_\calT^{-\alpha}$ for any $\alpha > 0$ as $n_k, n_\mathcal{T},Q \rightarrow \infty$, then $P(\widehat\Lambda_k = \boldsymbol{0}) \rightarrow 1$ for all $k \notin \mathcal{K}$ and $\widehat\Lambda_k \xrightarrow[]{p} \Lambda_k^{opt}$ for $k \in \mathcal{K}$. 
\end{lemma}

Lemma~\ref{adaptive lasso solution} implies that for sufficiently large sample sizes, the coefficients associated with the scores from ineligible sites are set to 0 with high probability, whereas for eligible sites the coefficients converge to the optimal weights $\Lambda_k^{opt},$ which maximizes the efficiency of the combined estimator. Hence the combined estimator $\widehat\theta_C$ enjoys the ``oracle" property in that it has the same asymptotic distribution as an oracle estimator having prior knowledge of $\mathcal{K}$. This is formalized in the following theorem.

\begin{theorem}[asymptotic normality and efficiency of combined estimator]\label{thm: asymptotic normality of combined} Suppose that Conditions~\ref{cond:target quad} to \ref{cond: spurious root} hold. Then, 
$$\sqrt{n_\mathcal{T}}(\widehat\theta_C - \theta_\mathcal{T}) \xrightarrow[]{d} N\left\{0, \left(A_\mathcal{T}\Sigma_{S,\mathcal{T}}^{-1}A_\mathcal{T} + \sum_{k\in \mathcal{K}} \rho_k A_k\Sigma_{S,k}^{-1}A_k\right)^{-1}\right\}$$
in distribution, as $n_\mathcal{T}, n_k, Q \rightarrow \infty.$
\end{theorem}

\noindent Again, the asymptotic variance of $\widehat\theta_C$ is the same as the inverse-variance-weighted estimator combining the $M$-estimators fitted locally in each eligible source site and the target site, which is the most efficient among all possible linear combinations. 

\section{Examples}\label{sec:example}
In this section, we give two examples on how the proposed procedure can be used, namely quantile regression and AUC maximization. We will verify the conditions required in Section~\ref{sec:theory} in these examples.

\subsection{Quantile regression}\label{subsec:quantile example}
Quantile regression estimates the conditional quantile of an outcome variable given a set of covariates, and it can be more robust to outliers compared to OLS. Let $X_k=(Z_k^\top, Y_k)^\top$ be an observation from site $k \in \{\mathcal{T}\} \cup \{1,\ldots,K\}$, where $Y_k \in \mathbb{R}$ is the outcome variable of interest and $Z_k \in \mathbb{R}^d$ is a $d$-dimensional covariate vector. Let 
$$F_{k,Y \mid Z}(y) = P_k(Y \leq y \mid Z) ~~\mbox{and}~~ Q_{k,Y\mid Z}(\tau) = \inf \{y: F_{k,Y|Z}(y)\geq \tau\}$$ 
be the conditional cumulative distribution function and the conditional $\tau$-th quantile, respectively, for site $k \in \{\mathcal{T}\} \cup \{1,\ldots,K\}.$ Throughout this subsecton, we assume that the conditional distribution $Y \mid Z$ in site $k$ is absolute continuous with a density function $p_{k,Y \mid Z}(\cdot)$ with respect to the Lebesgue measure, for $k \in \{\mathcal{T}\} \cup \{1,\ldots,K\}$. In quantile regressions, we assume that $$Q_{\mathcal{T},Y\mid Z}(\tau) = \beta_\mathcal{T}^\top Z$$ for some unknown vector $\beta_\mathcal{T} \in \mathbb{R}^d$. 

To make inference for quantile regression, we define the check function  
$$\xi_\tau(u) = u \left(\tau - \ind{u<0} \right).$$ 
It can be shown that $\beta_\mathcal{T}$ is the minimizer of the population objective function 
$$M_\mathcal{T}(\beta) = P_\mathcal{T}[\xi_\tau(Y_\mathcal{T} - \beta^\top Z_\mathcal{T})].$$ 
It is worth noting that even if the conditional quantile function is non-linear, $\beta_\mathcal{T}$ can still be defined as the minimizer of this objective function. We can minimize the following objective to estimate $\beta_\mathcal{T}$ in the target site. 
\begin{equation}\label{eq: quantile regression objective}
    M_{n,\mathcal{T}}(\beta) = n_{\mathcal{T}}^{-1}\sum_{i=1}^{n_\mathcal{T}} \xi_\tau(Y_{\mathcal{T},i} - \beta^\top Z_{\mathcal{T},i}).
\end{equation}
This objective function is simply an empirical average ($D=1$ in this example), but the presence of the indicator function makes it not differentiable with respect to $\beta$ at certain values. Similarly, let $\beta_k \in \mathbb{R}^d$ be the minimizer of $M_k(\beta) = P_k[\xi_\tau(Y_k - \beta^\top Z_k)]$ in the $k$-th source site, which can be estimated by minimizing the objective function in the source site
\begin{equation}
    M_{n,k}(\beta) = n_k^{-1}\sum_{i=1}^{n_k} \xi_\tau(Y_{k,i} - \beta^\top Z_{k,i}).
\end{equation}

The score functions $s_\mathcal{T}(\cdot;\beta)$ and $s_k(\cdot;\beta)$ can be chosen as $$s_\mathcal{T}(x;\beta) = s_k(x;\beta) = z\left(\ind{y < \beta^\top z} - \tau\right),$$
where $x=(z^\top, y)^\top$, Clearly, 
$$S_{n,\mathcal{T}} = n_\mathcal{T}^{-1}\sum_{i=1}^{n_\mathcal{T}}s_\mathcal{T}(Z_{\mathcal{T},i},Y_{\mathcal{T},i};\beta),$$ 
and
$$S_{n,k} = n_k^{-1}\sum_{i=1}^{n_k}s_k(Z_{k,i},Y_{k,i};\beta).$$ 
Furthermore, the second-derivative matrix $A_\mathcal{T}$ and $A_k$ are given by 
$$A_\mathcal{T}(\beta) = P_\mathcal{T}[p_{\mathcal{T},Y \mid Z}(\beta^\top Z)ZZ^\top]~\mbox{ and }~ A_k(\beta) = P_k[p_{k,Y\mid Z}(\beta^\top Z)ZZ^\top],$$
respectively. The following lemma establishes conditions to ensure the validity of Algorithm 1.

\begin{lemma}\label{lemma: quad quantile regression}
    Conditions~\ref{cond:target quad} to \ref{cond:empirical process} hold if the following set of assumptions hold for all sites $k \in \{\mathcal{T}\} \cup \{1,\ldots,K\}$:
    \begin{enumerate}[label=(\alph*)]
        \item $Z_k$ has finite second moment under $P_k$, that is $P_k[\|Z_k\|_2^2] < \infty$;
        \item for all $\beta_1$ and $\beta_2$ in a neighborhood of $\beta_\mathcal{T}$, $\|A_k(\beta_1) - A_k(\beta_2)\| \leq L\|\beta_1 - \beta_2\|_2$ for some $L>0$;
        \item there exists a function $h(z)$ such that $p_{k,Y \mid Z}(y \mid z) \leq h(z)$ for all $y$ and $P_k\left[\|Z_k\|_2^qh(Z_k)\right] < \infty$ for $1 \leq q \leq 5$.  
    \end{enumerate}
\end{lemma}

Assumption (b) is a Lipschitz continuity assumption on $A_\mathcal{T}(\beta)$ and $A_k(\beta)$ for $1\leq k \leq K$. Assumption (c) assumes that the conditional density of $p_{k,Y \mid Z}(y \mid z)$ is upper bounded by a suitably integrable function $h(z)$ uniformly in $y$. This assumption holds, for example, under a linear model $Y_k = Z_k^\top \beta_k + \epsilon_k,$ where $\epsilon_k$ has a bounded density function and $Z_k$ has a sufficient number of finite moments. With all these ingredients, we can obtain a consistent estimate of $\beta_\mathcal{T}$ and construct a corresponding 95\% CI borrowing information from source sites via Algorithm~\ref{alg:workflow}, where we take $\theta = \beta$.

\subsection{AUC maximization}\label{subsec:AUC example}
Let $X_\mathcal{T} = (Z_\mathcal{T}^\top, Y_\mathcal{T})^\top$ be an observation in the target site, where $Z_\mathcal{T} \in \mathbb{R}^{d+1}$ is a vector of predictors and $Y_\mathcal{T} \in \{0,1\}$ is a binary outcome. One approach \citep{pepe2006combining} of combining the predictors for classification is to find a linear combination $\beta^\top Z_\mathcal{T}$ that directly maximizes the area under the ROC curve (AUC). However, only the direction of $\beta$ is identifiable but not its magnitude. A commonly used strategy is to estimate the last $d$ components of $\beta$, which we denote as $\theta \in \mathbb{R}^d$, while placing a constraint on $\beta$ such as the first component of $\beta$ is 1 or the norm of $\beta$ is 1. For notational simplicity, we use $\beta(\theta)$ to denote the full vector of coefficients. In this case, maximizing the AUC is equivalent to minimizing the following objective function
\begin{equation}\label{eq: AUC objective}
    M_{n,\mathcal{T}}(\theta) = \frac{1}{n_\mathcal{T}(n_\mathcal{T}-1)}\sum_{i \neq j} \ind{Y_{\mathcal{T},i} > Y_{\mathcal{T},j}} \ind{Z_{\mathcal{T},i}^\top \beta(\theta) \leq Z_{\mathcal{T},j}^\top \beta(\theta) },
\end{equation}
and correspondingly we define the target parameter as $\theta_\mathcal{T} = \argmin_\theta P_\mathcal{T}[M_{n,\mathcal{T}}(\theta)]$. One important advantage of directly maximizing AUC compared to fitting a GLM is that the $M$-estimator associated with \eqref{eq: AUC objective} is agnostic to the link function. Moreover, although we have focused on binary outcomes here, our $M$-estimator is a special case of the maximum rank correlation estimator, which is broadly defined with the objective $M_{n,\mathcal{T}}(\theta)$ for continuous outcomes as well. Note that the presence of the indicator functions makes the objective function highly non-smooth. The source objective function and the source parameter are defined in the same way.
\begin{equation*}
    M_{n,k}(\theta) = \frac{1}{n_k(n_k-1)}\sum_{i \neq j} \ind{Y_{\mathcal{T},i} > Y_{\mathcal{T},j}} \ind{Z_{\mathcal{T},i}^\top \beta(\theta) \leq Z_{\mathcal{T},j}^\top \beta(\theta) }, \quad \theta_k = \argmin_\theta P_k[M_{n,k}(\theta)].
\end{equation*}

Before presenting the score functions and the second-derivative matrices, we introduce some more notations. First note that the objective functions $M_{n,\mathcal{T}}(\theta)$ and $M_{n,k}(\theta)$ are U-statistics of degree 2 with the kernel function $h(\cdot,\cdot;\theta)$ such that
\begin{equation*}
    h(x_1,x_2;\theta) = \frac{1}{2} \left[ \ind{y_1 > y_2} \ind{z_1^\top \beta(\theta) \leq z_2^\top \beta(\theta)} + \ind{y_2 > y_1} \ind{z_2^\top \beta(\theta) \leq z_1^\top \beta(\theta)} \right].
\end{equation*}
Define a function $\zeta_k(\cdot;\theta)$ for $k \in \{\mathcal{T}\} \cup \{1,\ldots,K\}$ such that
\begin{equation*}
    \zeta_k(x;\theta) = \frac{1}{2}P_k\left[\ind{y > Y_k} \ind{z^\top\beta(\theta) \leq Z_k^\top \beta(\theta)}\right] + \frac{1}{2}P_k\left[\ind{Y_k > y} \ind{Z_k^\top\beta(\theta) \leq z^\top \beta(\theta)}\right],
\end{equation*}
which can be succinctly written as $P_k h(x,\cdot;\theta)$ or $P_k h(\cdot,x;\theta)$. With these preparations, the following lemma establishes conditions to ensure the validity of Algorithm 1.

\begin{lemma}\label{lemma: quad auc maximization}
    Conditions~\ref{cond:target quad} to \ref{cond:empirical process} hold if the following assumptions hold for all sites $k \in \{\mathcal{T}\} \cup \{1,\ldots,K\}$:
    \begin{enumerate}[label=(\alph*)]
        \item all second-order partial derivatives of $\zeta_k(x;\theta)$ with respect to $\theta$ exist in a neighborhood of $\theta_\mathcal{T}$ for all $x$, and are square-integrable;
        \item $\|\nabla_2\zeta_k(x;\theta_1) - \nabla_2\zeta_k(x;\theta_2)\| \leq L(x)\|\theta_1 - \theta_2\|_2$ for some $P_k$-square-integrable function $L(x)$ and $\theta_1$ and $\theta_2$ in a neighborhood of $\theta_\mathcal{T}$, where $\nabla_q\zeta_k(x;\theta)$ denotes the $q$-th partial derivative of $\zeta_k$ with respect to $\theta$ for $q=1,2$;
        \item $P_k\left\{ \nabla_2 \zeta_k(\cdot;\theta_\calT) \right\}$ is positive definite for $k \in \{\calT\}\cup \mathcal{K}$;
        \item the class of functions $\{\nabla_1\zeta_k(\cdot;\theta): \|\theta - \theta_\calT\| \leq \delta\}$ is Donsker for some $\delta>0$ with an integrable envelope function $G(\cdot)$;
        \item the support of $Z_k$ is not contained in any proper linear subspace of $\mathbb{R}^{d+1}$, and there is at least one component of $Z_k$ that has an everywhere positive Lebesgue density conditional on the other components.
    \end{enumerate}
\end{lemma}

Under the assumption of Lemma~\ref{lemma: quad auc maximization}, the score functions in the target and source sites are the U-statistics corresponding to kernel: $s_\mathcal{T}(x_1,x_2;\theta) = \nabla_1 \zeta_\mathcal{T}(x_1;\theta) +\nabla_1 \zeta_\mathcal{T}(x_2;\theta) - P_\mathcal{T}\nabla_1\zeta_\mathcal{T}(\cdot;\theta)$ and $s_k(x_1,x_2;\theta) = \nabla_1 \zeta_k(x_1;\theta) +\nabla_1 \zeta_k(x_2;\theta) - P_k\nabla_1\zeta_k(\cdot;\theta)$, respectively. The matrices $A_\mathcal{T}(\theta)$ and $A_k(\theta)$ are given by $P_\mathcal{T}\nabla_2 \zeta_\mathcal{T}(\cdot;\theta)$ and $P_k\nabla_2 \zeta_k(\cdot;\theta)$, respectively.  When the outcome follows a correctly specified generalized regression model, $\zeta_k(\cdot;\theta)$ is a convex function of $\theta$, but its derivatives can have very complicated analytic forms \citep{sherman1993limiting}. More importantly, we oftentimes want to avoid to need of specifying the conditional distribution of $Y_k\mid Z_k.$ Fortunately, Algorithm~\ref{alg:workflow} provides a practical way to conduct statistical inference on $\theta_\mathcal{T}$ borrowing information from source sites without knowing the conditional distribution and relevant derivatives. 

\section{Numerical results}\label{sec:experiments}

\subsection{Simulation study}\label{subsec:simulation}
In this subsection, we investigate the performance of our proposed method in a set of simulation studies. We compare our approach with two alternatives that uses only the data from the target site and naively pools data from all sites, respectively. Specifically, we use $\widehat\theta_\calT$ defined in \eqref{eq:target-only estimator} as the target-only estimator and construct a corresponding Wald CI with the variance estimated by $n_\mathcal{T}^{-1}\widehat{A}_\mathcal{T}\widehat\Sigma_{S,\mathcal{T}}\widehat{A}_\mathcal{T}^{-1}$. For a reference pooling approach, we still combine $\widehat\theta_\calT$ with estimated scores $\{\widehat{S}_{n,k}: 1\leq k \leq K\}$ as in our proposal, but the weights for different source sites are estimated without the lasso penalty to rule out ineligible sites (full borrowing). 

We first focus on the quantile regression example in Section~\ref{subsec:quantile example}. For site $k \in \{\calT\} \cup \{1,\ldots,K\}$, the covariate vector $Z_k \in \mathbb{R}^5$ follows a standard multivariate normal distribution $N(\boldsymbol{0},\textnormal{I}_5)$, and the scalar outcome $Y_k$ is generated as $$Y_k \sim N\left(\beta_k^\top Z_k, \sigma_k^2\right).$$
The parameter of interest is the regression coefficient for median regression, i.e., $\tau=0.5$, in the target site. The true value of $\beta_\calT$ is set as $(-1,1,0.5,0,0)^\top$. We consider 3 experiment settings:
\begin{enumerate}
\item Setting I: $K=4$, $\sigma_k = \sigma_\calT = 1,$ $n_k = n_\calT = n$ for all $k$, two eligible sites, and two ineligible sites, where $\beta_3 =(-1.25, 0.75, 0.25, -0.25, -0.25)^\top$ and $\beta_4 = (-0.75, 1.25, 0.75, 0.25, 0.25)^\top$. This setting represents a moderate difference in all components of $\beta$ between eligible and ineligible sites. 
\item Setting II: $K=4$, $\sigma_k = \sigma_\calT = 1,$ $n_k = n_\calT = n$ for all $k$, two eligible sites, and two ineligible sites, where $\beta_3 = \beta_4 = (-0.5,1,0.5,0,0)^\top$. This setting represents a big difference in only one component of $\beta$ between eligible and ineligible sites.
\item Setting III:  $K=6$,  $\sigma_k= \sigma_\calT=1, k\neq 3,$ $\sigma_3 = 1.5,$ $n_k=n_\calT = n, k\neq 4,$ $n_4 = n/2,$ 
four eligible sites, and two ineligible sites, where $\beta_5 = (-0.5,1,0.5,0,0)^\top$ and $\beta_6 = (-0.7,0.7,0.2,0.3,-0.3)^\top.$ This setting represents a more realistic scenario. 
\end{enumerate}
For all three settings, we set the number of perturbations used in the Monte-Carlo approximation of the variance of the score to be 500, and $B_1$, the cardinality of the set $\{\widehat\theta_{\calT}^{*,j}:j \in \mathcal{S}\}$, to be 50.  The sample size  $n \in \{250,500,750,1000,1500,2000\}$.

\begin{figure}
    \centering
    \includegraphics[width=0.9\linewidth]{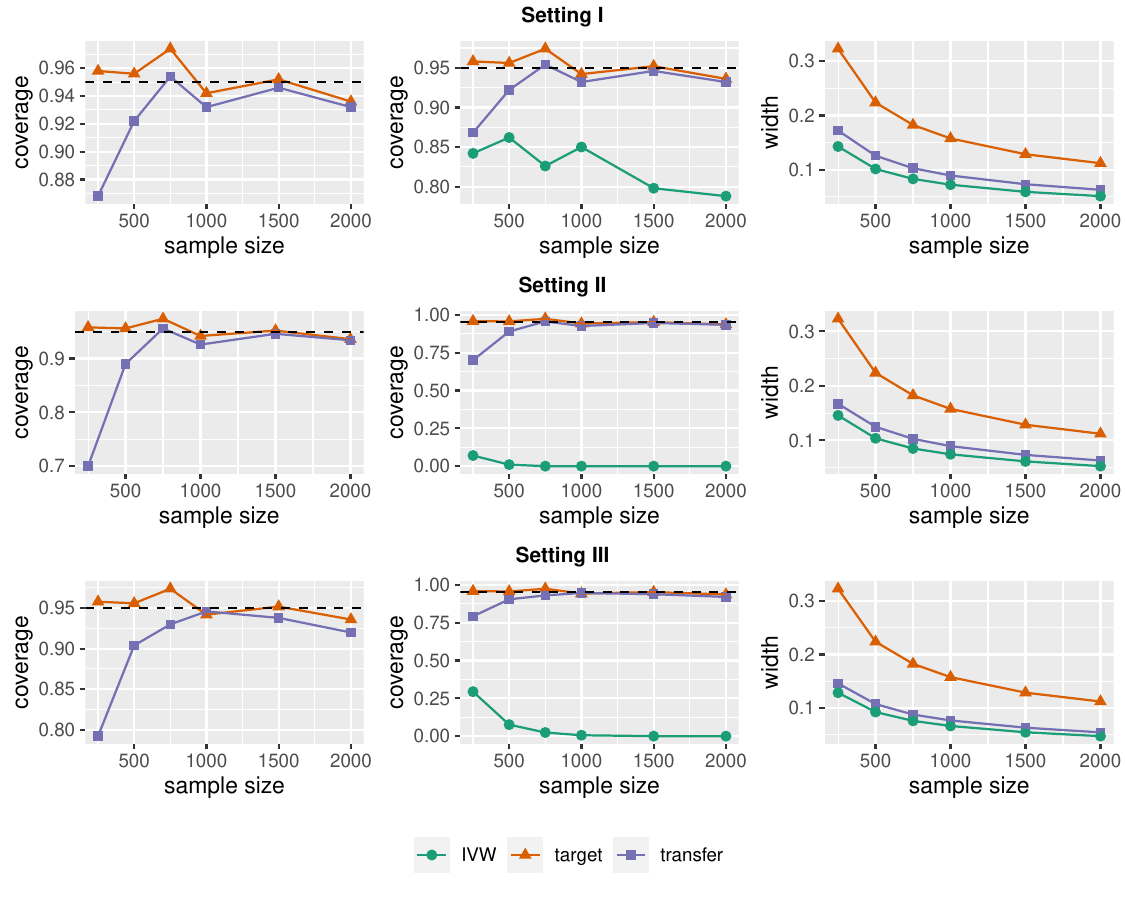}
    \caption{Quantile regression: empirical coverage level and width of $95\%$ CIs for $\beta_{\calT,1}$ under different simulation settings. Results are based on 500 simulation replications. The first column shows the coverage level of 95\% CIs based on target site data only (target) and the proposed adaptive combination (transfer) on a amplified scale;  the second column shows the coverage level of 95\% CIs constructed using all three approaches (target, adaptive transfer, and full borrowing); the third column shows the average width of the 95\% CIs constructed using all three approaches.}
    \label{fig: quantreg beta 1}
\end{figure}

Figure~\ref{fig: quantreg beta 1} shows the empirical coverage and average width of $95\%$ CI for $\beta_{\calT,1}$ over 500 simulation replications. CIs from both our target-only and transfer learning methods achieve the nominal coverage level with moderate sample sizes, and CIs based on the latter are substantially narrower reflecting the efficiency gain due to borrowing information from eligible source sites. When the sample size is small ($n=250$), non-zero weights are sometimes assigned to ineligible sites in the transfer learning approach, resulting in a moderate bias in the resulting estimator and under-coverage of the corresponding CI. However, the performance quickly improves with increasing sample size. In contrast, the naive borrowing approach does not account for the bias of the scores from ineligible sites, leading to a biased estimator and deteriorating coverage level of the corresponding CI as the sample size increases. This is especially notable in Settings II and III where the bias is severe and CI coverage eventually drops to near zero.  

\begin{figure}
    \centering
    \includegraphics[width=0.9\linewidth]{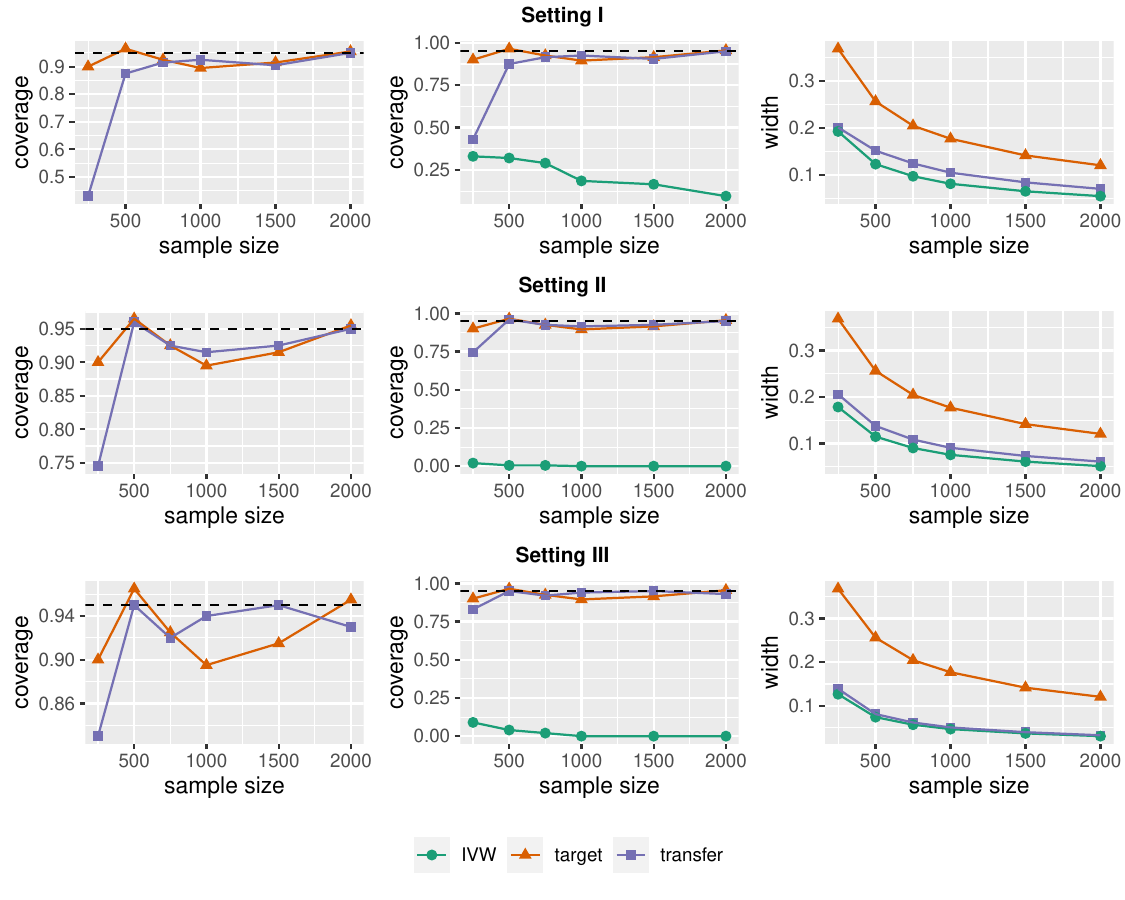}
    \caption{AUC maximization: coverage and width of $95\%$ CIs for $\beta_{\calT,2}$ under different simulation settings. Results are based on 200 simulation replications. The first column shows the coverage level of 95\% CIs based on target site data only (target) and the proposed adaptive combination (transfer) on a amplified scale;  the second column shows the coverage level of 95\% CIs constructed using all three approaches (target, adaptive transfer, and full borrowing); the third column shows the average width of the 95\% CIs constructed using all three approaches.}
    \label{fig: auc beta 2}
\end{figure}

Next, we turn to the AUC maximization example in Section~\ref{subsec:AUC example}.  We set the target parameter $\beta_\calT = (0.5, -0.5, 0.5, -0.5, 0)$. The covariate vector $Z_k \in \mathbb{R}^5$ is generated from $N(\boldsymbol{0},\sigma_k^2\Sigma)$ with $\Sigma_{ij} = 0.1^{|i-j|}$ for all sites $k \in \{\calT\} \cup \{1,\ldots,K\}$. In the target site, we generate binary outcome 
$$Y_\calT \sim \textnormal{Bernoulli}\left\{\frac{\exp(\beta_\calT^\top Z_\calT)}{1+\exp(\beta_\calT^\top Z_\calT)} \right\}.$$ 
We again explore 3 simulation settings for source sites, where the outcome is generated via
$$Y_k \sim \textnormal{Bernoulli}\left\{\frac{\exp(\beta_k^\top Z_k)}{1+\exp(\beta_k^\top Z_k)} \right\}.$$
Specifically,
\begin{enumerate}
\item Setting I: $K=4$, $\sigma_k = \sigma_\calT = 1.5$, $n_k = n_\calT = n$  for all $k$, two eligible sites and two ineligible sites, where $\beta_3 = (0.5, 0,0,0, 0.25)^\top$ and $\beta_4 = (0.5, 0,0,0, -0.25)^\top;$ 
\item Setting II: $K=6$, $\sigma_k = \sigma_\calT = 1.5$ for $k \neq 3,$ $\sigma_3= 1,$ $n_k = n_\calT = n$ for $k\neq 4,$ $n_4 = n/2,$ four eligible sites and two ineligible sites, where
$\beta_5 = (0.5, 0.5, 0.5,-0.5,0)^\top$ and $\beta_6 = (0.5, 0.25,-0.25,-0.5,0)^\top.$
This setting represents a realistic setting with varying sample size and covariate distribution across sites, and the difference in $\beta$ between eligible and ineligible sites is either concentrated on one component (site 5) or spread across several components (site 6). 
\item Setting III: the same as Setting II except that  $\sigma_k= 1.5,$ $n_k = n,$ and $Y_k \sim \mbox{Poisson}\left\{\exp\left(\beta_k^\top Z_k\right) \right\},$  $k\in \{3, 4\}.$  This setting represents a case with non-binary $Y$ in selected source sites. Note that it is meaningful to optimize the proposed objective function \eqref{eq: AUC objective}, which measures the discordance between an ordinal $Y$ and the linear combination $\beta^\top Z$. 
\end{enumerate}
We vary the sample size $n$ in $\{250,500,750,1000,1500,2000\}$ for all three settings. To speed up computation, we set both the number of perturbations used in the Monte-Carlo approximation of the variance of the score and $B_1$, the cardinality of the set $\{\widehat\theta_{\calT}^{*,j}:j \in \mathcal{S}\}$, to be 100. For $k \in \{\calT\} \cup \{1,\ldots,K\}$, we assume $\beta_{k,1}$ is positive and determine its value based on the other 4 components so that $\|\beta_k\|_2 = 1$. In Figure~\ref{fig: auc beta 2}, we present the empirical coverage and average width of resulting $95\%$ CI for $\beta_{\calT,2}$ over 200 simulation replications. Similar patterns are observed as in the quantile regression example: the proposed method is valid and more efficient than the analysis using the target site data only and the naive full borrowing method may introduce non-trivial biases by borrowing information from ineligible sites.

\subsection{Real data analysis}

The increasing availability of large-scale biobank data provides new opportunities to study the association between genetic variants and a wide range of phenotypes or diseases. For example, both the Massachusetts General and Brigham (MGB) Biobank \citep{castro2022mass} and the UK Biobank \citep{bycroft2018uk} contain rich genetic data that are linked to health records and phenotype information. However, the phenotype or disease outcome is often imperfectly measured, for example, through the occurrence of relevant diagnosis or billing codes or through patient self report. Moreover, the strength and nature of the associations between genetic variants and phenotypes may vary across racial and ethnic groups, and thus conclusions drawn from data on one racial group may not always generalize to another. Lastly, the sample size for some racial and ethnic groups of interest may be small and limits the power of the statistical analysis. In this section, we apply our proposed method to study the associations between risk of type 2 diabetes (T2D) and various genetic markers among the African American population in MGB, aiming to overcome the aforementioned challenges.

We have data from both the MGB Biobank and UK Biobank containing individual-level information on the minor allele frequency of 369 T2D-related single nucleotide polymorphisms (SNPs) reported by a previous study \citep{mahajan2018fine} and demographic variables including biological sex, racial group, and age. As the risk of T2D differs greatly by age and the association between T2D risk and SNPs may differ by various factors, we split the entire dataset by source (MGB vs UKB), age ($\geq 60$ versus $<60$) and race (white versus black). Other racial groups account for a small fraction of the total sample, and therefore we only focus on white and black. This grouping results in a total of eight datasets from two data sources, and we treat each set as a ``study site". The sample sizes for these eight sites are very different and given in Table~\ref{table: sample size}.

\begin{table}[h]
    \centering
    \begin{tabular}{lrrrr}
    \hline
         & data source & age group & racial group & sample size  \\
    \hline
        1 & MGB & $\geq 60$ & white & 9,584 \\
        2 & MGB & $\geq 60$ & black & 442 \\
        3 & MGB & $<60$ & white & 5,926 \\
        4 & MGB & $<60$ & black & 595 \\
        5 & UKB & $\geq 60$ & white & 204,186 \\
        6 & UKB & $\geq 60$ & black & 1585 \\
        7 & UKB & $<60$ & white & 255,060 \\
        8 & UKB & $<60$ & black & 6,057 \\
        \hline
    \end{tabular}
    \caption{Sample size in the 8 sites in the type 2 diabetes risk analysis.}
    \label{table: sample size}
\end{table}

\begin{figure}[h]
    \centering
    \includegraphics[width=0.9\linewidth]{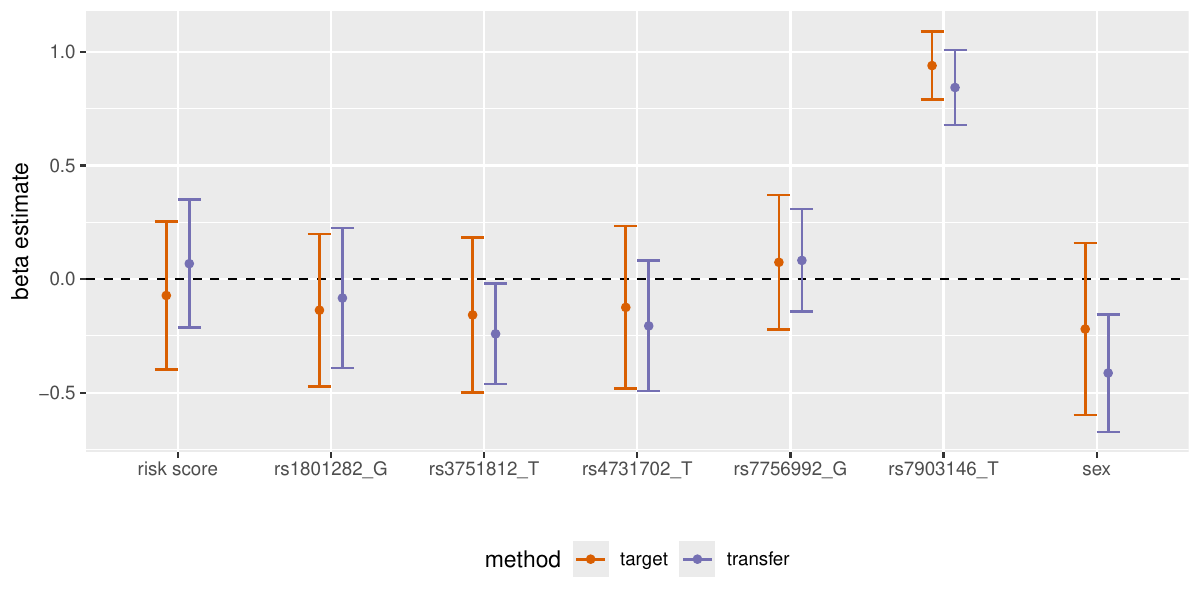}
    \caption{Association between type 2 diabetes and selected SNPs, risk score, and biological sex: comparing the target-only and transfer learning approaches. Dot represents the point estimate, and bars represent the associated confidence interval. Here,``rs1801282\_G" means the G allele at SNP rs1801282, and similarly for the other SNPs.}
    \label{fig: real data CI}
\end{figure}

Based on results reported by \citet{mahajan2018fine}, we select 5 SNPs (shown in Figure~\ref{fig: real data CI}) that demonstrated high marginal associations with T2D risk, and construct a polygenic risk score combining the remaining SNPs. It is worth noting, however, that the aforementioned findings in previous work were derived from European-descent individuals. For data from the MGB Biobank, the outcome of interest is the total count of the diagnosis codes involving T2D; and for data from the UK Biobank, the outcome of interest is a binary self-reported T2D status. While neither outcome measure is a gold-standard for the true T2D status of an individual, it is reasonable to expect that given the true disease status these measures are no longer dependent on the SNPs information. This motivates us to estimate the association by minimizing the discordance between the T2D outcome and a linear combination of selected risk factors as in \eqref{eq: AUC objective}. 

As seen in Table~\ref{table: sample size}, the sample size of African American patients aged 60 and above from MGB is disproportionately small. We purposely choose this group, i.e., site 2, as our target site, and the other groups as source sites from which we can potentially borrow information. We estimate the association between T2D risk and SNPs (minor allele frequency of 5 selected SNPs as well as the polygenic risk score) adjusting for biological sex by minimizing the discordance loss in \eqref{eq: AUC objective}. We compare the resulting point estimates and associated CIs using the target site data only as well as using the proposed approach in Figure~\ref{fig: real data CI}. 

In our analysis, two source sites, ``UKB $\ge 60$ Black'' and ``UKB $< 60$ Black'', receive nonzero weights, suggesting that the association between T2D risk and SNPs may be influenced more by race than by age. Generally, borrowing information from similar source sites narrows confidence intervals and increases power for detecting underlying associations. In this analysis, biological sex is coded as a binary variable, and our results indicate that males have a higher risk of T2D compared to females. This sex-based difference has been reported in previous studies \citep{kautzky2023sex}, but is not detected using only the target site data likely due to limited power.

The effect of the polygenic risk score is not statistically significant, although the direction of the point estimate differs depending on whether information is borrowed from source sites. The \texttt{T} allele at SNP \texttt{rs7903146} is associated with an increased risk of T2D. This SNP lies in the \texttt{TCF7L2} gene (transcription factor 7-like 2), which plays a role in blood glucose regulation, and has been consistently linked to T2D in prior research \citep{ding2018meta,del2021role}. SNP \texttt{rs3751812}, located in the \texttt{FTO} gene associated with fat mass and obesity, has been shown to elevate T2D risk in European populations. However, studies in African Americans have not reached a consistent conclusion \citep{bressler2010risk,hester2012implication}. In our analysis, the point estimate for this SNP is negative, with the confidence interval just excluding zero after borrowing information.

We do not observe significant associations between T2D risk and the other SNPs that were highly significant in prior studies based on European populations. Nonetheless, the signs of our point estimates are directionally consistent with those reported in the literature. Notably, the minor allele frequencies (MAFs) for these SNPs differ substantially between White and African American individuals in the MGB Biobank, except for \texttt{rs7903146}. Such differences in MAF may contribute to reduced power in detecting associations among African Americans. Alternatively, the genetic effects themselves may vary across racial groups, as frequently noted in the literature.

In summary, findings from this real-data example suggest that associations between type 2 diabetes (T2D) and genetic variants identified in a European-descent population may not directly generalize to African American patients in the MGB Biobank. This underscores the need for more inclusive genetic studies that prioritize diverse populations, as well as for statistical methods capable of effectively leveraging external data to study historically underrepresented population.

\section{Discussion}\label{sec:discussion}
In this paper, we develop a sampling-based federated learning framework for statistical inference on an $M$-estimator with a non-smooth objective function in a target population. Unlike other resampling methods that require repeatedly solving optimization problems, our approach instead only needs to evaluate the objective function on original and perturbed samples across different parameter values. To leverage information from similar source sites, we adaptively assign weights to each site with the dual goals of improving estimation precision and excluding ineligible sources.

As shown in Lemma~\ref{adaptive lasso solution}, when the sample size is sufficiently large, the data-adaptive weights from \eqref{eq: adaptive lasso} can effectively exclude all biased source sites and assign optimal weights to unbiased ones, thereby minimizing the variance of the resulting estimators. In practice, when the sample size is small, some biased sites may receive non-zero weights, introducing minor bias into the combined estimator and resulting in under-coverage of the corresponding confidence interval. This issue is akin to a type II error, which is unavoidable since no statistical test achieves perfect power with finite samples. To address this, our method can be extended by incorporating an additional sampling step—similar to the one proposed in \citet{guo2023robust}—to account for site selection uncertainty and attain uniformly valid coverage. However, this generally comes at the cost of reduced estimation precision, especially in the absence of external knowledge about which sources are unbiased. We leave further exploration of this direction for future work.

Throughout, we have implicitly assumed that the dimension of the parameter $d$ remains fixed as the sample size increases. In reality, the convergence behavior of the Markov chain may depend on this dimension—the number of steps required to reach the stationary distribution may grow with $d$. It would be interesting to investigate whether MCMC sampling remains effective when applied to penalized objective functions commonly used in high-dimensional settings. It is also of interest to extend the sampling-based inferential framework to semiparametric models, where an objective function can be defined by profiling out potentially infinite-dimensional nuisance parameters, analogous to the profile likelihood.

\bibliographystyle{apalike}
\bibliography{reference}

\appendix

\section{Useful lemmas with proofs}

\subsection{Asymptotic normality of $\widehat\theta_\calT$ and consistent estimation of its variance}

In this section, we introduce several lemmas showing that (a) the MCMC-based estimator is asymptotically normal with asymptotic variance taking a sandwich form (Lemma~\ref{MCMC estimator}); (b) this variance can be consistently estimated (Lemmas~\ref{second derivative estimator} and \ref{score variance estimator}.)

\begin{lemma}\label{MCMC estimator}
Suppose that $\tilde\theta_\calT$ is an asymptotically normal estimator of $\theta_\calT$. Under Conditions~\ref{cond:target quad} and \ref{cond:target tail}, $\widehat\theta_\mathcal{T} = \tilde{\theta}_\mathcal{T} + o_P(n_\mathcal{T}^{-1/2})$.
\end{lemma}


\begin{proof}[Proof of Lemma~\ref{MCMC estimator}]
Let us first recall the quadratic expansion $$R_n(\theta_1,\theta_2) = M_{n,\mathcal{T}}(\theta_1) - M_{n,\mathcal{T}}(\theta_2) - S_{n,\mathcal{T}}(\theta_2)(\theta_1 - \theta_2) - \frac{1}{2}(\theta_1 - \theta_2)^\top A_\mathcal{T}(\theta_1 - \theta_2),$$
and the density we are sampling from via the Markov chain Monte Carlo
$$L_\mathcal{T}(\theta) = C_n \exp\left\{-n_\mathcal{T}M_{n,\mathcal{T}}(\theta) + n_\mathcal{T}M_{n,\mathcal{T}}(\tilde\theta_\mathcal{T}) \right\}\ind{\|\theta\| \leq R},$$
where we use subscript $n$ to emphasize the dependence of the normalizing constant on the observed sample.

First, we argue that $S_{n,\calT}(\tilde\theta_\calT) = o_P(n_\calT^{-1/2}).$ Consider $\theta = \tilde\theta_\calT - A_\calT^{-1}S_{n,\calT}(\tilde\theta_\calT)$. 
\begin{equation*}
    M_{n,\mathcal{T}}(\theta) - M_{n,\mathcal{T}}(\tilde\theta_\calT) = -\frac{1}{2} S_{n,\calT}(\tilde\theta_\calT)^\top A_\calT^{-1}S_{n,\calT}(\tilde\theta_\calT) +  R_n\left(\tilde\theta_\calT - A_\calT^{-1}S_{n,\calT}(\tilde\theta_\calT),\tilde\theta_\calT\right).
\end{equation*}
As $\tilde\theta_\calT$ minimizes the objective function up to $o_P(n_\calT^{-1})$, the right-hand side of the above display is $o_P(n_\calT^{-1})$. Under Condition~\ref{cond:target quad}, the first term is the dominating term, which implies that $S_{n,\calT}(\tilde\theta_\calT) = o_P(n_\calT^{-1/2}).$ In the proof below, we will take $\theta_2 = \tilde\theta_\calT$, then $S_{n,\mathcal{T}}(\tilde\theta_\calT)(\theta_1 - \tilde\theta_\calT) = o_P(n_\calT^{-1/2}\|\theta_1 - \tilde\theta_\calT\|) = o_P(n_\calT^{-1} + \|\theta_1 - \tilde\theta_\calT\|^2)$ uniformly over $\theta_1$. Thus, this term can be absorbed into the remainder $R_n$. In the following, with a slight abuse of notation, we still write the remainder as $R_n$ but it now includes this term arising from the linear term in the expansion which is uniformly small.

As $\theta_\mathcal{T}$ is a finite-dimensional parameter, it suffices to show that $\widehat\theta_{\mathcal{T},j} = \tilde{\theta}_{\mathcal{T},j} + o_P(n_\mathcal{T}^{-1/2})$ for all $j$. The difference between $\widehat\theta_{\mathcal{T},j}$ and $\tilde\theta_{\mathcal{T},j}$ is given by
\begin{align*}
    \widehat\theta_{\mathcal{T},j} - \tilde\theta_{\mathcal{T},j} &= C_n\int_{\|\theta\|\leq R} (\theta_j - \tilde\theta_{\mathcal{T},j})\exp\left\{-n_\mathcal{T}M_{n,\mathcal{T}}(\theta) + n_\mathcal{T}M_{n,\mathcal{T}}(\tilde\theta_\mathcal{T})\right\}d\theta. \\
    &= n_\mathcal{T}^{-d/2}C_n\int_{\|u + n_\calT^{1/2}\tilde\theta_\calT\|\leq n_\mathcal{T}^{1/2}R} \frac{u_j}{\sqrt{n_\mathcal{T}}} \exp\left\{-n_\mathcal{T}M_{n,\mathcal{T}}\left(\tilde\theta_\mathcal{T}+\frac{u}{\sqrt{n_\mathcal{T}}}\right) + n_\mathcal{T}M_{n,\mathcal{T}}(\tilde\theta_\mathcal{T})\right\}du,
\end{align*}
where we change variable and let $u = \sqrt{n_\mathcal{T}}(\theta - \tilde\theta_\mathcal{T})$.

The normalizing constant $C_n$ is such that
\begin{align*}
    C_n^{-1} &= \int_{\|\theta\| \leq R} \exp\left\{-n_\mathcal{T}M_{n,\mathcal{T}}(\theta) + n_\mathcal{T}M_{n,\mathcal{T}}(\tilde\theta_\mathcal{T})\right\}d\theta \\
    &= n_\mathcal{T}^{-d/2}\int_{\|u + n_\calT^{1/2}\tilde\theta_\calT\|\leq n_\mathcal{T}^{1/2}R} \exp\left\{-n_\mathcal{T}M_{n,\mathcal{T}}\left(\tilde\theta_\mathcal{T}+\frac{u}{\sqrt{n_\mathcal{T}}}\right) + n_\mathcal{T}M_{n,\mathcal{T}}(\tilde\theta_\mathcal{T})\right\}du,
\end{align*}
using the same change of variable. Suppose for now that
\begin{equation}\label{norm constant}
    n_\mathcal{T}^{-d/2}C_n(2\pi)^{d/2}\left\{\textnormal{det}(A)\right\}^{1/2} \xrightarrow[]{p} 1, \quad \textnormal{ as } n \rightarrow \infty,
\end{equation}
and $n_\mathcal{T}^{-d/2}C_n$ is $O_P(1)$, it then suffices to show that $I_j = o_P(1)$ with $I_j$ defined as 
\begin{align}\label{difference in mean}
    I_j &= \int_{\|u + n_\calT^{1/2}\tilde\theta_\calT\|\leq n_\mathcal{T}^{1/2}R} u_j \exp\left\{-n_\mathcal{T}M_{n,\mathcal{T}}\left(\tilde\theta_\mathcal{T}+\frac{u}{\sqrt{n_\mathcal{T}}}\right) + n_\mathcal{T}M_{n,\mathcal{T}}(\tilde\theta_\mathcal{T})\right\}du \nonumber \\
    &= \int_{\|u + n_\calT^{1/2}\tilde\theta_\calT\|\leq n_\mathcal{T}^{1/2}R} u_j \exp\left\{-\frac{1}{2}u^\top A_\mathcal{T}u - n_\mathcal{T}R_n\left(\tilde\theta_\mathcal{T}+\frac{u}{\sqrt{n_\mathcal{T}}}\right)\right\}du.
\end{align}
We will establish \eqref{norm constant} later and focus on \eqref{difference in mean} now. 

By definition, showing $I_j = o_P(1)$ is equivalent to showing that for any $\epsilon>0$, there exists $N>0$ such that for any $n_\mathcal{T} \geq N$, $P(I_j > \epsilon) < \epsilon$. Given any $\epsilon>0$, we will split the integral $I_j$ into three terms and show that each one of them is small for sufficiently large $n_\calT$. Let $\Lambda_{\min}(A_\mathcal{T})$ denote the minimum eigenvalue of $A_\mathcal{T}$, which is assumed to be positive. By Condition~\ref{cond:target quad}, there exists a sufficiently small $C_3$ such that for sufficiently large $n_\mathcal{T}$,
\begin{equation*}
    P\left(\sup_{\|\theta_1 - \theta_2\|_2 \leq C_3, \ \|\theta_2 - \theta_\mathcal{T}\|_2 \leq \delta_{n_\calT}} \frac{|R_n(\theta_1, \theta_2)|}{\|\theta_1 - \theta_2\|_2^2 + n_\mathcal{T}^{-1}} > \frac{1}{8}\Lambda_{\min}(A_\mathcal{T}) \right) < \frac{\epsilon}{2}.
\end{equation*}
Since $\tilde\theta_\mathcal{T}$ is asymptotically normal, for sufficiently large $n_\mathcal{T}$, $P(\|\tilde\theta_\mathcal{T} - \theta_\mathcal{T}\|_2 > \delta_{n_\calT}) < \epsilon/2$ as $n_\calT^{1/2}\delta_{n_\calT} \rightarrow \infty$. As a result, for sufficiently large $n_\mathcal{T}$, with probability at most $\epsilon$,
\begin{equation}\label{choice of C3}
    \sup_{\|\theta - \tilde\theta_\mathcal{T}\|_2 \leq C_3} \frac{|R_n(\theta,\tilde\theta_\mathcal{T})|}{\|\theta-\tilde\theta_\mathcal{T}\|_2^2 + n_{\mathcal{T}}^{-1}} > \frac{1}{8}\Lambda_{\min}(A_\mathcal{T}).
\end{equation}
At the same time, by dominated convergence theorem,
\begin{equation*}
    \lim_{U\rightarrow \infty} \int \ind{\|u\|_2 > U}u_j \exp\left\{-\frac{1}{4}u^\top A_\mathcal{T} u\right\}du =0,
\end{equation*}
as the integrand converges to 0 pointwise and the envelop function $|u_j|\exp(-u^\top A_\mathcal{T}u/4)$ is integrable. As a result, there exists a large constant $C_2 > 1$ such that $\int_{\|u\| > C_2}u_j\exp(-u^\top A_\mathcal{T}u/4)du < \epsilon$. We will split the integral using the constants $C_2$ and $C_3$.

First, consider the integral
\begin{equation*}
    I_{j,1} = \int_{C_2 < \|u\| \leq n_\mathcal{T}^{1/2}C_3} u_j \exp\left\{-\frac{1}{2}u^\top A_\mathcal{T}u - n_\mathcal{T}R_n\left(\tilde\theta_\mathcal{T}+\frac{u}{\sqrt{n_\mathcal{T}}}\right)\right\}du.
\end{equation*}
We have
\begin{align*}
    P(I_{j,1} < \epsilon) &> P\left(-\frac{1}{2}u^\top A_\mathcal{T}u - n_\mathcal{T}R_n\left(\tilde\theta_\mathcal{T}+\frac{u}{\sqrt{n_\mathcal{T}}}\right) \leq -\frac{1}{4}u^\top A_\mathcal{T}u, \forall C_2 \leq \|u\|_2 \leq C_3n_\mathcal{T}^{1/2}\right) \\
    &> P\left(\sup_{C_2 \leq \|u\|_2 \leq C_3n_\mathcal{T}^{1/2}}\frac{n_\mathcal{T}|R_n(\tilde\theta_\mathcal{T} + u/\sqrt{n_\mathcal{T}})|}{\|u\|_2^2 + 1} \leq \frac{1}{8}\Lambda_{\min}(A_\mathcal{T})\right) \\
    &= P\left(\sup_{C_2n_{\mathcal{T}}^{-1/2} \leq \|\theta - \tilde\theta_\mathcal{T}\|_2 \leq C_3} \frac{|R_n(\theta,\tilde\theta_\mathcal{T})|}{\|\theta-\tilde\theta_\mathcal{T}\|_2^2 + n_{\mathcal{T}}^{-1}} \leq \frac{1}{8}\Lambda_{\min}(A_\mathcal{T}) \right) > 1-\epsilon,
\end{align*}
for sufficiently large $n_\mathcal{T}$.

Next, consider the integral
\begin{align*}
    I_{j,2} &= \int_{n_\mathcal{T}^{1/2}C_3 < \|u\|,\|u + n_\calT^{1/2}\tilde\theta_\calT\|\leq n_\mathcal{T}^{1/2}R} u_j \exp\left\{-\frac{1}{2}u^\top A_\mathcal{T}u - n_\mathcal{T}R_n\left(\tilde\theta_\mathcal{T}+\frac{u}{\sqrt{n_\mathcal{T}}}\right)\right\}du \\
    &= \int_{n_\mathcal{T}^{1/2}C_3 < \|u\|,\|u + n_\calT^{1/2}\tilde\theta_\calT\|\leq n_\mathcal{T}^{1/2}R} u_j \exp\left\{ - n_\mathcal{T}M_{n,\mathcal{T}}\left(\tilde\theta_\mathcal{T}+\frac{u}{\sqrt{n_\mathcal{T}}}\right) + n_\mathcal{T}M_{n,\mathcal{T}}(\tilde\theta_\mathcal{T})\right\}du \\
    &= n_{\mathcal{T}}^{(d+1)/2}\int_{C_3 < \|\theta-\tilde\theta_\mathcal{T}\|, \|\theta\| \leq R} (\theta_j - \tilde\theta_{\mathcal{T},j})\exp\{- n_\mathcal{T}M_{n,\mathcal{T}}(\theta) + n_\mathcal{T}M_{n,\mathcal{T}}(\tilde\theta_\mathcal{T})\}d\theta.
\end{align*}
Now that $C_3$ has been chosen (independent of $n_\calT$) and fixed, by Condition~\ref{cond:target tail}, there exists $\epsilon_1 > 0$ such that
\begin{equation}\label{choice of epsilon1}
    P\left(\inf_{\|\theta-\tilde\theta_\mathcal{T}\|_2 > C_3} \left\{M_{n,\mathcal{T}}(\theta) - M_{n,\mathcal{T}}(\tilde\theta_\mathcal{T})\right\} \geq \epsilon_1 \right) > 1-\epsilon,
\end{equation}
for sufficiently large $n_\mathcal{T}$. Thus, for sufficiently large $n_\mathcal{T}$, with probability at least $1-\epsilon$,
\begin{align*}
    I_{j,2} &= n_{\mathcal{T}}^{(d+1)/2}\int_{C_3 < \|\theta-\tilde\theta_\mathcal{T}\|, \|\theta\| \leq R} (\theta_j - \tilde\theta_{\mathcal{T},j})\exp\{- n_\mathcal{T}M_{n,\mathcal{T}}(\theta) + n_\mathcal{T}M_{n,\mathcal{T}}(\tilde\theta_\mathcal{T})\}d\theta \\
    &\leq n_{\mathcal{T}}^{(d+1)/2}\int_{C_3 < \|\theta-\tilde\theta_\mathcal{T}\|, \|\theta\| \leq R} (\theta_j - \tilde\theta_{\mathcal{T},j})\exp\{- n_\mathcal{T}\epsilon_1\}d\theta \\
    &\leq n_{\mathcal{T}}^{(d+1)/2}\int_{C_3 < \|\theta-\tilde\theta_\mathcal{T}\|, \|\theta\| \leq R} \left(R+\|\tilde\theta_\calT\|\right)\exp\{- n_\mathcal{T}\epsilon_1\}d\theta,
\end{align*}
which is smaller than $\epsilon$ with high probability for sufficiently large $n_\mathcal{T}$. That is, $P(I_{j,2} < \epsilon) > 1 - \epsilon$ for sufficiently large $n_\mathcal{T}$.

Lastly, we consider the integral
\begin{align*}
    I_{j,3} &= \int_{\|u\| \leq C_2} u_j \exp\left\{-\frac{1}{2}u^\top A_\mathcal{T}u - n_\mathcal{T}R_n\left(\tilde\theta_\mathcal{T}+\frac{u}{\sqrt{n_\mathcal{T}}}\right)\right\}du \\
    &= \int_{\|u\| \leq C_2} u_j \exp\left\{-\frac{1}{2}u^\top A_\mathcal{T}u\right\}\left(\exp\left\{ - n_\mathcal{T}R_n\left(\tilde\theta_\mathcal{T}+\frac{u}{\sqrt{n_\mathcal{T}}}\right)\right\}-1\right)du
\end{align*}
where the second line follows from $\int_{\|u\| \leq C_2} u_j\exp(-u^\top A_\mathcal{T} u/2)du =0$. Note that $|e^x - 1| \leq |2x|$ for $|x| \leq \delta_1$ for some positive constant $\delta_1$. Again by Condition~\ref{cond:target quad}, there exists sufficiently small $C_4$ such that
\begin{equation*}
    P\left(\sup_{\|\theta - \tilde\theta_\mathcal{T}\|_2 \leq C_4,\|\tilde\theta_\mathcal{T} - \theta_\calT\|_2 \leq \delta_{n_\calT}}\frac{|R_n(\theta,\tilde\theta_\mathcal{T})|}{\|\theta - \tilde\theta_\mathcal{T}\|_2^2 + n_\mathcal{T}^{-1}} > \min\left\{\frac{\delta_1}{2C_2^2}, \frac{\epsilon}{2\int|u_j|(\|u\|^2+1)\exp(-u^\top A_\mathcal{T} u /2) du}\right\} \right)  < \frac{\epsilon}{2}.
\end{equation*}
Moreover, for sufficiently large $n_\mathcal{T}$, $P(\|\tilde\theta_\mathcal{T}-\theta_\mathcal{T}\| > \delta_{n_\calT}) < \epsilon/2$. Hence, with probability at most $\epsilon$,
\begin{equation*}
    \sup_{\|\theta - \tilde\theta_\mathcal{T}\|_2 \leq C_4}\frac{|R_n(\theta,\tilde\theta_\mathcal{T})|}{\|\theta - \tilde\theta_\mathcal{T}\|_2^2 + n_\mathcal{T}^{-1}} > \min\left\{\frac{\delta_1}{2C_2^2}, \frac{\epsilon}{2\int|u_j|(\|u\|^2+1)\exp(-u^\top A_\mathcal{T} u /2) du}\right\},
\end{equation*}
for sufficiently large $n_\mathcal{T}$. Note that as $C_2$ and $C_4$ are chosen independent of $n_\mathcal{T}$, for sufficiently large $n_\mathcal{T}$, $C_4n_{\mathcal{T}}^{1/2} > C_2$, and therefore with probability at most $\epsilon$,
\begin{equation*}
    \sup_{\|u\|_2 \leq C_2}\frac{n_\mathcal{T}|R_n(\tilde\theta_\mathcal{T} + u/\sqrt{n_\mathcal{T}})|}{\|u\|_2^2 + 1} > \min\left\{\frac{\delta_1}{2C_2^2}, \frac{\epsilon}{2\int|u_j|(\|u\|^2+1)\exp(-u^\top A_\mathcal{T} u /2) du}\right\}.
\end{equation*}
Now, under the complement of the event in the above display, $n_\mathcal{T}|R_n(\tilde\theta_\mathcal{T} + u/\sqrt{n_\mathcal{T}})| \leq \delta_1(\|u\|_2^2+1)/(2C_2^2) \leq \delta_1$, and therefore
\begin{align*}
    I_{j,3} &= \int_{\|u\|_2 \leq c_2} u_j \exp\left\{-\frac{1}{2}u^\top A_\mathcal{T}u\right\}\left(\exp\left\{ - n_\mathcal{T}R_n\left(\tilde\theta_\mathcal{T}+\frac{u}{\sqrt{n_\mathcal{T}}}\right)\right\}-1\right)du \\
    &\leq \int_{\|u\|_2 \leq c_2} 2|u_j| \exp\left\{-\frac{1}{2}u^\top A_\mathcal{T}u\right\}\left| n_\mathcal{T}R_n\left(\tilde\theta_\mathcal{T}+\frac{u}{\sqrt{n_\mathcal{T}}}\right)\right|du \\
    &= \int_{\|u\|_2 \leq c_2} 2|u_j|(\|u\|_2^2+1) \exp\left\{-\frac{1}{2}u^\top A_\mathcal{T}u\right\}\frac{n_\mathcal{T}|R_n(\tilde\theta_\mathcal{T} + u/\sqrt{n_\mathcal{T}})|}{\|u\|_2^2 + 1}du \\
    &< \epsilon.
\end{align*}
Thus, we have $P(I_{j,3} < \epsilon) > 1-\epsilon$. Combining our conclusions on $I_{j,1}$, $I_{j,2}$ and $I_{j,3}$, we get $P(I_j > 3\epsilon) < 3\epsilon$, as a result of union bound and the fact that $I_j = \sum_{l=1}^3I_{j,l}$.

To finish the proof, it remains to show \eqref{norm constant}, which is equivalent to showing
\begin{equation*}
    \int_{\|u + n_\calT^{1/2}\tilde\theta_\calT\|\leq n_\mathcal{T}^{1/2}R}\frac{1}{(2\pi)^{d/2}\{\det(A_\mathcal{T})\}^{1/2}}\exp\left\{-n_\mathcal{T}M_{n,\mathcal{T}}\left(\tilde\theta_\mathcal{T} + \frac{u}{\sqrt{n_\mathcal{T}}}\right) + n_\mathcal{T}M_{n,\mathcal{T}}(\tilde\theta_\mathcal{T})\right\}du \xrightarrow[]{p} 1.
\end{equation*}
As we have shown, for sufficiently large $n_\mathcal{T}$, \eqref{choice of C3} holds with probability at most $\epsilon$. By dominated convergence theorem, there exists constant $\tilde{C}_2 > 1$ such that $\int_{\|u\| > \tilde{C}_2} \exp(-u^\top A_\mathcal{T} u/4)du < \epsilon (2\pi)^{d/2}\{\det(A_\mathcal{T})\}^{1/2}/2$. Then, similar to how we study $I_{j,1}$, we can show that
\begin{equation*}
    P\left(\int_{\tilde{C}_2 < \|u\| \leq n_\mathcal{T}^{1/2}C_3}\frac{1}{(2\pi)^{d/2}\{\det(A_\mathcal{T})\}^{1/2}}\exp\left\{-n_\mathcal{T}M_{n,\mathcal{T}}\left(\tilde\theta_\mathcal{T} + \frac{u}{\sqrt{n_\mathcal{T}}}\right) + n_\mathcal{T}M_{n,\mathcal{T}}(\tilde\theta_\mathcal{T})\right\}du > \epsilon \right) < \epsilon.
\end{equation*}
Moreover,
\begin{align*}
    &\int_{n_\mathcal{T}^{1/2}C_3 < \|u\|,\|u + n_\calT^{1/2}\tilde\theta_\calT\|\leq n_\mathcal{T}^{1/2}R} \frac{1}{(2\pi)^{d/2}\{\det(A_\mathcal{T})\}^{1/2}} \exp\left\{ - n_\mathcal{T}M_{n,\mathcal{T}}\left(\tilde\theta_\mathcal{T}+\frac{u}{\sqrt{n_\mathcal{T}}}\right) + n_\mathcal{T}M_{n,\mathcal{T}}(\tilde\theta_\mathcal{T})\right\}du \\
    &= n_{\mathcal{T}}^{d/2}\int_{C_3 < \|\theta-\tilde\theta_\mathcal{T}\|, \|\theta\| \leq R} \frac{1}{(2\pi)^{d/2}\{\det(A_\mathcal{T})\}^{1/2}}\exp\{- n_\mathcal{T}M_{n,\mathcal{T}}(\theta) + n_\mathcal{T}M_{n,\mathcal{T}}(\tilde\theta_\mathcal{T})\}d\theta \\
    &\leq n_{\mathcal{T}}^{d/2}\int_{C_3 < \|\theta-\tilde\theta_\mathcal{T}\|, \|\theta\| \leq R} \frac{1}{(2\pi)^{d/2}\{\det(A_\mathcal{T})\}^{1/2}}\exp\{-n_\mathcal{T}\epsilon_1\}d\theta 
\end{align*}
with probability at least $1-\epsilon$ for sufficiently large $n_\mathcal{T}$. Thus, similar to how we bound $I_{j,2}$, we can show that this integral converges to 0 in probability.

It remains to show that
\begin{align*}
    &\int_{\|u\| \leq \tilde{C}_2}\frac{1}{(2\pi)^{d/2}\{\det(A_\mathcal{T})\}^{1/2}} \exp\left\{-\frac{1}{2}u^\top A_\mathcal{T}u - n_\mathcal{T}R_n\left(\tilde\theta_\mathcal{T}+\frac{u}{\sqrt{n_\mathcal{T}}}\right)\right\} du \\
    &\xlongrightarrow[]{p}  \int \frac{1}{(2\pi)^{d/2}\{\det(A_\mathcal{T})\}^{1/2}}\exp\left\{-\frac{1}{2}u^\top A_\mathcal{T}u\right\}du=1.
\end{align*}
As $\int_{\|u\| > \tilde{C}_2}\exp(-u^\top A_\mathcal{T} u/2)du/((2\pi)^{d/2}\{\det(A_\mathcal{T})\}^{1/2}) < \epsilon/2$, it suffices to show that
\begin{equation*}
    P\left(\int_{\|u\| \leq \tilde{C}_2}\frac{\exp\left\{-\frac{1}{2}u^\top A_\mathcal{T}u\right\}}{(2\pi)^{d/2}\{\det(A_\mathcal{T})\}^{1/2}} \left(\exp\left\{ - n_\mathcal{T}R_n\left(\tilde\theta_\mathcal{T}+\frac{u}{\sqrt{n_\mathcal{T}}}\right)\right\} -1\right)du > \frac{\epsilon}{2}\right) < \epsilon,
\end{equation*}
for sufficiently large $n_\mathcal{T}$. But this can be shown in the same way as we study the integral $I_{j,3}$ previously, and hence we omit the details here.

Finally, we note that the above argument established Lemma~\ref{MCMC estimator} with $B = \infty$ where $B$ is the sample size of the MCMC sample. In practice, we can choose $B$ to be arbitrarily large so that the Monte Carlo approximation error is arbitrarily small. To rigorously analyze the effect of $B$, we let $\mathcal{O}_\calT$ denote the data in the source site.
\begin{equation*}
    P\left(\left|\frac{1}{B}\sum_{b=1}^B \widehat\theta_{\calT,j}^{*,b} - \widehat\theta_{\calT,j}\right| > \epsilon n_\calT^{-1/2} \Big\rvert \mathcal{O}_\calT \right) \leq \frac{\Var\left(\widehat\theta_{\calT,j}^{*,1} \rvert \mathcal{O}_\calT\right)}{\epsilon^2 n_\calT^{-1}B} \leq \frac{R^2}{\epsilon^2 n_\calT^{-1}B}
\end{equation*}
by Chebyshev's inequality and the fact that the density is 0 outside of a ball of radius $R$. Marginalizing over $\mathcal{O}_\calT$, we get that 
\begin{equation*}
    P\left(\left|\frac{1}{B}\sum_{b=1}^B \widehat\theta_{\calT,j}^{*,b} - \widehat\theta_{\calT,j}\right| > \epsilon n_\calT^{-1/2} \right) \leq \frac{R^2}{\epsilon^2 n_\calT^{-1}B},
\end{equation*}
which converges to 0 for any $\epsilon$ is $B \gg n_\calT$. Hence, with $B \gg n_\calT$, the Monte Carlo error is indeed negligible. 
\end{proof}


\begin{lemma}\label{second derivative estimator}
Suppose that $\tilde\theta_\calT$ is an asymptotic normal estimator of $\theta_\calT$. Under Conditions~\ref{cond:target quad} and \ref{cond:target tail}, $\widehat A_\mathcal{T}^{-1} = A_\mathcal{T}^{-1} + o_P(1)$.
\end{lemma}

\begin{proof}[Proof of Lemma~\ref{second derivative estimator}]
We show that the covariance matrix of the MCMC sample $\{\widehat\theta_{\mathcal{T}}^{*,j}\}_{j=1}^B$ (scaled by $n_\mathcal{T}$) serves as a consistent estimator of the inverse of the second derivative matrix, that is, $\Gamma_\mathcal{T} = A_\mathcal{T}^{-1}$. We start by decomposing the second moment as follows:
\begin{align*}
    &\quad \int_{\left\|\theta\right\|_2 \leq R} \theta_k\theta_l C_n \exp\left\{-n_\mathcal{T}M_{n,\mathcal{T}}(\theta) +n_\mathcal{T} M_{n,\mathcal{T}}(\tilde{\theta}_\mathcal{T})\right\}d\theta \\
    &= C_n n_\mathcal{T}^{-d/2}\int_{\|u + n_\calT^{1/2}\tilde\theta_\calT\|\leq n_\mathcal{T}^{1/2}R} \left(\tilde\theta_{\mathcal{T},k} + \frac{u_k}{\sqrt{n_\mathcal{T}}}\right) \left(\tilde\theta_{\mathcal{T},l} + \frac{u_l}{\sqrt{n_\mathcal{T}}}\right) \exp\left\{-n_\mathcal{T}M_{n,\mathcal{T}}\left(\tilde\theta + \frac{u}{\sqrt{n_\mathcal{T}}}\right) + n_\mathcal{T}M_{n,\mathcal{T}}(\tilde\theta_\mathcal{T})\right\}du \\
    &= \tilde\theta_{\mathcal{T},k} \tilde\theta_{\mathcal{T},l} + \frac{C_n n_\mathcal{T}^{-d/2}}{\sqrt{n_\mathcal{T}}}\left(\tilde\theta_{\mathcal{T},k} I_l + \tilde\theta_{\mathcal{T},l} I_k\right) \\
    &\quad + \frac{C_n n_\mathcal{T}^{-d/2}}{n_\mathcal{T}}\int_{\|u + n_\calT^{1/2}\tilde\theta_\calT\|\leq n_\mathcal{T}^{1/2}R} u_k u_l \exp\left\{-n_\mathcal{T}M_{n,\mathcal{T}}\left(\tilde\theta_\mathcal{T} + \frac{u}{\sqrt{n_\mathcal{T}}}\right) + n_\mathcal{T}M_{n,\mathcal{T}}(\tilde\theta)\right\}du,
\end{align*}
where the normalizing constant $C_n$ and the integrals $I_k$ and $I_l$ are defined as in the proof of Lemma~\ref{MCMC estimator}. Let $\widehat\Gamma_{k,l}$ denote the $(k,l)$-th entry of the covariance matrix of the MCMC samples. Then, 
\begin{align*}
    \widehat\Gamma_{k,l} &= \tilde\theta_{\mathcal{T},k} \tilde\theta_{\mathcal{T},l} + \frac{C_n n_\mathcal{T}^{-d/2}}{\sqrt{n_\mathcal{T}}}\left(\tilde\theta_{\mathcal{T},k} I_l + \tilde\theta_{\mathcal{T},l} I_k\right) - \widehat\theta_{\mathcal{T},k} \widehat\theta_{\mathcal{T},l} \\
    &\quad + \frac{C_n n_\mathcal{T}^{-d/2}}{n_\mathcal{T}}\int_{\|u + n_\calT^{1/2}\tilde\theta_\calT\|\leq n_\mathcal{T}^{1/2}R} u_k u_l \exp\left\{-n_\mathcal{T}M_{n,\mathcal{T}}\left(\tilde\theta_\mathcal{T} + \frac{u}{\sqrt{n_\mathcal{T}}}\right) + n_\mathcal{T}M_{n,\mathcal{T}}(\tilde\theta)\right\}du \\
    &= \tilde\theta_{\mathcal{T},k} \tilde\theta_{\mathcal{T},l} + \tilde\theta_{\mathcal{T},k} (\widehat\theta_{\mathcal{T},l} - \tilde\theta_{\mathcal{T},l})+ \tilde\theta_{\mathcal{T},l} (\widehat\theta_{\mathcal{T},k} - \tilde\theta_{\mathcal{T},k}) - \widehat\theta_{\mathcal{T},k} \widehat\theta_{\mathcal{T},l} \\
    &\quad + \frac{C_n n_\mathcal{T}^{-d/2}}{n_\mathcal{T}}\int_{\|u + n_\calT^{1/2}\tilde\theta_\calT\|\leq n_\mathcal{T}^{1/2}R} u_k u_l \exp\left\{-n_\mathcal{T}M_{n,\mathcal{T}}\left(\tilde\theta_\mathcal{T} + \frac{u}{\sqrt{n_\mathcal{T}}}\right) + n_\mathcal{T}M_{n,\mathcal{T}}(\tilde\theta)\right\}du \\
    &= -(\widehat\theta_{\mathcal{T},l} - \tilde\theta_{\mathcal{T},l}) (\widehat\theta_{\mathcal{T},k} - \tilde\theta_{\mathcal{T},k}) \\
    &\quad + \frac{C_n n_\mathcal{T}^{-d/2}}{n_\mathcal{T}}\int_{\|u + n_\calT^{1/2}\tilde\theta_\calT\|\leq n_\mathcal{T}^{1/2}R} u_k u_l \exp\left\{-n_\mathcal{T}M_{n,\mathcal{T}}\left(\tilde\theta_\mathcal{T} + \frac{u}{\sqrt{n_\mathcal{T}}}\right) + n_\mathcal{T}M_{n,\mathcal{T}}(\tilde\theta)\right\}du,
\end{align*}
where the second equality follows from the definition of $I_k$ and $I_l$. Lemma~\ref{MCMC estimator} implies that $(\widehat\theta_{\mathcal{T},l} - \tilde\theta_{\mathcal{T},l}) (\widehat\theta_{\mathcal{T},k} - \tilde\theta_{\mathcal{T},k}) = o_P(n_\mathcal{T}^{-1})$, and we have already shown that $C_n n_\mathcal{T}^{-d/2} = O_P(1)$. Hence, to prove Lemma~\ref{second derivative estimator}, it suffices to show that
\begin{equation*}
    I_{kl} = C_n^{-1} n_\mathcal{T}^{d/2} (\Gamma_\mathcal{T})_{k,l} + o_P(1),
\end{equation*}
where
\begin{equation*}
    I_{kl} = \int_{\|u + n_\calT^{1/2}\tilde\theta_\calT\|\leq n_\mathcal{T}^{1/2}R} u_k u_l \exp\left\{-n_\mathcal{T}M_{n,\mathcal{T}}\left(\tilde\theta_\mathcal{T} + \frac{u}{\sqrt{n_\mathcal{T}}}\right) + n_\mathcal{T}M_{n,\mathcal{T}}(\tilde\theta)\right\}du.
\end{equation*}
Note that
\begin{equation*}
    C_n^{-1} n_\mathcal{T}^{d/2}(\Gamma_\mathcal{T})_{k,l} = \int u_k u_l \frac{C_n^{-1} n_\mathcal{T}^{d/2}}{(2\pi)^{d/2}\{\det(A_\mathcal{T})\}^{1/2}} \exp\left\{-\frac{1}{2}u^\top A_\mathcal{T}u\right\}du  \xrightarrow[]{p} \int u_k u_l \exp\left\{-\frac{1}{2}u^\top A_\mathcal{T}u\right\}du.
\end{equation*}
Moreover, $\int_{\|u + n_\calT^{1/2}\tilde\theta_\calT\| > n_\mathcal{T}^{1/2}R} u_k u_l \exp\{-u^\top A_\mathcal{T}u/2\}du = o_P(1)$ by dominated convergence theorem. Therefore, it suffices to show that
\begin{equation*}
    I_{kl} - \int_{\|u + n_\calT^{1/2}\tilde\theta_\calT\|\leq n_\mathcal{T}^{1/2}R} u_k u_l \exp\left\{-\frac{1}{2}u^\top A_\mathcal{T}u\right\}du = o_P(1).
\end{equation*}

The rest of the proof is similar to the proof of Lemma~\ref{MCMC estimator} where we bound a similar integral. Specifically, for sufficiently large $n_\mathcal{T}$, with probability at most $\epsilon$,
\begin{equation*}
    \sup_{\|\theta - \tilde\theta_\mathcal{T}\|_2 \leq C_3} \frac{|R_n(\theta,\tilde\theta_\mathcal{T})|}{\|\theta-\tilde\theta_\mathcal{T}\|_2^2 + n_{\mathcal{T}}^{-1}} > \frac{1}{8}\Lambda_{\min}(A_\mathcal{T}),
\end{equation*}
where $C_3$ is defined as in the proof of Lemma~\ref{MCMC estimator}. Let $C_2^\prime > 1$ be suffiently large such that $\int_{\|u\|> C_2^\prime} |u_k u_l| \exp\{-u^\top A_\mathcal{T} u/4\}du < \epsilon/2$. Then, with probability at most $\epsilon$,
\begin{multline*}
    \Big|\int_{C_2^\prime < \|u\| \leq n_\mathcal{T}^{1/2}C_3} u_k u_l \exp\left\{-n_\mathcal{T}M_{n,\mathcal{T}}\left(\tilde\theta_\mathcal{T} + \frac{u}{\sqrt{n_\mathcal{T}}}\right) + n_\mathcal{T}M_{n,\mathcal{T}}(\tilde\theta)\right\}du - \\
    \int_{C_2^\prime < \|u\| \leq n_\mathcal{T}^{1/2}C_3} u_k u_l \exp\left\{-\frac{1}{2}u^\top A_\mathcal{T}u\right\}du \Big| > \epsilon.
\end{multline*}
Moreover, \eqref{choice of epsilon1} implies that, with probability at least $1-\epsilon$
\begin{align*}
    &\int_{n_\mathcal{T}^{1/2}C_3 < \|u\|, \|u + n_\calT^{1/2}\tilde\theta_\calT\|\leq n_\mathcal{T}^{1/2}R} u_k u_l \exp\left\{-n_\mathcal{T}M_{n,\mathcal{T}}\left(\tilde\theta_\mathcal{T} + \frac{u}{\sqrt{n_\mathcal{T}}}\right) + n_\mathcal{T}M_{n,\mathcal{T}}(\tilde\theta)\right\}du \\
    &\leq n_{\mathcal{T}}^{(d+2)/2}\int_{C_3 < \|\theta-\tilde\theta_\mathcal{T}\|, \|\theta\| \leq R} |\theta_k - \tilde\theta_{\mathcal{T},k}||\theta_l - \tilde\theta_{\mathcal{T},l}|\exp\{- n_\mathcal{T}M_{n,\mathcal{T}}(\theta) + n_\mathcal{T}M_{n,\mathcal{T}}(\tilde\theta_\mathcal{T})\}d\theta \\
    &\leq n_{\mathcal{T}}^{(d+1)/2}\int_{C_3 < \|\theta-\tilde\theta_\mathcal{T}\|,\|\theta\| \leq R} |\theta_k - \tilde\theta_{\mathcal{T},k}||\theta_l - \tilde\theta_{\mathcal{T},l}|\exp\{- n_\mathcal{T}\epsilon_1\}d\theta \\
    &\leq n_{\mathcal{T}}^{(d+1)/2}\int_{C_3 < \|\theta-\tilde\theta_\mathcal{T}\|,\|\theta\| \leq R} (R+\|\theta_\calT\|)^2\exp\{- n_\mathcal{T}\epsilon_1\}d\theta,
\end{align*}
which is upper bounded by $\epsilon/2$ for sufficiently large $n_\mathcal{T}$ with high probability. At the same time, as $ n_\mathcal{T}^{1/2}C_3 > C_2^\prime$, $\int_{ n_\mathcal{T}^{1/2}C_3 < \|u\|, \|u + n_\calT^{1/2}\tilde\theta_\calT\|\leq n_\mathcal{T}^{1/2}R} u_k u_l \exp\left\{-\frac{1}{2}u^\top A_\mathcal{T}u\right\}du < \epsilon/2$. Hence with probability at most $\epsilon$,
\begin{multline*}
    \Big|\int_{n_\mathcal{T}^{1/2}C_3 < \|u\|,\|u + n_\calT^{1/2}\tilde\theta_\calT\|\leq n_\mathcal{T}^{1/2}R} u_k u_l \exp\left\{-n_\mathcal{T}M_{n,\mathcal{T}}\left(\tilde\theta_\mathcal{T} + \frac{u}{\sqrt{n_\mathcal{T}}}\right) + n_\mathcal{T}M_{n,\mathcal{T}}(\tilde\theta)\right\}du - \\
    \int_{n_\mathcal{T}^{1/2}C_3 < \|u\|, \|u + n_\calT^{1/2}\tilde\theta_\calT\|\leq n_\mathcal{T}^{1/2}R} u_k u_l \exp\left\{-\frac{1}{2}u^\top A_\mathcal{T}u\right\}du \Big| > \epsilon.
\end{multline*}
It remains to study the integral
\begin{align*}
    &\int_{ \|u\| \leq C_2^\prime} u_k u_l \exp\left\{-n_\mathcal{T}M_{n,\mathcal{T}}\left(\tilde\theta_\mathcal{T} + \frac{u}{\sqrt{n_\mathcal{T}}}\right) + n_\mathcal{T}M_{n,\mathcal{T}}(\tilde\theta)\right\}du -
    \int_{ \|u\| \leq C_2^\prime} u_k u_l \exp\left\{-\frac{1}{2}u^\top A_\mathcal{T}u\right\}du \\
    &= \int_{\|u\| \leq C_2^\prime}u_ku_l \exp\left\{-\frac{1}{2}u^\top A_\mathcal{T}u\right\} \left(\exp\left\{ - n_\mathcal{T}R_n\left(\tilde\theta_\mathcal{T}+\frac{u}{\sqrt{n_\mathcal{T}}}\right)\right\} -1\right)du.
\end{align*}
But we can bound this integral in a similar way as we study the integral $I_{j,3}$ in the proof of Lemma~\ref{MCMC estimator}. The details are omitted here for brevity.

Again, the above argument established Lemma~\ref{second derivative estimator} for $B = \infty$. In practice, $B$ can be made arbitrarily large to make the Monte Carlo error arbitrarily small. Similar argument as in the proof of Lemma~\ref{MCMC estimator} based on Chebyshev's inequality can be used to explicitly analyze the Monte Carlo error for finite $B$.
\end{proof}


\begin{lemma}\label{score variance estimator}
Suppose that $\tilde\theta_\calT$ is an asymptotic normal estimator of $\theta_\calT$. Under Conditions~\ref{cond:target quad} to \ref{cond:target perturb quad}, $\widehat\Sigma_{S,\mathcal{T}} = \textnormal{Var}[\sqrt{n_\mathcal{T}}S_{n,\mathcal{T}}(\theta_\mathcal{T})] + o_P(1)$.
\end{lemma}

\begin{proof}[Proof of Lemma~\ref{score variance estimator}]
Recall that for a given $\theta$, we have defined $V_\mathcal{T}(\theta) = \textnormal{Var}_W[M^\dagger_{n,\mathcal{T}}(\theta) - M^\dagger_{n,\mathcal{T}}(\widehat\theta_\mathcal{T})]$. By definition,
\begin{equation*}
     M^\dag_{n,\mathcal{T}}(\theta) - M^\dag_{n,\mathcal{T}}(\widehat\theta_\mathcal{T}) = S^\dag_{n,\mathcal{T}}(\widehat\theta_\mathcal{T})(\theta- \widehat\theta_\mathcal{T}) + \frac{1}{2}(\theta- \widehat\theta_\mathcal{T})^\top A_{\mathcal{T}} (\theta- \widehat\theta_\mathcal{T}) + R_n^\dagger(\theta,\widehat\theta_\mathcal{T}).
\end{equation*}
Note that the quadratic term $(\theta- \widehat\theta_\mathcal{T})^\top A_{\mathcal{T}} (\theta- \widehat\theta_\mathcal{T})$ does not depend on the weights $W$, and hence 
\begin{align}\label{eq: var decomp}
    V_\mathcal{T}(\theta) &= \textnormal{Var}_W\left[M^\dagger_{n,\mathcal{T}}(\theta) - M^\dagger_{n,\mathcal{T}}(\widehat\theta_\mathcal{T})\right] \nonumber \\
    &= \textnormal{Var}_W\left[S^\dag_{n,\mathcal{T}}(\widehat\theta_\mathcal{T})(\theta- \widehat\theta_\mathcal{T}) + R_n^\dagger(\theta,\widehat\theta_\mathcal{T})\right] \nonumber \\
    &= (\theta - \widehat\theta_\mathcal{T})^\top \Var_W[S^\dagger_{n,\mathcal{T}}(\widehat\theta_\mathcal{T})](\theta - \widehat\theta_\mathcal{T}) + 2\Cov_W\left(S^\dag_{n,\mathcal{T}}(\widehat\theta_\mathcal{T})(\theta- \widehat\theta_\mathcal{T}), R_n^\dagger(\theta,\widehat\theta_\mathcal{T})\right) + \Var_W\left[R_n^\dagger(\theta,\widehat\theta_\mathcal{T})\right]
\end{align}
For given $\theta$, let $\Theta(\theta) \in \mathbb{R}^{d(d+1)/2}$ denote the vector obtained by vectorizing the upper triangular part (including the diagonal) of $(\theta - \widehat\theta_\mathcal{T})(\theta - \widehat\theta_\mathcal{T})^\top$ and let $\Theta(\theta)_{k,l}$ denote the component of $\Theta(\theta)$ that corresponds to the $(k,l)$-th entry in the matrix $(\theta - \widehat\theta_\mathcal{T})(\theta - \widehat\theta_\mathcal{T})^\top$ for $k \leq l$. For notational convenience, we will often write $\Theta(\theta)$ simply as $\Theta$. Let $\gamma \in \mathbb{R}^{d(d+1)/2}$ denote the vector whose $(k,l)$-th component is such that 
\begin{equation*}
    \left(\Var_W[S^\dagger_{n,\mathcal{T}}(\widehat\theta_\mathcal{T})]\right)_{k,l} = 
    \begin{cases}
    \gamma_{k,k}, & \text{if } 1\leq k=l \leq d\\
    \gamma_{k,l}/2, &\text{if } 1 \leq k < l \leq d \\
    \gamma_{l,k}/2, &\text{if } 1 \leq l < k \leq d
    \end{cases}
\end{equation*}
Moreover, define an error term $\varepsilon(\theta)$ as
\begin{equation*}
    \varepsilon(\theta) = 2\Cov_W\left(S^\dag_{n,\mathcal{T}}(\widehat\theta_\mathcal{T})(\theta- \widehat\theta_\mathcal{T}), R_n^\dagger(\theta,\widehat\theta_\mathcal{T})\right) + \Var_W\left[R_n^\dagger(\theta,\widehat\theta_\mathcal{T})\right].
\end{equation*}
Then, \eqref{eq: var decomp} can be compactly written as $V_\mathcal{T}(\theta) = \gamma^\top \Theta + \varepsilon(\theta)$. Recall that we use $\widehat\Theta_\mathcal{T}^{*,j}$ to denote $\Theta(\widehat\theta_\mathcal{T}^{*,j})$, for $j \in \mathcal{S}$ where $\mathcal{S}$ is an index set of the MCMC samples that are sufficiently close to $\widehat\theta_\mathcal{T}$. Then,
\begin{equation*}
    V(\widehat\theta_\mathcal{T}^{*,j}) = \gamma^\top \widehat\Theta_\mathcal{T}^{*,j} + \varepsilon(\widehat\theta_\mathcal{T}^{*,j}), \quad \forall j \in \mathcal{S}.
\end{equation*}
Without loss of generality, we assume $\sum_{j \in \mathcal{S}} (\widehat\Theta_\mathcal{T}^{*,j})_{k,l}(\widehat\Theta_\mathcal{T}^{*,j})_{u,v} = 0$ if $k \neq u$ or $l \neq v$. Note that this can be easily achieved by defining a new set of points of cardinality $2^d|\mathcal{S}|$ from $\{\widehat\theta_\mathcal{T}^{*,j}: j \in \mathcal{S}\}$ by symmetrizing each component of $\widehat\theta_\mathcal{T}^{*,j}$ around $\widehat\theta_\mathcal{T}$ and using them in the linear regression. Then, the off-diagonal elements of the Gram matrix in the linear regression are all 0. Hence, we have
\begin{equation*}
    \widehat\gamma_{k,l} = \frac{\sum_{j \in \mathcal{S}}(\widehat\Theta_\mathcal{T}^{*,j})_{k,l} V(\widehat\theta_\mathcal{T}^{*,j})}{\sum_{j \in \mathcal{S}}(\widehat\Theta_\mathcal{T}^{*,j})_{k,l}^2} = \gamma_{k,l} + \frac{\sum_{j \in \mathcal{S}}(\widehat\Theta_\mathcal{T}^{*,j})_{k,l} \varepsilon(\widehat\theta_\mathcal{T}^{*,j})}{\sum_{j \in \mathcal{S}}(\widehat\Theta_\mathcal{T}^{*,j})_{k,l}^2},
\end{equation*}
and hence the error can be written as
\begin{align*}
    \left|\widehat\gamma_{k,l} - \gamma_{k,l}\right| &= \left|\frac{\sum_{j \in \mathcal{S}}(\widehat\Theta_\mathcal{T}^{*,j})_{k,l} \varepsilon(\widehat\theta_\mathcal{T}^{*,j})}{\sum_{j \in \mathcal{S}}(\widehat\Theta_\mathcal{T}^{*,j})_{k,l}^2}\right| \\
    &= \frac{\left|\sum_{j \in \mathcal{S}}(\widehat\theta_\mathcal{T}^{*,j} - \widehat\theta_\mathcal{T})_{k}(\widehat\theta_\mathcal{T}^{*,j} - \widehat\theta_\mathcal{T})_{l} \cdot \varepsilon(\widehat\theta_\mathcal{T}^{*,j})\right|}{\sum_{j \in \mathcal{S}}(\widehat\theta_\mathcal{T}^{*,j} - \widehat\theta_\mathcal{T})_{k}^2(\widehat\theta_\mathcal{T}^{*,j} - \widehat\theta_\mathcal{T})_{l}^2} \\
    &\leq \sqrt{\frac{\sum_{j \in \mathcal{S}}\varepsilon^2(\widehat\theta_\mathcal{T}^{*,j})}{\sum_{j \in \mathcal{S}}(\widehat\theta_\mathcal{T}^{*,j} - \widehat\theta_\mathcal{T})_{k}^2(\widehat\theta_\mathcal{T}^{*,j} - \widehat\theta_\mathcal{T})_{l}^2 }},
\end{align*}
where the last line follows from Cauchy-Schwarz inequality. It suffices to study the two ratios
\begin{equation*}
    \pi_1 = \frac{\sum_{j \in \mathcal{S}} \Cov_W^2\left(S^\dag_{n,\mathcal{T}}(\widehat\theta_\mathcal{T})(\widehat\theta_\mathcal{T}^{*,j}- \widehat\theta_\mathcal{T}), R_n^\dagger(\widehat\theta_\mathcal{T}^{*,j},\widehat\theta_\mathcal{T})\right)}{\sum_{j \in \mathcal{S}}(\widehat\theta_\mathcal{T}^{*,j} - \widehat\theta_\mathcal{T})_{k}^2(\widehat\theta_\mathcal{T}^{*,j} - \widehat\theta_\mathcal{T})_{l}^2 } \quad \textnormal{and} \quad 
    \pi_2 = \frac{\sum_{j \in \mathcal{S}} \Var_W^2\left[R_n^\dagger(\widehat\theta_\mathcal{T}^{*,j},\widehat\theta_\mathcal{T})\right]}{\sum_{j \in \mathcal{S}}(\widehat\theta_\mathcal{T}^{*,j} - \widehat\theta_\mathcal{T})_{k}^2(\widehat\theta_\mathcal{T}^{*,j} - \widehat\theta_\mathcal{T})_{l}^2 }
\end{equation*}
separately, since
\begin{equation*}
    \varepsilon^2(\widehat\theta_\mathcal{T}^{*,j}) \leq 8\Cov_W^2\left(S^\dag_{n,\mathcal{T}}(\widehat\theta_\mathcal{T})(\widehat\theta_\mathcal{T}^{*,j}- \widehat\theta_\mathcal{T}), R_n^\dagger(\widehat\theta_\mathcal{T}^{*,j},\widehat\theta_\mathcal{T})\right) + \Var_W^2\left[R_n^\dagger(\widehat\theta_\mathcal{T}^{*,j},\widehat\theta_\mathcal{T})\right].
\end{equation*}

We first show that $n_\mathcal{T}^2\pi_1 = o_P(1)$, that is, for any $\epsilon>0$ and $\delta>0$, there exists a sufficiently large $N$ such that $P(n_\mathcal{T}^2\pi_1 > \tilde{C}\epsilon) < \delta$ for $n_\mathcal{T} > N$, where $\tilde{C}$ is a constant that we will specify later. By Cauchy-Schwarz inequality,
\begin{align*}
    n_\mathcal{T}^2\pi_1 &\leq \frac{ \sum_{j \in \mathcal{S}} n_\mathcal{T}^2\Var_W[S^\dag_{n,\mathcal{T}}(\widehat\theta_\mathcal{T})(\widehat\theta_\mathcal{T}^{*,j}- \widehat\theta_\mathcal{T})] \Var_W[R_n^\dagger(\widehat\theta_\mathcal{T}^{*,j},\widehat\theta_\mathcal{T})] }{ \sum_{j \in \mathcal{S}}(\widehat\theta_\mathcal{T}^{*,j} - \widehat\theta_\mathcal{T})_{k}^2(\widehat\theta_\mathcal{T}^{*,j} - \widehat\theta_\mathcal{T})_{l}^2 } \\
    &\leq \frac{ \sum_{j \in \mathcal{S}} n_\mathcal{T}(\widehat\theta_\mathcal{T}^{*,j}- \widehat\theta_\mathcal{T})^\top \Var_W[\sqrt{n_\mathcal{T}} S^\dag_{n,\mathcal{T}}(\widehat\theta_\mathcal{T})](\widehat\theta_\mathcal{T}^{*,j}- \widehat\theta_\mathcal{T}) \ \mathbb{E}_W[R_n^\dagger(\widehat\theta_\mathcal{T}^{*,j},\widehat\theta_\mathcal{T})^2] }{ \sum_{j \in \mathcal{S}}(\widehat\theta_\mathcal{T}^{*,j} - \widehat\theta_\mathcal{T})_{k}^2(\widehat\theta_\mathcal{T}^{*,j} - \widehat\theta_\mathcal{T})_{l}^2 }.
\end{align*}
Let $\Lambda_{\max}$ denote the maximum eigenvalue of the covariance matrix of the score function $\Var[\sqrt{n_\mathcal{T}}S_{n,\mathcal{T}}(\theta_\mathcal{T})]$. Suppose for now that $\Var_W[\sqrt{n_\mathcal{T}} S^\dag_{n,\mathcal{T}}(\widehat\theta_\mathcal{T})] \xrightarrow[]{p} \Var[\sqrt{n_\mathcal{T}}S_{n,\mathcal{T}}(\theta_\mathcal{T})]$ (we will establish this convergence later,) then for sufficiently large $n_\mathcal{T}$, with probability at least $1-\delta/2$,
\begin{align*}
    n_\mathcal{T}^2 \pi_1 &\leq \frac{\sum_{j \in \mathcal{S}} 1.1\Lambda_{\max}n_\mathcal{T}\|\widehat\theta_\mathcal{T}^{*,j} - \widehat\theta_\mathcal{T}\|_2^2 \ \mathbb{E}_W[R_n^\dagger(\widehat\theta_\mathcal{T}^{*,j},\widehat\theta_\mathcal{T})^2]}{\sum_{j \in \mathcal{S}}(\widehat\theta_\mathcal{T}^{*,j} - \widehat\theta_\mathcal{T})_{k}^2(\widehat\theta_\mathcal{T}^{*,j} - \widehat\theta_\mathcal{T})_{l}^2} \\
    &\leq 1.1\Lambda_{\max}C_1^2  \frac{\sum_{j \in \mathcal{S}} \mathbb{E}_W[R_n^\dagger(\widehat\theta_\mathcal{T}^{*,j},\widehat\theta_\mathcal{T})^2]}{\sum_{j \in \mathcal{S}}(\widehat\theta_\mathcal{T}^{*,j} - \widehat\theta_\mathcal{T})_{k}^2(\widehat\theta_\mathcal{T}^{*,j} - \widehat\theta_\mathcal{T})_{l}^2}
\end{align*}
where the last line follows from the fact that $\|\widehat\theta_{\mathcal{T}}^{*,j} - \widehat\theta_\mathcal{T}\|_2 \leq C_1 n_\mathcal{T}^{-1/2}$ for all $j \in \mathcal{S}$. In addition, by Condition~\ref{cond:target perturb quad}, for sufficiently large $n_\mathcal{T}$, 
\begin{equation*}
    P\left(\sup_{\|\theta_1 - \theta_2\|\leq \delta_{n_\calT}, \|\theta_2 - \theta_\mathcal{T}\| \leq \delta_{n_\calT}} \frac{\mathbb{E}_W[R_n^\dagger(\theta_1,\theta_2)^2]}{\{\|\theta_1 - \theta_2\|_2^2 + n_\mathcal{T}^{-1}\}^2} > \epsilon \right) < \frac{\delta}{4}.
\end{equation*}
Lemma~\ref{MCMC estimator} implies that for sufficiently large $n_\mathcal{T}$, $P(\|\widehat\theta_\mathcal{T} - \theta_\mathcal{T}\| \geq \delta_{n_\calT}) < \delta/4$. Moreover, $\|\widehat\theta_{\mathcal{T}}^{*,j} - \widehat\theta_\mathcal{T}\|_2 \leq C_1 n_\mathcal{T}^{-1/2} < \delta_{n_\calT} $ for all $j \in \mathcal{S}$ for sufficiently large $n_\mathcal{T}$. Therefore, for sufficiently large $n_\mathcal{T}$,
\begin{equation*}
    P\left(\sup_{j \in \mathcal{S}} \frac{\mathbb{E}_W[R_n^\dagger(\widehat\theta_\mathcal{T}^{*,j}, \widehat\theta_\mathcal{T})^2]}{\{\|\widehat\theta_\mathcal{T}^{*,j} - \widehat\theta_\mathcal{T}\|_2^2 + n_\mathcal{T}^{-1}\}^2} > \epsilon \right) < \frac{\delta}{2}.
\end{equation*}
Hence, with probability at least $1-\delta$,
\begin{align*}
    n_\mathcal{T}^2\pi_1 &\leq 1.1 \Lambda_{\max}C_1^2 \epsilon \frac{\sum_{j \in \mathcal{S}} \{\|\widehat\theta_\mathcal{T}^{*,j} - \widehat\theta_\mathcal{T}\|_2^2 + n_\mathcal{T}^{-1}\}^2 }{ \sum_{j \in \mathcal{S}}(\widehat\theta_\mathcal{T}^{*,j} - \widehat\theta_\mathcal{T})_{k}^2(\widehat\theta_\mathcal{T}^{*,j} - \widehat\theta_\mathcal{T})_{l}^2 } \\
    &= 1.1 \Lambda_{\max}C_1^2 \epsilon \frac{|\mathcal{S}|^{-1}\sum_{j \in \mathcal{S}} \{\|\sqrt{n_\mathcal{T}}(\widehat\theta_\mathcal{T}^{*,j} - \widehat\theta_\mathcal{T})\|_2^2 + 1\}^2 }{ |\mathcal{S}|^{-1}\sum_{j \in \mathcal{S}}\{\sqrt{n_\mathcal{T}}(\widehat\theta_\mathcal{T}^{*,j} - \widehat\theta_\mathcal{T})_{k}\}^2\{\sqrt{n_\mathcal{T}}(\widehat\theta_\mathcal{T}^{*,j} - \widehat\theta_\mathcal{T})_{l}\}^2 }.
\end{align*}
Note that $\widehat\theta_\mathcal{T}^{*,j}$ were obtained from an i.i.d. sample from the density function $L_{n,\mathcal{T}}$, and hence by law of large number, the averages in the numerator and denominator converge to the corresponding expectations under the density proportional to $L_{n,\mathcal{T}}$ truncated at $C_1n_\mathcal{T}^{-1/2}$, when the sample size of the MCMC sample $B$ is large so that $|\mathcal{S}|$ is large. The fact that $|\mathcal{S}|$ is large relies on the region $|\theta - \widehat\theta_\calT| \leq C_1n_\mathcal{T}^{-1/2}$ having a probability bounded away from 0 under $L_{n,\mathcal{T}}$, which can be shown using similar argument as in the proof of Lemma~\ref{MCMC estimator}. Again, similar to the proof of Lemma~\ref{MCMC estimator} and Lemma~\ref{second derivative estimator}, we can show that these expectations converge to the corresponding expectations of a truncated normal distribution. Thus, when $B$ is large enough, for sufficiently large $n_\mathcal{T}$,
\begin{equation*}
    P\left(n_\mathcal{T}^2\pi_1 > 1.1 \Lambda_{\max} C_1^2 C_2 \epsilon\right) < \delta,
\end{equation*}
where $C_2$ is a constant (defined in terms of $A_\mathcal{T}$) related to the moments of truncated normal distribution. The ratio $\pi_2$ can be studied in a similar fashion by noting that for given $\epsilon > 0$, 
\begin{align*}
    n_\mathcal{T}^2\pi_2 &\leq \frac{n_\mathcal{T}^2\sum_{j \in \mathcal{S}} \mathbb{E}_W^2\left[R_n^\dagger(\widehat\theta_\mathcal{T}^{*,j},\widehat\theta_\mathcal{T})^2\right]}{\sum_{j \in \mathcal{S}}(\widehat\theta_\mathcal{T}^{*,j} - \widehat\theta_\mathcal{T})_{k}^2(\widehat\theta_\mathcal{T}^{*,j} - \widehat\theta_\mathcal{T})_{l}^2 } \leq \epsilon^2 \frac{n_\mathcal{T}^2\sum_{j \in \mathcal{S}} \{\|\widehat\theta_\mathcal{T}^{*,j} - \widehat\theta_\mathcal{T}\|_2^2 + n_\mathcal{T}^{-1}\}^4}{\sum_{j \in \mathcal{S}}(\widehat\theta_\mathcal{T}^{*,j} - \widehat\theta_\mathcal{T})_{k}^2(\widehat\theta_\mathcal{T}^{*,j} - \widehat\theta_\mathcal{T})_{l}^2 },
\end{align*}
with high probability for sufficiently large $n_\mathcal{T}$. 

So far, we have shown that $n_\mathcal{T}^2\pi_1 = o_P(1)$ and $n_\mathcal{T}^2\pi_2 = o_P(1)$. Then,
\begin{equation*}
    n_\mathcal{T}|\widehat\gamma_{k,l} - \gamma_{k,l}| \leq \sqrt{8n_\mathcal{T}^2\pi_1 + 2n_\mathcal{T}^2\pi_2} = o_P(1),
\end{equation*}
which implies that
\begin{equation*}
    \widehat\Sigma_{S,\mathcal{T}} - \Var_W[\sqrt{n_\mathcal{T}}S_{n,\mathcal{T}}^\dagger(\widehat\theta_\mathcal{T})] = o_P(1).
\end{equation*}

Recall that
\begin{align*}
    S_{n,\mathcal{T}}(\theta) &= \binom{n_\mathcal{T}}{D}^{-1}\sum_{1\leq i_1 < i_2 < \cdots < i_D \leq n_\mathcal{T}}s_\mathcal{T}(X_{\mathcal{T},i_1},\ldots, X_{\mathcal{T},i_D};\theta); \\
    S^\dagger_{n,\mathcal{T}}(\theta) &= \binom{n_\mathcal{T}}{D}^{-1} \sum_{1\leq i_1 < i_2 < \cdots < i_D \leq n_\mathcal{T}} W_{k,i_1} W_{k,i_2} \cdots W_{k,i_D} s_\mathcal{T}(X_{\mathcal{T},i_1},\ldots,X_{\mathcal{T},i_D};\theta).
\end{align*}
To finish the proof, it remains to show that
\begin{equation*}
    \Var_W[\sqrt{n_\mathcal{T}}S_{n,\mathcal{T}}^\dagger(\widehat\theta_\mathcal{T})] - \Var[\sqrt{n_\mathcal{T}}S_{n,\mathcal{T}}(\theta_\mathcal{T})] = o_P(1).
\end{equation*}
Under continuity of $\Var[\sqrt{n_\mathcal{T}}S_{n,\mathcal{T}}(\theta)]$ at $\theta_\calT$, it suffices to show that $$\Var_W[\sqrt{n_\mathcal{T}}S_{n,\mathcal{T}}^\dagger(\widehat\theta_\mathcal{T})] - \Var[\sqrt{n_\mathcal{T}}S_{n,\mathcal{T}}(\widehat\theta_\mathcal{T})] = o_P(1).$$
This has been shown in \citet{han2022multiplier}.
\end{proof}

\subsection{U-processes with degenerate kernel functions}

In this section, we introduce results on U-processes with degenerate kernel functions, which will be useful in later proofs. Here, a function $g$ of the $D$-tuple $(x_1,x_2,\ldots,x_D)$ is called $P$-degenerate if $Pg(x_1,\ldots,x_{k-1},\cdot,x_{k+1},\ldots,x_D) = 0$ for all $(x_1,\ldots,x_{k-1},x_{k+1},\ldots,x_D)$ and $1\leq k \leq D$. Proofs of these results can be found in \cite{sherman1993limiting} and \cite{sherman1994maximal}. 

\begin{theorem}[Theorem 3 in \cite{sherman1993limiting}]\label{thm: sherman}
    Let $\mathcal{G} = \{g(\cdot,\cdot,\phi): \phi \in \Phi \}$ be a class of $P$-degenerate functions. Let $Q$ denote the product measure $P \times P$. Suppose there exists a point $\phi_0$ in $\Phi$ for which $g(\cdot,\cdot,\phi_0)$ is the constant function 0. If (i) $\mathcal{G}$ is Euclidean for a constant envelope, and (ii) $Qg(\cdot,\cdot,\phi)^2 \rightarrow 0$ as $\phi \rightarrow \phi_0$, then $U_n g(\cdot,\cdot,\phi) = o_P(1/n)$ uniformly over $o_P(1)$ neighborhoods of $\phi_0$. 
\end{theorem}

\begin{theorem}[Corollary 8 in Section 6 of \cite{sherman1994maximal}]\label{thm: sherman stronger}
    Let $\mathcal{G}$ be a class of $P$-degenerate functions on $\mathcal{S}^k$, $k\geq 1$. Suppose $\mathcal{G}$ has the form $\{g(\cdot,\phi): \phi \in \Phi\}$, and that $\phi_0$ is a point in $\Phi$ for which $g(\cdot,\phi_0) \equiv 0$. Let $P^k = P \times \cdots \times P$ ($k$ factors). If (i) $\mathcal{G}$ is Euclidean for an envelope $G$ satisfying $P^k G^2 < \infty$; (ii) $P^k|g(\cdot,\phi)| \rightarrow 0$ as $\phi \rightarrow \phi_0$, then uniformly over $o_P(1)$ neighborhoods of $\phi_0$, $U_n^k g(\cdot,\phi) = o_P(1/n^{k/2})$.
\end{theorem}

\subsection{Consistent estimation in the source site}

In this section, we show that (a) the regression-based estimate motivated by \eqref{eq: source quadratic approx} of the score vector in the source site, $\widehat S_{n,k}(\widehat\theta_\calT)$, achieves negligible error (Lemma~\ref{consistent estimation source score}); (b) the regression-based estimate of the second-derivative matrix, $\widehat A_k$, is consistent (Lemma~\ref{consistent estimation source score}); and (c) the perturbation-based estimate of the variance of the score, $\widehat\Sigma_{S,k}$ is consistent (Lemma~\ref{score variance estimator source}.)

\begin{lemma}\label{consistent estimation source score}
    Under Conditions~\ref{cond:target quad} to \ref{cond:sample size ratio}, for $1\leq k \leq K$, $\widehat S_{n,k}(\widehat\theta_\calT) - S_{n,k}(\widehat\theta_\calT) = o_P(n_k^{-1/2})$ and $\widehat A_k - A_k(\theta_\mathcal{T}) = o_P(1)$.
\end{lemma}

\begin{proof}[Proof of Lemma~\ref{consistent estimation source score}]
Recall that the target site broadcast $\{\widehat\theta_{\mathcal{T}}^{*,j}: j \in \mathcal{S}\}$, a collection of parameter values close to $\widehat\theta_\calT$, to the source sites. For each $j \in \mathcal{S}$, we defined the vector $\widehat\delta_\mathcal{T}^{*,j} = \widehat\theta_\mathcal{T}^{*,j}- \widehat\theta_\mathcal{T}$ and $\widehat\Theta_\mathcal{T}^{*,j}$ as the vectorized upper triangular part of $(\widehat\theta_\mathcal{T}^{*,j}- \widehat\theta_\mathcal{T})(\widehat\theta_\mathcal{T}^{*,j}- \widehat\theta_\mathcal{T})^\top$. Then,
\begin{align*}
    M_{n,k}(\widehat\theta_\mathcal{T}^{*,j}) - M_{n,k}(\widehat\theta_\mathcal{T}) &= S_{n,k}(\widehat\theta_\mathcal{T})^\top(\widehat\theta_\mathcal{T}^{*,j} - \widehat\theta_\mathcal{T}) + \frac{1}{2}(\widehat\theta_\mathcal{T}^{*,j} - \widehat\theta_\mathcal{T})^\top A_k(\theta_\mathcal{T})(\widehat\theta_\mathcal{T}^{*,j}-\widehat\theta_\mathcal{T}) + R_{n,k}(\widehat\theta_\mathcal{T}^{*,j},\widehat\theta_\mathcal{T}) \\
    &= S_{n,k}(\widehat\theta_\mathcal{T})^\top \widehat\delta_\mathcal{T}^{*,j} + \beta^{(k)}(\theta_\mathcal{T})^\top \widehat\Theta_\mathcal{T}^{*,j} + R_{n,k}(\widehat\theta_\mathcal{T}^{*,j},\widehat\theta_\mathcal{T}). 
\end{align*}
Again, without loss of generality, we assume that both $\sum_{j \in \mathcal{S}}\widehat\delta_\mathcal{T}^{*,j}(\widehat\delta_\mathcal{T}^{*,j})^\top$ and $\sum_{j \in S}\widehat\Theta_\mathcal{T}^{*,j}(\widehat\Theta_\mathcal{T}^{*,j})^\top$ are diagonal matrices, and $\sum_{j \in S} \widehat\delta_\mathcal{T}^{*,j}(\widehat\Theta_\mathcal{T}^{*,j})^\top$ is the zero matrix. We approximately calculate $S_{n,k}(\widehat\theta_\mathcal{T})$ and $A_k(\theta_\mathcal{T})$ by regressing $M_{n,k}(\widehat\theta_\mathcal{T}^{*,j}) - M_{n,k}(\widehat\theta_\mathcal{T})$ against $(\widehat\delta_\mathcal{T}^{*,j}, \widehat\Theta_\mathcal{T}^{*,j})$. 

\textbf{Properties of $\widehat S_{n,k}(\widehat\theta_\mathcal{T})$.} The coefficient corresponding to $\widehat\delta_\mathcal{T}^{*,j}$, $\widehat\alpha^{(k)}$, can be written as
\begin{equation}\label{eq:alpha hat}
    \widehat\alpha^{(k)}_u = \left(S_{n,k}(\widehat\theta_\mathcal{T})\right)_u + \frac{\sum_{j \in \mathcal{S}}R_{n,k}(\widehat\theta_\mathcal{T}^{*,j},\widehat\theta_\mathcal{T})(\widehat\delta_\mathcal{T}^{*,j})_u}{\sum_{j \in \mathcal{S}}(\widehat\delta_\mathcal{T}^{*,j})_u^2}.
\end{equation}
Hence, by applying Cauchy-Schwarz inequality, we can bound the (scaled) approximation error by
\begin{align*}
    n_k^{1/2}\left|\left(\widehat\alpha^{(k)} - S_{n,k}(\widehat\theta_\mathcal{T})\right)_u\right| &= \frac{\left|\sum_{j \in \mathcal{S}}n_k^{1/2}R_{n,k}(\widehat\theta_\mathcal{T}^{*,j},\widehat\theta_\mathcal{T})(\widehat\delta_\mathcal{T}^{*,j})_u\right|}{\sum_{j \in \mathcal{S}}(\widehat\delta_\mathcal{T}^{*,j})_u^2} \leq \sqrt{\frac{\sum_{j \in \mathcal{S}}n_k R_{n,k}^2(\widehat\theta_\mathcal{T}^{*,j},\widehat\theta_\mathcal{T})}{\sum_{j \in \mathcal{S}}(\widehat\theta_\mathcal{T}^{*,j} - \widehat\theta_\mathcal{T})_u^2}}.
\end{align*}

For any $\epsilon>0$, under Condition~\ref{cond:source quad}, for sufficiently large $n_k$, with probability at least $1-\epsilon/2$, $\left|R_{n,k}(\theta_1,\theta_2)\right| \leq \epsilon\left(\|\theta_1 - \theta_2\|_2^2 + n_k^{-1}\right)$ uniformly for $\|\theta_1-\theta_2\|_2 \leq \delta_{n_k}$, $\|\theta_2 - \theta_\mathcal{T}\|_2 \leq \delta_{n_k}$. Moreover, as Lemma~\ref{MCMC estimator} implies that $\widehat\theta_\mathcal{T}$ is an asymptotically normal estimator of $\theta_\mathcal{T}$ and $\|\widehat\theta_\mathcal{T}^{*,j} - \widehat\theta_\mathcal{T}\|_2 \leq C_1n_\mathcal{T}^{-1/2}$, for sufficiently large $n_\mathcal{T}$, $\|\widehat\theta_\mathcal{T} - \theta_\mathcal{T}\|_2 \leq \delta_{n_k}$ and $\|\widehat\theta_\mathcal{T}^{*,j} - \widehat\theta_\mathcal{T}\|_2 \leq \delta_{n_k}$ with probability at least $1-\epsilon/2$. As $\widehat\theta_\mathcal{T}$ is independent of the source site sample, we have that for sufficiently large $n_k$ and $n_\mathcal{T}$,
\begin{equation}
    R_{n,k}(\widehat\theta_\mathcal{T}^{*,j},\widehat\theta_\mathcal{T}) \leq \epsilon\left(\|\widehat\theta_\mathcal{T}^{*,j} - \widehat\theta_\mathcal{T}\|_2^2 + n_k^{-1}\right) \leq C\epsilon\left(n_\mathcal{T}^{-1} + n_k^{-1}\right),
\end{equation}
with probability at least $1-\epsilon$. Thus, with probability at least $1-\epsilon$,
\begin{align*}
    n_k^{1/2}\left|\left(\widehat\alpha^{(k)} - S_{n,k}(\widehat\theta_\mathcal{T})\right)_u\right| &\leq \epsilon \sqrt{\frac{\sum_{j \in \mathcal{S}}n_k\left(\|\widehat\theta_\mathcal{T}^{*,j}-\widehat\theta_\mathcal{T}\|_2^2 + n_k^{-1}\right)^2}{\sum_{j \in \mathcal{S}}(\widehat\theta_\mathcal{T}^{*,j} - \widehat\theta_\mathcal{T})_u^2}} \\
    &\leq \epsilon \left(C_1^2n_kn_\mathcal{T}^{-1} + 1\right)^{1/2} \sqrt{\frac{\sum_{j \in \mathcal{S}}\left(n_\mathcal{T}\|\widehat\theta_\mathcal{T}^{*,j}-\widehat\theta_\mathcal{T}\|_2^2 + n_\mathcal{T}n_k^{-1}\right)}{\sum_{j \in \mathcal{S}}n_\mathcal{T}(\widehat\theta_\mathcal{T}^{*,j} - \widehat\theta_\mathcal{T})_u^2}} \leq \tilde{C}\epsilon.
\end{align*}
for some constant $\tilde{C}$ that involves the second moment of a truncated normal distribution and $m_1,m_2$ in Condition~\ref{cond:sample size ratio}. The above argument applies to each element of $\widehat\alpha^{(k)}$, and therefore $\widehat\alpha^{(k)} - S_{n,k}(\widehat\theta_\mathcal{T}) = o_P(n_k^{-1/2})$.

\textbf{Properties of $\widehat A_k$.} Given a value of $\theta$, we define a vector $\beta^{(k)}(\theta)$ corresponding to the vectorized upper triangular part of $A_k(\theta)$, that is, a vector whose $(u,v)$-th entry satisfies the following:
\begin{equation}
    (A_k(\theta))_{u,v} = 
    \begin{cases}
    2(\beta^{(k)}(\theta))_{u,u}, & \text{if } 1\leq u=v \leq d; \\
    (\beta^{(k)}(\theta))_{u,v}, &\text{if } 1 \leq u < v \leq d; \\
    (\beta^{(k)}(\theta))_{v,u}, &\text{if} 1 \leq v < u \leq d.
    \end{cases}
\end{equation}
For the ease of notation, we write $\beta^{(k)}(\theta_\mathcal{T})$ simply as $\beta^{(k)}$ when it does not cause any confusion. Similar to \eqref{eq:alpha hat}, we have
\begin{equation}\label{eq:beta hat}
    \widehat\beta^{(k)}_{u,v} = \beta^{(k)}_{u,v} + \frac{\sum_{j \in \mathcal{S}} R_{n,k}(\widehat\theta_\mathcal{T}^{*,j},\widehat\theta_\mathcal{T})(\widehat\Theta_{\mathcal{T}}^{*,j})_{u,v}}{\sum_{j \in \mathcal{S}}(\widehat\Theta_{\mathcal{T}}^{*,j})_{u,v}^2},
\end{equation}
and hence for any given $\epsilon>0$,
\begin{align*}
    |\widehat\beta^{(k)}_{u,v} - \beta^{(k)}_{u,v}| &= \frac{\left|\sum_{j \in \mathcal{S}} R_{n,k}(\widehat\theta_\mathcal{T}^{*,j},\widehat\theta_\mathcal{T})(\widehat\Theta_{\mathcal{T}}^{*,j})_{u,v}\right|}{\sum_{j \in \mathcal{S}}(\widehat\Theta_{\mathcal{T}}^{*,j})_{u,v}^2} \\
    &\leq \sqrt{\frac{\sum_{j \in \mathcal{S}} R_{n,k}^2(\widehat\theta_\mathcal{T}^{*,j},\widehat\theta_\mathcal{T})}{\sum_{j \in \mathcal{S}} (\widehat\Theta_\mathcal{T}^{*,j})_{u,v}^2}} = \sqrt{\frac{\sum_{j \in \mathcal{S}} R_{n,k}^2(\widehat\theta_\mathcal{T}^{*,j},\widehat\theta_\mathcal{T})}{ \sum_{j \in \mathcal{S}} (\widehat\theta_\mathcal{T}^{*,j} - \widehat\theta_\mathcal{T})_u^2(\widehat\theta_\mathcal{T}^{*,j} - \widehat\theta_\mathcal{T})_v^2 }} \\
    &\leq \epsilon \sqrt{ \frac{ \sum_{j \in \mathcal{S}} \left( \|\widehat\theta_\mathcal{T}^{*,j} - \widehat\theta_\mathcal{T}\|_2^2 + n_k^{-1}\right)^2 }{ \sum_{j \in \mathcal{S}} (\widehat\theta_\mathcal{T}^{*,j} - \widehat\theta_\mathcal{T})_u^2(\widehat\theta_\mathcal{T}^{*,j} - \widehat\theta_\mathcal{T})_v^2 } } = \epsilon \sqrt{ \frac{ \sum_{j \in \mathcal{S}} \left( n_\mathcal{T}\|\widehat\theta_\mathcal{T}^{*,j} - \widehat\theta_\mathcal{T}\|_2^2 + n_\mathcal{T}n_k^{-1}\right)^2 }{ \sum_{j \in \mathcal{S}} n_\mathcal{T}(\widehat\theta_\mathcal{T}^{*,j} - \widehat\theta_\mathcal{T})_u^2 n_\mathcal{T}(\widehat\theta_\mathcal{T}^{*,j} - \widehat\theta_\mathcal{T})_v^2 } } \leq \tilde{C}\epsilon,
\end{align*}
where the last line holds with high probability for sufficiently large $n_k$ and $n_\mathcal{T}$, and $\tilde{C}$ is some constant involving the fourth-moment of a truncated normal distribution and $m_1$ in Condition~\ref{cond:sample size ratio}. The above holds for each entry of $\widehat\beta^{(k)}$, and hence this shows that $\widehat\beta^{(k)} - \beta^{(k)} = o_P(1)$ and consequently 
$\widehat A_k - A_k(\theta_\mathcal{T}) = o_P(1)$.

\end{proof}

\begin{lemma}\label{score variance estimator source}
    Under Conditions~\ref{cond:target quad} to \ref{cond:sample size ratio}, $\widehat\Sigma_{S,k} = \textnormal{Var}[\sqrt{n_k}S_{n,k}(\theta_\mathcal{T})] + o_P(1)$.
\end{lemma}

This result can be proved in a similar manner as in the proof of Lemma~\ref{score variance estimator}, and we omit the details here for brevity.

\subsection{Sufficient condition for Condition~\ref{cond:target perturb quad} based on uniform integrability}

Here, we introduce a general sufficient condition for the quadratic approximation condition on the perturbed target objective function. This condition is based on uniform integrability of the remainder $R_n^\dagger(\theta_1, \theta_2).$ However, in many cases, it is more straightforward to directly analyze the second moment of $R_n^\dagger(\theta_1, \theta_2)$, as we do for the examples in Section~\ref{sec:example}.

\begin{lemma}\label{convergence in moment}
Suppose that, for any $\epsilon > 0$, there exists a sufficiently small $\delta > 0$ such that
\begin{enumerate}[label=(\alph*)]
    \item \begin{equation*}
    \limsup_{n_\mathcal{T} \rightarrow \infty} P\left(\sup_{\|\theta_1 - \theta_2\|_2 \leq \delta, \ \|\theta_2 - \theta_\mathcal{T}\|_2 \leq \delta} \frac{|R_n^\dagger(\theta_1, \theta_2)|}{\|\theta_1 - \theta_2\|_2^2 + n_\mathcal{T}^{-1}} > \epsilon \right) < \epsilon.
    \end{equation*}
    \item 
    \begin{equation*}
    \lim_{U\rightarrow \infty} \left(\sup_{n_\mathcal{T} \in \mathbb{N}}\mathbb{E}_{X_\mathcal{T},W_\mathcal{T}}\left[\sup_{\|\theta_1 - \theta_2\|_2 \leq \delta, \ \|\theta_2 - \theta_\mathcal{T}\|_2 \leq \delta}\left\{\frac{R_n^\dagger(\theta_1, \theta_2)}{\|\theta_1 - \theta_2\|_2^2 + n_\mathcal{T}^{-1}}\right\}^2 \ind{\frac{|R_n^\dagger(\theta_1, \theta_2)|}{\|\theta_1 - \theta_2\|_2^2 + n_\mathcal{T}^{-1}} > \sqrt{U}}\right]\right) = 0.
    \end{equation*}
\end{enumerate}    
Then, Condition~\ref{cond:target perturb quad} holds. 
\end{lemma}

\begin{proof}[Proof of Lemma~\ref{convergence in moment}]
Fix a given $\epsilon > 0$. There exists a $\delta_1 > 0$ such that, by (b) in Condition~\ref{cond:target perturb quad}
\begin{equation*}
    \lim_{U\rightarrow \infty} \left(\sup_{n_\mathcal{T} \in \mathbb{N}}\mathbb{E}_{X,W}\left[\sup_{\|\theta_1 - \theta_2\|_2 \leq \delta_1, \ \|\theta_2 - \theta_\mathcal{T}\|_2 \leq \delta_1}\left\{\frac{R_n^\dagger(\theta_1, \theta_2)}{\|\theta_1 - \theta_2\|_2^2 + n_\mathcal{T}^{-1}}\right\}^2 \ind{\frac{|R_n^\dagger(\theta_1, \theta_2)|}{\|\theta_1 - \theta_2\|_2^2 + n_\mathcal{T}^{-1}} > \sqrt{U}}\right]\right) = 0,
\end{equation*}
which implies that there exists $\bar{U}>0$ such that
\begin{equation}\label{UI}
    \mathbb{E}_{X,W}\left[\sup_{\|\theta_1 - \theta_2\|_2 \leq \delta_1, \ \|\theta_2 - \theta_\mathcal{T}\|_2 \leq \delta_1}\left\{\frac{R_n^\dagger(\theta_1, \theta_2)}{\|\theta_1 - \theta_2\|_2^2 + n_\mathcal{T}^{-1}}\right\}^2 \ind{\frac{|R_n^\dagger(\theta_1, \theta_2)|}{\|\theta_1 - \theta_2\|_2^2 + n_\mathcal{T}^{-1}} > \sqrt{\bar{U}}}\right] < \frac{\epsilon^2}{8}
\end{equation}
for all $n_\mathcal{T}$; and by (a) in Condition~\ref{cond:target perturb quad}, 
\begin{equation*}
    \limsup_{n_\mathcal{T} \rightarrow \infty} P\left(\sup_{\|\theta_1 - \theta_2\|_2 \leq \delta_1, \ \|\theta_2 - \theta_\mathcal{T}\|_2 \leq \delta_1} \frac{|R_n^\dagger(\theta_1, \theta_2)|}{\|\theta_1 - \theta_2\|_2^2 + n_\mathcal{T}^{-1}} > \sqrt{\frac{\epsilon}{2}} \right) < \frac{\epsilon^2}{8\bar{U}},
\end{equation*}
which implies that for all $n_\mathcal{T} \geq N_1$
\begin{equation}\label{ratio small 1}
    P\left(\sup_{\|\theta_1 - \theta_2\|_2 \leq \delta_1, \ \|\theta_2 - \theta_\mathcal{T}\|_2 \leq \delta_1} \frac{|R_n^\dagger(\theta_1, \theta_2)|}{\|\theta_1 - \theta_2\|_2^2 + n_\mathcal{T}^{-1}} > \sqrt{\frac{\epsilon}{2}} \right) < \frac{\epsilon^2}{8\bar{U}},
\end{equation}
for some $N_1 \in \mathbb{N}$. 
We take $\delta = \delta_1$ and consider $n_\mathcal{T} \geq N_1$. To start, we decompose the expectation into several terms as follows
\begin{align*}
    \frac{\mathbb{E}_W[\{R_n^\dagger(\theta_1, \theta_2)\}^2]}{\{\|\theta_1 - \theta_2\|_2^2 + n_\mathcal{T}^{-1}\}^2} &= \mathbb{E}_W\left[\left\{\frac{R_n^\dagger(\theta_1,\theta_2)}{\|\theta_1-\theta_2\|_2^2 + n_\mathcal{T}^{-1}}\right\}^2\ind{\frac{|R_n^\dagger(\theta_1,\theta_2)|}{\|\theta_1-\theta_2\|_2^2 + n_\mathcal{T}^{-1}} > \sqrt{\frac{\epsilon}{2}}}\right] \\
    &\quad + \mathbb{E}_W\left[\left\{\frac{R_n^\dagger(\theta_1,\theta_2)}{\|\theta_1-\theta_2\|_2^2 + n_\mathcal{T}^{-1}}\right\}^2\ind{\frac{|R_n^\dagger(\theta_1,\theta_2)|}{\|\theta_1-\theta_2\|_2^2 + n_\mathcal{T}^{-1}} \leq \sqrt{\frac{\epsilon}{2}}}\right] \\
    &= \mathbb{E}_W\left[\left\{\frac{R_n^\dagger(\theta_1,\theta_2)}{\|\theta_1-\theta_2\|_2^2 + n_\mathcal{T}^{-1}}\right\}^2\ind{\frac{|R_n^\dagger(\theta_1,\theta_2)|}{\|\theta_1-\theta_2\|_2^2 + n_\mathcal{T}^{-1}} > \sqrt{\frac{\epsilon}{2}}}\ind{\frac{|R_n^\dagger(\theta_1,\theta_2)|}{\|\theta_1-\theta_2\|_2^2 + n_\mathcal{T}^{-1}} > \sqrt{\bar{U}}}\right] \\
    &\quad + \mathbb{E}_W\left[\left\{\frac{R_n^\dagger(\theta_1,\theta_2)}{\|\theta_1-\theta_2\|_2^2 + n_\mathcal{T}^{-1}}\right\}^2\ind{\frac{|R_n^\dagger(\theta_1,\theta_2)|}{\|\theta_1-\theta_2\|_2^2 + n_\mathcal{T}^{-1}} > \sqrt{\frac{\epsilon}{2}}}\ind{\frac{|R_n^\dagger(\theta_1,\theta_2)|}{\|\theta_1-\theta_2\|_2^2 + n_\mathcal{T}^{-1}} \leq \sqrt{\bar{U}}}\right] \\
    &\quad +  \mathbb{E}_W\left[\left\{\frac{R_n^\dagger(\theta_1,\theta_2)}{\|\theta_1-\theta_2\|_2^2 + n_\mathcal{T}^{-1}}\right\}^2\ind{\frac{|R_n^\dagger(\theta_1,\theta_2)|}{\|\theta_1-\theta_2\|_2^2 + n_\mathcal{T}^{-1}} \leq \sqrt{\frac{\epsilon}{2}}}\right] \\
    &\leq \mathbb{E}_W\left[\left\{\frac{R_n^\dagger(\theta_1,\theta_2)}{\|\theta_1-\theta_2\|_2^2 + n_\mathcal{T}^{-1}}\right\}^2\ind{\frac{|R_n^\dagger(\theta_1,\theta_2)|}{\|\theta_1-\theta_2\|_2^2 + n_\mathcal{T}^{-1}} > \sqrt{\bar{U}}}\right] \\
    &\quad + \bar{U}\mathbb{E}_W\left[\ind{\frac{|R_n^\dagger(\theta_1,\theta_2)|}{\|\theta_1-\theta_2\|_2^2 + n_\mathcal{T}^{-1}} > \sqrt{\frac{\epsilon}{2}}}\right] + \frac{\epsilon}{2}.
\end{align*}
Therefore, 
\begin{align*}
    &\sup_{\|\theta_1 - \theta_2\|_2 \leq \delta, \ \|\theta_2 - \theta_\mathcal{T}\|_2 \leq \delta} \frac{\mathbb{E}_W[\{R_n^\dagger(\theta_1, \theta_2)\}^2]}{\{\|\theta_1 - \theta_2\|_2^2 + n_\mathcal{T}^{-1}\}^2} \\
    &\leq \frac{\epsilon}{2} + \bar{U} \sup_{\|\theta_1 - \theta_2\|_2 \leq \delta, \ \|\theta_2 - \theta_\mathcal{T}\|_2 \leq \delta} \mathbb{E}_W\left[\ind{\frac{|R_n^\dagger(\theta_1,\theta_2)|}{\|\theta_1-\theta_2\|_2^2 + n_\mathcal{T}^{-1}} > \sqrt{\frac{\epsilon}{2}}}\right] \\
    &\quad + \sup_{\|\theta_1 - \theta_2\|_2 \leq \delta, \ \|\theta_2 - \theta_\mathcal{T}\|_2 \leq \delta} \mathbb{E}_W\left[\left\{\frac{R_n^\dagger(\theta_1,\theta_2)}{\|\theta_1-\theta_2\|_2^2 + n_\mathcal{T}^{-1}}\right\}^2\ind{\frac{|R_n^\dagger(\theta_1,\theta_2)|}{\|\theta_1-\theta_2\|_2^2 + n_\mathcal{T}^{-1}} > \sqrt{\bar{U}}}\right].
\end{align*}
And hence,
\begin{align*}
    &P\left(\sup_{\|\theta_1 - \theta_2\|_2 \leq \delta, \ \|\theta_2 - \theta_\mathcal{T}\|_2 \leq \delta} \frac{\mathbb{E}_W[\{R_n^\dagger(\theta_1, \theta_2)\}^2]}{\{\|\theta_1 - \theta_2\|_2^2 + n_\mathcal{T}^{-1}\}^2} > \epsilon \right) \\
    &\leq \underbrace{P\left(\sup_{\|\theta_1 - \theta_2\|_2 \leq \delta, \ \|\theta_2 - \theta_\mathcal{T}\|_2 \leq \delta} \mathbb{E}_W\left[\ind{\frac{|R_n^\dagger(\theta_1,\theta_2)|}{\|\theta_1-\theta_2\|_2^2 + n_\mathcal{T}^{-1}} > \sqrt{\frac{\epsilon}{2}}}\right] > \frac{\epsilon}{4\bar{U}}\right)}_{\textnormal{prob.1}} \\
    &\quad + \underbrace{P\left( \sup_{\|\theta_1 - \theta_2\|_2 \leq \delta, \ \|\theta_2 - \theta_\mathcal{T}\|_2 \leq \delta} \mathbb{E}_W\left[\left\{\frac{R_n^\dagger(\theta_1,\theta_2)}{\|\theta_1-\theta_2\|_2^2 + n_\mathcal{T}^{-1}}\right\}^2\ind{\frac{|R_n^\dagger(\theta_1,\theta_2)|}{\|\theta_1-\theta_2\|_2^2 + n_\mathcal{T}^{-1}} > \sqrt{\bar{U}}}\right] > \frac{\epsilon}{4}\right)}_{\textnormal{prob.2}}.
\end{align*}
We will show that both prob.1 and prob.2 are upper bounded by $\epsilon/2$. 

We analyze prob.1 first. 
\begin{align*}
    &P\left(\sup_{\|\theta_1 - \theta_2\|_2 \leq \delta, \ \|\theta_2 - \theta_\mathcal{T}\|_2 \leq \delta} \mathbb{E}_W\left[\ind{\frac{|R_n^\dagger(\theta_1,\theta_2)|}{\|\theta_1-\theta_2\|_2^2 + n_\mathcal{T}^{-1}} > \sqrt{\frac{\epsilon}{2}}}\right] > \frac{\epsilon}{4\bar{U}}\right) \\
    &\leq P\left( \mathbb{E}_W\left[\ind{\sup_{\|\theta_1 - \theta_2\|_2 \leq \delta, \ \|\theta_2 - \theta_\mathcal{T}\|_2 \leq \delta}\frac{|R_n^\dagger(\theta_1,\theta_2)|}{\|\theta_1-\theta_2\|_2^2 + n_\mathcal{T}^{-1}} > \sqrt{\frac{\epsilon}{2}}}\right] > \frac{\epsilon}{4\bar{U}}\right) \\
    &= P\left( P_W\left(\sup_{\|\theta_1 - \theta_2\|_2 \leq \delta, \ \|\theta_2 - \theta_\mathcal{T}\|_2 \leq \delta}\frac{|R_n^\dagger(\theta_1,\theta_2)|}{\|\theta_1-\theta_2\|_2^2 + n_\mathcal{T}^{-1}} > \sqrt{\frac{\epsilon}{2}}\right) > \frac{\epsilon}{4\bar{U}}\right) \\
    &\leq \left(\frac{\epsilon}{4\bar{U}}\right)^{-1}\mathbb{E}_X\left[P_W\left(\sup_{\|\theta_1 - \theta_2\|_2 \leq \delta, \ \|\theta_2 - \theta_\mathcal{T}\|_2 \leq \delta}\frac{|R_n^\dagger(\theta_1,\theta_2)|}{\|\theta_1-\theta_2\|_2^2 + n_\mathcal{T}^{-1}} > \sqrt{\frac{\epsilon}{2}}\right)\right] \\
    &= \left(\frac{\epsilon}{4\bar{U}}\right)^{-1}P\left(\sup_{\|\theta_1 - \theta_2\|_2 \leq \delta, \ \|\theta_2 - \theta_\mathcal{T}\|_2 \leq \delta}\frac{|R_n^\dagger(\theta_1,\theta_2)|}{\|\theta_1-\theta_2\|_2^2 + n_\mathcal{T}^{-1}} > \sqrt{\frac{\epsilon}{2}}\right) \\
    &\leq \left(\frac{\epsilon}{4\bar{U}}\right)^{-1}\left(\frac{\epsilon^2}{8\bar{U}}\right) = \frac{\epsilon}{2},
\end{align*}
where the second inequality is a result of applying Markov's inequality and the third inequality follows from \eqref{ratio small 1}. Next, we analyze prob.2. Applying Markov's inequality, we get
\begin{align*}
    &P\left( \sup_{\|\theta_1 - \theta_2\|_2 \leq \delta, \ \|\theta_2 - \theta_\mathcal{T}\|_2 \leq \delta} \mathbb{E}_W\left[\left\{\frac{R_n^\dagger(\theta_1,\theta_2)}{\|\theta_1-\theta_2\|_2^2 + n_\mathcal{T}^{-1}}\right\}^2\ind{\frac{|R_n^\dagger(\theta_1,\theta_2)|}{\|\theta_1-\theta_2\|_2^2 + n_\mathcal{T}^{-1}} > \sqrt{\bar{U}}}\right] > \frac{\epsilon}{4}\right) \\
    &\leq P\left( \mathbb{E}_W\left[\sup_{\|\theta_1 - \theta_2\|_2 \leq \delta, \ \|\theta_2 - \theta_\mathcal{T}\|_2 \leq \delta} \left\{\frac{R_n^\dagger(\theta_1,\theta_2)}{\|\theta_1-\theta_2\|_2^2 + n_\mathcal{T}^{-1}}\right\}^2\ind{\frac{|R_n^\dagger(\theta_1,\theta_2)|}{\|\theta_1-\theta_2\|_2^2 + n_\mathcal{T}^{-1}} > \sqrt{\bar{U}}}\right] > \frac{\epsilon}{4}\right) \\
    &\leq \left(\frac{\epsilon}{4}\right)^{-1}\mathbb{E}_X\left[\mathbb{E}_W\left[\sup_{\|\theta_1 - \theta_2\|_2 \leq \delta, \ \|\theta_2 - \theta_\mathcal{T}\|_2 \leq \delta} \left\{\frac{R_n^\dagger(\theta_1,\theta_2)}{\|\theta_1-\theta_2\|_2^2 + n_\mathcal{T}^{-1}}\right\}^2\ind{\frac{|R_n^\dagger(\theta_1,\theta_2)|}{\|\theta_1-\theta_2\|_2^2 + n_\mathcal{T}^{-1}} > \sqrt{\bar{U}}}\right]\right] \\
    &= \left(\frac{\epsilon}{4}\right)^{-1}\mathbb{E}_{X,W}\left[\sup_{\|\theta_1 - \theta_2\|_2 \leq \delta, \ \|\theta_2 - \theta_\mathcal{T}\|_2 \leq \delta} \left\{\frac{R_n^\dagger(\theta_1,\theta_2)}{\|\theta_1-\theta_2\|_2^2 + n_\mathcal{T}^{-1}}\right\}^2\ind{\frac{|R_n^\dagger(\theta_1,\theta_2)|}{\|\theta_1-\theta_2\|_2^2 + n_\mathcal{T}^{-1}} > \sqrt{\bar{U}}}\right] \\
    &\leq \left(\frac{\epsilon}{4}\right)^{-1}\left(\frac{\epsilon^2}{8}\right) = \frac{\epsilon}{2},
\end{align*}
where the last inequality follows from \eqref{UI}. 

Combining our results so far, we have found sufficiently small $\delta > 0$ such that for sufficiently large $n_\mathcal{T}$,
\begin{equation*}
    P\left(\sup_{\|\theta_1 - \theta_2\|_2 \leq \delta, \ \|\theta_2 - \theta_\mathcal{T}\|_2 \leq \delta} \frac{\mathbb{E}_W[\{R_n^\dagger(\theta_1, \theta_2)\}^2]}{\{\|\theta_1 - \theta_2\|_2^2 + n_\mathcal{T}^{-1}\}^2} > \epsilon \right) < \epsilon.
\end{equation*}
Note that here $\delta$ is chosen based on $\epsilon$ independent of $n_\calT$, and we have proved a stronger version of Condition~\ref{cond:target perturb quad}. Indeed, for $n_\calT$ sufficiently large, $\delta_{n_\calT} \leq \delta$.
\end{proof}

\section{Proof of theorems and lemmas in Section~\ref{sec:theory}}

\begin{proof}[Proof of Theorem~\ref{thm:normality target estimator}]
The asymptotic normality of $\widehat\theta_\mathcal{T}$ is a direct result of the asymptotic normality of $\tilde\theta_\mathcal{T}$ combined with Lemma~\ref{MCMC estimator}. The consistency of the estimated variance is a result of Lemma~\ref{second derivative estimator} and Lemma~\ref{score variance estimator}. 
\end{proof}

\begin{proof}[Proof of Theorem~\ref{thm: test consistency}]
To show this result, we first establish an asymptotic expansion of $S_{n,k}(\widehat\theta_\calT)$. We then combine this expansion result with Lemmas~\ref{consistent estimation source score} and \ref{score variance estimator source} and apply Slutsky's theorem.

\textbf{An asymptotic expansion of $S_{n,k}(\widehat\theta_\mathcal{T})$.} We establish an asymptotic expansion of $S_{n,k}(\widehat\theta_\mathcal{T})$ and subsequently its asymptotic normality. Recall that for a generic distribution $Q$ and a generic function $f$ of $(X_1,\ldots,X_D)$ indexed by parameter $\theta$, we define $Qf(\theta) = \EE_Q[f(X_1,\ldots,X_D;\theta)]$. Furthermore, we use $U_{n,k}f(\theta)$ to denote the $U$-process indexed by $\theta$ such that $$U_{n,k} f(\theta) = \binom{n_k}{D}^{-1}\sum_{1\leq i_1 < i_2 < \cdots < i_D \leq n_k} f(X_{k,i_1},\ldots,X_{k,i_D};\theta).$$ With this notation, we can write $S_{n,k}(\theta)$ as $U_{n,k}s_k(\theta)$ and $\EE_{P_k}[s_k(X_1,\ldots,X_D;\theta)]$ as $P_ks_k(\theta)$.

\begin{align}\label{expansion of Snk first part}
    &S_{n,k}(\widehat\theta_\mathcal{T}) - P_ks_k(\theta_\mathcal{T}) \nonumber \\
    =& U_{n,k} s_k(\widehat\theta_\mathcal{T}) - P_k s_k(\theta_\mathcal{T}) \nonumber \\
    =& P_k\left\{ s_k(\widehat\theta_\mathcal{T}) - s_k(\theta_\mathcal{T})\right\} + (U_{n,k} - P_k) s_k(\theta_\mathcal{T}) + (U_{n,k} - P_k)\left\{ s_k(\widehat\theta_\mathcal{T}) - s_k(\theta_\mathcal{T})\right\}.
\end{align}
We study the first term on the second line more closely. Applying a Taylor expansion
\begin{align}\label{expansion of Snk second part}
    P_k\left\{ s_k(\widehat\theta_\mathcal{T}) - s_k(\theta_\mathcal{T})\right\} &= A_k(\bar\theta_\mathcal{T})(\widehat\theta_\mathcal{T} - \theta_\mathcal{T}) \nonumber \\
    &= A_k(\theta_\mathcal{T})(\widehat\theta_\mathcal{T} - \theta_\mathcal{T}) + \left(A_k(\bar\theta_\mathcal{T}) - A_k(\theta_\mathcal{T})\right)(\widehat\theta_\mathcal{T} - \theta_\mathcal{T}) \nonumber \\
    &= A_k(\theta_\mathcal{T})(\widehat\theta_\mathcal{T} - \theta_\mathcal{T}) + o_P(n_\mathcal{T}^{-1/2}),
\end{align}
for $\bar\theta_\mathcal{T}$ in between $\theta_\mathcal{T}$ and $\widehat\theta_\mathcal{T}$. Here, the last line follows from the continuity of $A_k(\theta)$ at $\theta_\mathcal{T}$ under Condition~\ref{cond:continuity of A} and the asymptotic normality of $\widehat\theta_\mathcal{T}$ from Theorem~\ref{thm:normality target estimator} which together imply that $\|(A_k(\bar\theta_\mathcal{T}) - A_k(\theta_\mathcal{T}))(\widehat\theta_\mathcal{T} - \theta_\mathcal{T})\|_2 \leq \|A_k(\bar\theta_\mathcal{T}) - A_k(\theta_\mathcal{T})\|_2\|\widehat\theta_\mathcal{T} - \theta_\mathcal{T}\|_2 = o_p(1)O_p(n_{\mathcal{T}}^{-1/2}) = o_P(n_\mathcal{T}^{-1/2})$. Furthermore, Condition~\ref{cond:empirical process} implies that $(U_{n,k} - P_k)\{s_k(\widehat\theta_\mathcal{T}) - s_k(\theta_\mathcal{T})\} = o_P(n_k^{-1/2})$. Combining this with \eqref{expansion of Snk first part} and \eqref{expansion of Snk second part}, we get that
\begin{equation}\label{expansion of Snk}
    S_{n,k}(\widehat\theta_\mathcal{T}) - P_ks_k(\theta_\mathcal{T}) = A_k(\theta_\mathcal{T})(\widehat\theta_\mathcal{T} - \theta_\mathcal{T}) + (U_{n,k} - P_k) s_k(\theta_\mathcal{T}) + o_P(n_\mathcal{T}^{-1/2}) + o_P(n_k^{-1/2}).
\end{equation}
By standard theory on U-statistics, $n_k^{1/2}(U_{n,k} - P_k) s_k(\theta_\mathcal{T}) \xrightarrow{d}{} N(0,\Var[n_k^{1/2}S_{n,k}(\theta_\mathcal{T})])$. Moreover, $\widehat\theta_\mathcal{T}$ is obtained using only target site sample, and is thus independent of the source site sample. This implies that $A_k(\theta_\mathcal{T})(\widehat\theta_\mathcal{T} - \theta_\mathcal{T})$ and $(U_{n,k} - P_k) s_k(\theta_\mathcal{T})$ are independent. Applying Lemma~\ref{MCMC estimator}, we have that
\begin{equation*}
    n_k^{1/2}\left\{S_{n,k}(\widehat\theta_\mathcal{T}) - P_ks_k(\theta_\mathcal{T})\right\} \xrightarrow{d}{} N\left(0, \Var[n_k^{1/2}S_{n,k}(\theta_\mathcal{T})] + \rho_kA_k(\theta_\mathcal{T}A_\mathcal{T}^{-1}\textnormal{Var}[\sqrt{n}S_{n,\mathcal{T}}(\theta_\mathcal{T})]A_\mathcal{T}^{-1}A_k(\theta_\mathcal{T})\right).
\end{equation*}
Furthermore, as $\widehat S_{n,k}(\widehat\theta_\mathcal{T}) - S_{n,k}(\widehat\theta_\mathcal{T}) = o_P(n_k^{-1/2})$ by Lemma~\ref{consistent estimation source score},
\begin{equation}\label{distribution of test stat}
    n_k^{1/2}\left\{\widehat S_{n,k}(\widehat\theta_\mathcal{T}) - P_ks_k(\theta_\mathcal{T})\right\} \xrightarrow{d}{} N\left(0, \Var[n_k^{1/2}S_{n,k}(\theta_\mathcal{T})] + \rho_kA_k(\theta_\mathcal{T})A_\mathcal{T}^{-1}\textnormal{Var}[\sqrt{n}S_{n,\mathcal{T}}(\theta_\mathcal{T})]A_\mathcal{T}^{-1}A_k(\theta_\mathcal{T})\right).
\end{equation}
Applying Lemmas~\ref{second derivative estimator}, \ref{score variance estimator}, \ref{consistent estimation source score} and \ref{score variance estimator source}, we can also show that the asymptotic variance in \eqref{distribution of test stat} can be consistently estimated by $\widehat\Sigma_{S,k} + n_kn_\mathcal{T}^{-1} \widehat A_k (\widehat A_\mathcal{T})^{-1} \widehat\Sigma_{S,\calT}(\widehat A_\calT)^{-1}\widehat A_k$.

\textbf{Combining with previous results.} Finally, by Condition~\ref{cond: spurious root}, for $A_k(\theta_\calT)$ positive definite, $P_ks_k(\theta_\calT) = P_ks_k(\theta_k) = 0$ if and only if $k \in \mathcal{K}$. Applying Slutsky's theorem and continuous mapping theorem, we get that $T_k \xrightarrow[]{d} \chi^2_d$ for $k \in \mathcal{K}$, while it diverges to infinity for $k \notin \mathcal{K}$. 

\textbf{Local power analysis of a test using $T_k$.} In fact, as a byproduct of the asymptotic expansion of $S_{n,k}(\widehat\theta_\calT)$ (and also $\widehat S_{n,k}(\widehat\theta_\calT)$), we get a local power analysis in testing the hypothesis $\theta_\calT = \theta_k$ using $T_k$ as the test statistic. Consider the local alternative $\theta_k = \theta_\mathcal{T} + h n_k^{-1/2}$. Now we compare the proposed test based on $\widehat S_{n,k}(\widehat\theta_\mathcal{T})$ with a Wald test using asymptotically normal estimators of both $\theta_k$ and $\theta_\mathcal{T}$. In particular, suppose the estimators $\widehat\theta_k$ and $\widehat\theta_\mathcal{T}$ satisfy the following
\begin{align*}
    \sqrt{n_k}(\widehat\theta_k - \theta_k) &\xrightarrow{d}{} N\left(0, (A_k)^{-1}\textnormal{Var}[\sqrt{n_k}S_{n,k}(\theta_k)](A_k)^{-1}\right), \\
    \sqrt{n_\mathcal{T}}(\widehat{\theta}_\mathcal{T} - \theta_\mathcal{T}) &\xrightarrow[]{d} N\left(0, (A_\mathcal{T})^{-1}\textnormal{Var}[\sqrt{n}S_{n,\mathcal{T}}(\theta_\mathcal{T})](A_\mathcal{T})^{-1}\right),
\end{align*}
and are mutually independent. This implies that
\begin{equation*}
    \sqrt{n_k}\left(\widehat\theta_k - \widehat\theta_\mathcal{T} \right) \xrightarrow[]{d} N\left(h, \ (A_k)^{-1}\textnormal{Var}[\sqrt{n_k}S_{n,k}(\theta_k)](A_k)^{-1} + n_kn_\mathcal{T}^{-1}(A_\mathcal{T})^{-1}\textnormal{Var}[\sqrt{n}S_{n,\mathcal{T}}(\theta_\mathcal{T})](A_\mathcal{T})^{-1}\right).
\end{equation*}
Now, recall that 
\begin{equation*}
    A_k(\theta) = \frac{\partial}{\partial\theta}\EE\left[s_k(X_{k,1},\ldots,X_{k,D}; \theta)\right] = \frac{\partial}{\partial \theta}P_ks_k(\theta).
\end{equation*}
Using a first-order Taylor expansion, $P_ks_k(\theta_k) - P_ks_k(\theta_\mathcal{T})  \approx A_k(\theta_\mathcal{T})(\theta_k - \theta_\mathcal{T}) = A_k(\theta_\mathcal{T}) h n_k^{-1/2}$. Then, combined with \eqref{distribution of test stat}, we get
\begin{equation*}
    n_k^{1/2}S_{n,k}(\widehat\theta_\mathcal{T}) \xrightarrow{d}{} N\left(-A_k(\theta_\mathcal{T})h, \Var[n_k^{1/2}S_{n,k}(\theta_\mathcal{T})] + \rho_kA_k(\theta_\mathcal{T})A_\mathcal{T}^{-1}\textnormal{Var}[\sqrt{n}S_{n,\mathcal{T}}(\theta_\mathcal{T})]A_\mathcal{T}^{-1}A_k(\theta_\mathcal{T})\right).
\end{equation*}
Therefore, if we use $\chi^2_d(c)$ to denote a non-central chi-squared distribution with $d$ degrees of freedom and non-centrality parameter $c$, we get
\begin{align*}
    T_k &\xrightarrow{d}{} \chi^2_d \left(h^\top A_k(\theta_\mathcal{T}) \left\{ \Var[n_k^{1/2}S_{n,k}(\theta_\mathcal{T})] + \rho_kA_k(\theta_\mathcal{T})A_\mathcal{T}^{-1}\textnormal{Var}[\sqrt{n}S_{n,\mathcal{T}}(\theta_\mathcal{T})]A_\mathcal{T}^{-1}A_k(\theta_\mathcal{T}) \right\}^{-1} A_k(\theta_\mathcal{T})h\right) \\
    &= \chi^2_d\left(h^\top \left\{ A_k(\theta_\mathcal{T})^{-1}\Var[n_k^{1/2}S_{n,k}(\theta_\mathcal{T})]A_k(\theta_\mathcal{T})^{-1} + \rho_kA_\mathcal{T}^{-1}\textnormal{Var}[\sqrt{n}S_{n,\mathcal{T}}(\theta_\mathcal{T})]A_\mathcal{T}^{-1} \right\}^{-1} h\right)
\end{align*}
Define the Wald-test statistic,
\begin{equation*}
    W_k =  n_k\left(\widehat\theta_k - \widehat\theta_\mathcal{T} \right)^\top \left\{A_k^{-1}\textnormal{Var}[\sqrt{n_k}S_{n,k}(\theta_k)]A_k^{-1} + \rho_kA_\mathcal{T}^{-1}\textnormal{Var}[\sqrt{n}S_{n,\mathcal{T}}(\theta_\mathcal{T})]A_\mathcal{T}^{-1}\right\}^{-1}\left(\widehat\theta_k - \widehat\theta_\mathcal{T} \right).
\end{equation*}
Then,
\begin{equation*}
    W_k \xrightarrow{d}{} \chi_d^2\left(h^\top \left\{A_k^{-1}\textnormal{Var}[\sqrt{n_k}S_{n,k}(\theta_\calT)]A_k^{-1} + \rho_kA_\mathcal{T}^{-1}\textnormal{Var}[\sqrt{n}S_{n,\mathcal{T}}(\theta_\mathcal{T})]A_\mathcal{T}^{-1}\right\}^{-1} h\right) 
\end{equation*}
Thus, $T_k$ and $W_k$ are asymptotically equivalent under local alternatives.

\end{proof}

\begin{proof}[Proof of Lemma~\ref{adaptive lasso solution}]
The proof follows similar arguments as in \citet{knight2000asymptotics} and \citet{zou2006adaptive}. Recall that the weights are chosen via solving the following penalized regression problem:
\begin{equation*}
    (\widehat\Lambda_1,\ldots,\widehat\Lambda_K) = \argmin_{\Lambda_k \in \mathbb{R}^{d\times d}, 1\leq k \leq K} \left\{\frac{1}{Q}\sum_{q=1}^Q \left\|\widehat\theta_\mathcal{T}^{(q)} - \sum_{k=1}^K \Lambda_k \widehat S_{n,k}^{(q)}\right\|_2^2 + \lambda\sum_{k=1}^K \frac{1}{p_k}\left\|\Lambda_k\right\|_1 \right\},
\end{equation*}
where the vector $(\widehat\theta_\mathcal{T}^{(q)}, \widehat S_{n,1}^{(q)}, \cdots, \widehat S_{n,K}^{(q)})^\top $ are i.i.d. following $N(0,\widehat\Omega)$ conditioning on the data. Equivalently, we can let $(\widehat\theta_\mathcal{T}^{(q)}, \widehat S_{n,1}^{(q)}, \cdots, \widehat S_{n,K}^{(q)})^\top = \widehat\Omega^{1/2}\varepsilon^{(q)},$ where $\{\varepsilon^{(1)},\ldots,\varepsilon^{(Q)}\}$ is an i.i.d. sample from $N(0,I_{d(K+1)})$ independent of all target and source data. 

\textbf{Some more notations.} To simplify notation, we collect all the weight matrices $\Lambda_k$ and define a matrix $\Lambda \in \mathbb{R}^{d(K+1)\times d}$ such that
\begin{equation*}
    \Lambda^\top = 
    \begin{pmatrix}
        I_d & -\Lambda_1 & \cdots & -\Lambda_K
    \end{pmatrix}.
\end{equation*}
We also collect the second derivative matrices in all source sites at $\theta_\calT$ to form the matrix $A_1^K = (A_1, \ldots, A_K)^\top \in \mathbb{R}^{dK \times d}$, and collect the variance of score in all source sites to form the block diagonal matrix $\Sigma_1^K = \textnormal{diag}(\rho_1^{-1}\Sigma_{S,1},\ldots,\rho_K^{-1}\Sigma_{S,K}) \in \mathbb{R}^{dK \times dK}$ where $\rho_k$ is the sample size ratio between source site $k$ and target site.
With these population quantities defined, we can now form the following population analogue of $\widehat\Omega$ defined in \eqref{eq: large normal vector}:
\begin{equation*}
    \Omega = 
    \begin{pmatrix}
        (A_\mathcal{T})^{-1}\Sigma_{S,\mathcal{T}}(A_\mathcal{T})^{-1} & ( A_\mathcal{T})^{-1} \Sigma_{S,\mathcal{T}} (A_\mathcal{T})^{-1} (A_1^K)^\top \\
        A_1^K (A_\mathcal{T})^{-1} \Sigma_{S,\mathcal{T}} (A_\mathcal{T})^{-1} & \Sigma_1^K + A_1^K (A_\mathcal{T})^{-1} \Sigma_{S,\mathcal{T}} (A_\mathcal{T})^{-1} (A_1^K)^\top 
    \end{pmatrix}.
\end{equation*}

\textbf{Probabilistic limit of the squared error.} In what follows, we will establish that for fixed $\Lambda$, as $n_\calT, n_k,$ and $Q$ all tend to infinity,
\begin{equation*}
   \frac{1}{Q}\sum_{q=1}^Q \left\|\widehat\theta_\mathcal{T}^{(q)} - \sum_{k=1}^K \Lambda_k \widehat S_{n,k}^{(q)}\right\|_2^2 = \frac{1}{Q}\sum_{q=1}^Q \left\|\Lambda^\top \widehat\Omega^{1/2}\varepsilon^{(q)}\right\|_2^2 = \frac{1}{Q}\sum_{q=1}^Q (\varepsilon^{(q)})^\top \widehat\Omega^{1/2} \Lambda \Lambda^\top \widehat\Omega^{1/2}\varepsilon^{(q)} \xrightarrow[]{p} \textnormal{tr}(\Lambda^\top \Omega \Lambda),
\end{equation*}
where the first two equalities hold via simply algebra and the convergence in probability is what we will show now. Note that by continuous mapping theorem $\textnormal{tr}(\Lambda^\top \widehat\Omega \Lambda) - \textnormal{tr}(\Lambda^\top \Omega \Lambda) = o_P(1)$, hence it suffices to show that $\sum_{q=1}^Q (\varepsilon^{(q)})^\top \widehat\Omega^{1/2} \Lambda \Lambda^\top \widehat\Omega^{1/2}\varepsilon^{(q)}/Q - \textnormal{tr}(\Lambda^\top \widehat\Omega \Lambda) = o_P(1)$, or equivalently, 
\begin{equation*}
    \lim P \left(\left|\frac{1}{Q}\sum_{q=1}^Q (\varepsilon^{(q)})^\top \widehat\Omega^{1/2} \Lambda \Lambda^\top \widehat\Omega^{1/2}\varepsilon^{(q)} - \textnormal{tr}(\Lambda^\top \widehat\Omega \Lambda)\right| > \epsilon \right) = 0,
\end{equation*}
for any positive $\epsilon$. Let $\mathcal{O}$ denote the observed data in the target and all source sites. By Chebyshev's inequality,
\begin{equation*}
   P \left(\left|\frac{1}{Q}\sum_{q=1}^Q (\varepsilon^{(q)})^\top \widehat\Omega^{1/2} \Lambda \Lambda^\top \widehat\Omega^{1/2}\varepsilon^{(q)} - \textnormal{tr}(\Lambda^\top \widehat\Omega \Lambda)\right| > \epsilon \mid \mathcal{O} \right) \leq \frac{m_4(\widehat\Omega)}{Q\epsilon^2} \leq \frac{2m_4(\Omega)}{Q\epsilon^2},
\end{equation*}
where $m_4(\widehat\Omega)$ denote the variance of $(\varepsilon^{(q)})^\top \widehat\Omega^{1/2} \Lambda \Lambda^\top \widehat\Omega^{1/2}\varepsilon^{(q)}$ conditioning on $\mathcal{O}$ and $m_4(\Omega)$ denote the variance of $(\varepsilon^{(q)})^\top \Omega^{1/2} \Lambda \Lambda^\top \Omega^{1/2}\varepsilon^{(q)}$. Here, the second inequality holds for sufficiently large $n_\calT$ and $n_k$ by continuous mapping theorem. Marginalizing over $\mathcal{Q}$, we get 
\begin{equation*}
   P \left(\left|\frac{1}{Q}\sum_{q=1}^Q (\varepsilon^{(q)})^\top \widehat\Omega^{1/2} \Lambda \Lambda^\top \widehat\Omega^{1/2}\varepsilon^{(q)} - \textnormal{tr}(\Lambda^\top \widehat\Omega \Lambda)\right| > \epsilon \right) \leq \frac{2m_4(\Omega)}{Q\epsilon^2} \rightarrow 0,
\end{equation*}
for any $\epsilon$ as $Q \rightarrow \infty.$ 

\textbf{Probabilistic limit of the penalty.} To start, we note that $p_k \sim \textnormal{Uniform}(0,1)$ for $k \in \mathcal{K}$, and hence $\lambda p_k^{-1} = o_P(1)$. For $k \notin \mathcal{K}$, $T_k \xrightarrow[]{p} \infty$ and
\begin{equation*}
    p_k = P(\chi^2_d > T_k) = P(\chi^2_d - d> T_k - d) \leq \exp\{-(T_k-d)/8\},
\end{equation*}
as a $\chi^2_d$ random variable is sub-exponential, which implies that $p_k^{-1} \geq \exp\{(T_k-d)/8\}$. As $T_k$ is of order $n_k$, $\lambda p_k^{-1} \xrightarrow[]{p} +\infty$ for $\lambda$ decaying with $n_k$ at any polynomial rate. This implies that the penalty term $\lambda\sum_{k=1}^K \frac{1}{p_k}\left\|\Lambda_k\right\|_1$ converges in probability to 0 if $\Lambda_k = 0$ for all $k \notin \mathcal{K}$ and to $+\infty$ otherwise.

\textbf{Probabilistic limit of the weights.} The objective function in the penalized regression in \eqref{eq: adaptive lasso} is convex in the weights $\Lambda_k$. This, combined with the pointwise convergence for fixed $\Lambda$, implies uniform convergence (see \citet{knight2000asymptotics}.) Following the epi-convergence results as in \citet{knight2000asymptotics} (see Theorem 1 therein), we get that the minimizer $\widehat\Lambda_k$ converges to the minimizer of the probabilistic limit of the objective function, denoted as $\Lambda_k^{opt}$. This minimizer must have $\Lambda_k^{opt} = 0$ for $k \notin \mathcal{K}$, as otherwise the probabilistic limit is infinity. Therefore, we now minimize $\textnormal{tr}(\Lambda^\top \Omega \Lambda)$ subject to the constraint that $\Lambda_k = 0$ for $k \notin \mathcal{K}$.

Without loss of generality, assume that $\mathcal{K} = \{1,\ldots,|\mathcal{K}|\}$, and for simplicity we define $J = |\mathcal{K}|$. Define the following submatrix of $\Omega$
\begin{equation*}
    \Omega_J = 
    \begin{pmatrix}
        (A_\mathcal{T})^{-1} \Sigma_{S,\mathcal{T}} (A_\mathcal{T})^{-1} & (A_\mathcal{T})^{-1}\Sigma_{S,\mathcal{T}}(A_\mathcal{T})^{-1} (A_1^J)^\top \\
        A_1^J (A_\mathcal{T})^{-1} \Sigma_{S,\mathcal{T}} (A_\mathcal{T})^{-1} & \Sigma_1^J + A_1^J (A_\mathcal{T})^{-1} \Sigma_{S,\mathcal{T}} (A_\mathcal{T})^{-1} (A_1^J)^\top
    \end{pmatrix}
\end{equation*}
with $A_1^J = (A_1, \ldots, A_J)^\top \in \mathbb{R}^{dJ \times d}$, and $\Sigma_1^J = \textnormal{diag}(\rho_1^{-1}\Sigma_{S,1},\ldots,\rho_J^{-1}\Sigma_{S,J}) \in \mathbb{R}^{dJ \times dJ}$. When $\Lambda_k = 0$ for $k > J$, $\textnormal{tr}(\Lambda^\top \Omega \Lambda) = \textnormal{tr}(\Lambda_J^\top \Omega_J \Lambda_J)$. Setting its derivative to 0, we get
\begin{align}\label{optimal weight}
    &\quad ((\Lambda_1^{opt})^\top \ \cdots \ (\Lambda_J^{opt})^\top)^\top \nonumber \\
    &= \left\{ \Sigma_1^J +   A_1^J (  A_\mathcal{T})^{-1} \Sigma_{S,\mathcal{T}}(  A_\mathcal{T})^{-1}(  A_1^J)^\top \right\}^{-1}   A_1^J ( A_\mathcal{T})^{-1} \Sigma_{S,\mathcal{T}}(  A_\mathcal{T})^{-1} \nonumber \\
    &= ( \Sigma_1^J)^{-1}  A_1^J (  A_\mathcal{T})^{-1} \Sigma_{S,\mathcal{T}}(  A_\mathcal{T})^{-1} \nonumber \\
    &\quad - ( \Sigma_1^J)^{-1}   A_1^J \left\{  A_\mathcal{T} ( \Sigma_{S,\mathcal{T}})^{-1}   A_\mathcal{T} + (  A_1^J)^\top ( \Sigma_1^J)^{-1}   A_1^J \right\}^{-1}(  A_1^J)^\top ( \Sigma_1^J)^{-1}  A_1^J (  A_\mathcal{T})^{-1} \Sigma_{S,\mathcal{T}}(  A_\mathcal{T})^{-1} \nonumber \\ 
    &= ( \Sigma_1^J)^{-1}  A_1^J (  A_\mathcal{T})^{-1} \Sigma_{S,\mathcal{T}}(  A_\mathcal{T})^{-1} - ( \Sigma_1^J)^{-1}  A_1^J (  A_\mathcal{T})^{-1} \Sigma_{S,\mathcal{T}}(  A_\mathcal{T})^{-1} \nonumber \\
    &\quad + ( \Sigma_1^J)^{-1}   A_1^J \left\{  A_\mathcal{T} ( \Sigma_{S,\mathcal{T}})^{-1}   A_\mathcal{T} + (  A_1^J)^\top ( \Sigma_1^J)^{-1}   A_1^J \right\}^{-1}   A_\mathcal{T} ( \Sigma_{S,\mathcal{T}})^{-1}   A_\mathcal{T} (  A_\mathcal{T})^{-1} \Sigma_{S,\mathcal{T}}(  A_\mathcal{T})^{-1} \nonumber \\
    &= ( \Sigma_1^J)^{-1}   A_1^J \left\{ A_\mathcal{T} ( \Sigma_{S,\mathcal{T}})^{-1}   A_\mathcal{T} + (  A_1^J)^\top ( \Sigma_1^J)^{-1}  A_1^J \right\}^{-1} \nonumber \\
    &= ( \Sigma_1^J)^{-1}   A_1^J \left\{ A_\mathcal{T} ( \Sigma_{S,\mathcal{T}})^{-1}   A_\mathcal{T} + \sum_{j=1}^J \rho_j A_j ( \Sigma_{S,j})^{-1}  A_j \right\}^{-1},
\end{align}
where we used the Woodbury matrix identity \citep{woodbury1949stability}
\begin{align*}
    &\quad \left\{ \Sigma_1^J +   A_1^J (  A_\mathcal{T})^{-1} \Sigma_{S,\mathcal{T}}(  A_\mathcal{T})^{-1}(  A_1^J)^\top \right\}^{-1} \nonumber \\
    &= ( \Sigma_1^J)^{-1} - ( \Sigma_1^J)^{-1}   A_1^J \left\{ A_\mathcal{T} ( \Sigma_{S,\mathcal{T}})^{-1}   A_\mathcal{T} + (  A_1^J)^\top ( \Sigma_1^J)^{-1}   A_1^J \right\}^{-1}(  A_1^J)^\top ( \Sigma_1^J)^{-1}.
\end{align*}

\textbf{Selection consistency.} The previous arguments showed that for $k \notin \mathcal{K}$, $\widehat\Lambda_k \xrightarrow[]{p} 0$. We now show that, in fact, with high probability $\widehat\Lambda_k$ is exactly 0, that is $P(\widehat\Lambda_k \neq 0) \rightarrow 0$. If $\widehat\Lambda_{k,ij} \neq 0$, the KKT condition implies that
\begin{equation}\label{eq: KKT}
    \left|\frac{1}{Q}\sum_{q=1}^{Q}2\left(\widehat\theta_{\calT,i}^{(q)} - \sum_{l=1}^K \widehat \Lambda_{l,i.}\widehat S_{n,l}^{(q)}\right)\left(S_{n,k}^{(q)}\right)_j\right| = \left|2\widehat\Lambda_{i.}\frac{1}{Q}\sum_{q=1}^{Q}\widehat\Omega^{1/2}\varepsilon^{(q)}\left(\widehat\Omega^{1/2}\varepsilon^{(q)}\right)_{kj}\right| = \frac{\lambda}{p_k},
\end{equation}
where the subscript $i.$ refers to the $i$-th row of a matrix. As $\widehat\Lambda_k \xrightarrow[]{p} \Lambda_k^{opt}$, and applying Chebyshev's inequality again we see that $\sum_{q=1}^{Q}\widehat\Omega^{1/2}\varepsilon^{(q)}(\widehat\Omega^{1/2}\varepsilon^{(q)})_{kj}/Q \xrightarrow[]{p} \EE_\varepsilon[\Omega^{1/2}\varepsilon(\Omega^{1/2}\varepsilon)_{kj}]$, we see that the middle term in \eqref{eq: KKT} converges in probability to a constant, that is, it is $O_P(1)$. On the other hand, when $k\notin \mathcal{K}$, $\lambda p_k^{-1}\xrightarrow[]{p} +\infty$. Hence, $P(\widehat\Lambda_{k,ij} \neq 0) \leq P(\eqref{eq: KKT} \textnormal{ holds}) \rightarrow 0$. This argument holds for each entry of $\widehat\Lambda_k$, and hence a union bound yields the desired result.

\end{proof}

\begin{proof}[Proof of Theorem~\ref{thm: asymptotic normality of combined}]
As in the proof of Lemma~\ref{adaptive lasso solution}, without loss of generality, suppose that $\mathcal{K} = \{1,\ldots,J\}$ where $J = |\mathcal{K}|$. Recall that the combined estimator $\widehat\theta_C = \widehat\theta_\mathcal{T} - \sum_{k=1}^K \widehat\Lambda_k \widehat S_{n,k}(\widehat\theta_\mathcal{T})$ and construct a hypothetical (requiring prior knowledge on $\mathcal{K}$) intermediate estimator $\bar\theta_C = \widehat\theta_\mathcal{T} - \sum_{j=1}^J \widehat\Lambda_j \widehat S_{n,j}(\widehat\theta_\mathcal{T})$. By Lemma~\ref{adaptive lasso solution}, as the number of sites $K$ is fixed, we have $P(\widehat\Lambda_k = \boldsymbol{0} \ \forall k \notin \mathcal{K}) \rightarrow 1$. This implies that with high probability $\widehat\theta_C$ and $\bar\theta_C$ coincide, and hence $n_\mathcal{T}^{1/2}(\widehat\theta_C - \bar\theta_C) \xrightarrow[]{p} 0$. By Slutsky's theorem, it suffices to show that $\bar\theta_C$ converges in distribution.

Recall the asymptotic normality result for $\widehat S_{n,k}(\widehat\theta_\mathcal{T})$ in \eqref{distribution of test stat}. This, combined with the asymptotic normality of $\widehat\theta_\mathcal{T}$ in Theorem~\ref{thm:normality target estimator}, leads to
\begin{equation*}
    \sqrt{n_\mathcal{T}}\left\{
    \begin{pmatrix}
        \widehat\theta_\mathcal{T} \\
        \widehat S_{n,1}(\widehat\theta_\mathcal{T}) \\
        \vdots \\
        \widehat S_{n,J}(\widehat\theta_\mathcal{T}) 
    \end{pmatrix}
    -
    \begin{pmatrix}
        \theta_\mathcal{T} \\
        \boldsymbol{0} \\
        \vdots \\
        \boldsymbol{0}
    \end{pmatrix}\right\}
    \xrightarrow[]{d}
    N\left(\boldsymbol{0}, \Omega_J\right),
\end{equation*}
where $\Omega_J$ is defined above display \eqref{optimal weight}. Applying Lemma~\ref{adaptive lasso solution}, we get
\begin{align*}
    \widehat\Lambda^\top_J &= 
    \begin{pmatrix}
        I_d & -\widehat\Lambda_1 & \cdots & -\widehat\Lambda_J
    \end{pmatrix}
    \xrightarrow[]{p}
    (\Lambda^{opt}_J)^\top = 
    \begin{pmatrix}
        I_d & -\Lambda_1^{opt} & \cdots & -\Lambda_J^{opt}
    \end{pmatrix}.
\end{align*}
Also note that
\begin{equation*}
    \bar\theta_C = \widehat\Lambda_J^\top \begin{pmatrix}
        \widehat\theta_\mathcal{T} \\
        \widehat S_{n,1}(\widehat\theta_\mathcal{T}) \\
        \vdots \\
        \widehat S_{n,J}(\widehat\theta_\mathcal{T}) 
    \end{pmatrix}.
\end{equation*}
By Slutsky's theorem
\begin{equation*}
    \sqrt{n_\mathcal{T}}\left\{
    \widehat\Lambda_J^\top
    \begin{pmatrix}
        \widehat\theta_\mathcal{T} \\
        \widehat S_{n,1}(\widehat\theta_\mathcal{T}) \\
        \vdots \\
        \widehat S_{n,J}(\widehat\theta_\mathcal{T}) 
    \end{pmatrix}
    -
    \widehat\Lambda_J^\top
    \begin{pmatrix}
        \theta_\mathcal{T} \\
        \boldsymbol{0} \\
        \vdots \\
        \boldsymbol{0}
    \end{pmatrix}\right\}
    \xrightarrow[]{d}
    N\left(\boldsymbol{0}, (\Lambda_J^{opt})^\top \Omega_J \Lambda_J^{opt}\right),
\end{equation*}
that is
\begin{equation*}
    \sqrt{n_\mathcal{T}}\left(\bar\theta_C - \theta_\mathcal{T}\right) \xrightarrow[]{d} N\left(\boldsymbol{0}, (\Lambda_J^{opt})^\top \Omega_J \Lambda_J^{opt}\right).
\end{equation*}

Combining all results and after simplification, we get
\begin{equation*}
    \sqrt{n_\mathcal{T}}\left(\widehat\theta_C - \theta_\mathcal{T}\right) \xrightarrow[]{d} N\left(0, (A_\mathcal{T}\Sigma_{S,\mathcal{T}}^{-1}A_\mathcal{T} + \sum_{k\in \mathcal{K}} \rho_k A_k\Sigma_{S,k}^{-1}A_k)^{-1}\right).
\end{equation*}
\end{proof}

\section{Proof of lemmas in Section~\ref{sec:example}}

\begin{proof}[Proof of Lemma~\ref{lemma: quad quantile regression}]
Define the function $\xi_{\tau,\beta}: (z,y) \mapsto \xi_\tau(y-\beta^\top z)$ and the function $s_{\tau,\beta}: (z,y) \mapsto z(\ind{y<\beta^\top z} - \tau)$. Recall that in the quantile regression example, we have
\begin{equation*}
    M_{n,\mathcal{T}}(\beta) = n_{\mathcal{T}}^{-1}\sum_{i=1}^{n_\mathcal{T}} (Y_{\mathcal{T},i} - \beta^\top Z_{\mathcal{T},i})\left(\tau - \ind{Y_{\mathcal{T},i} < \beta^\top Z_{\mathcal{T},i}} \right) = n_{\mathcal{T}}^{-1}\sum_{i=1}^{n_\mathcal{T}} \xi_{\tau,\beta}(Z_{\mathcal{T},i}, Y_{\mathcal{T},i}),
\end{equation*}
and
\begin{equation*}
    S_{n,\mathcal{T}}(\beta) = n_\mathcal{T}^{-1}\sum_{i=1}^{n_\mathcal{T}} Z_{\mathcal{T},i}\left(\ind{Y_{\mathcal{T},i} < \beta^\top Z_{\mathcal{T},i}} -\tau \right) =  n_{\mathcal{T}}^{-1}\sum_{i=1}^{n_\mathcal{T}} s_{\tau,\beta}(Z_{\mathcal{T},i}, Y_{\mathcal{T},i}).
\end{equation*}
Furthermore, for a generic function $f$ of $(Z,Y)$, we define $P_{n,\mathcal{T}}f = n_\mathcal{T}^{-1}\sum_{i=1}^{n_\mathcal{T}} f(Z_{\mathcal{T},i}, Y_{\mathcal{T},i})$ and use $P_\mathcal{T}f$ as a shorthand notation for $E_\mathcal{T}[f(Z_\mathcal{T},Y_\mathcal{T})]$. 

\textbf{Verification of Condition~\ref{cond:target quad}.} First, we apply simple algebra and simplify the expression $\xi_{\tau,\beta_1}(Z,Y)-\xi_{\tau,\beta_2}(Z,Y) - (\beta_1 - \beta_2)^\top s_{\tau,\beta_2}(Z,Y)$ for generic $\beta_1$, $\beta_2$, $Z$ and $Y$. To start, we note that this quantity simplifies to 0 when $Y < \min\{\beta_1^\top Z, \beta_2^\top Z\}$ or when $Y \geq \max\{\beta_1^\top Z, \beta_2^\top Z\}$. When $\beta_1^\top Z \leq Y < \beta_2^\top Z$,
\begin{align*}
    &\quad \xi_{\tau,\beta_1}(Z,Y)-\xi_{\tau,\beta_2}(Z,Y) - (\beta_1 - \beta_2)^\top s_{\tau,\beta_2}(Z,Y) \\
    &= (Y-\beta_1^\top Z)(\tau - \ind{Y < \beta_1^\top Z}) - (Y-\beta_2^\top Z)(\tau - \ind{Y < \beta_2^\top Z}) \\
    &\quad - (\ind{Y<\beta_2^\top Z} -\tau)Z^\top(\beta_1 - \beta_2) \\
    &= \tau\left( Y - \beta_1^\top Z\right) - (\tau - 1)\left( Y - \beta_2^\top Z\right) - (1-\tau)Z^\top (\beta_1 - \beta_2) \\
    &= Y - \beta_1^\top Z = \left|Y - \beta_1^\top Z\right|;
\end{align*}
and when $\beta_2^\top Z \leq Y < \beta_1^\top Z$,
\begin{align*}
    &\quad \xi_{\tau,\beta_1}(Z,Y)-\xi_{\tau,\beta_2}(Z,Y) - (\beta_1 - \beta_2)^\top s_{\tau,\beta_2}(Z,Y) \\
    &= (Y-\beta_1^\top Z)(\tau - \ind{Y < \beta_1^\top Z}) - (Y-\beta_2^\top Z)(\tau - \ind{Y < \beta_2^\top Z}) \\
    &\quad - (\ind{Y<\beta_2^\top Z} -\tau)Z^\top(\beta_1 - \beta_2) \\
    &= (\tau-1)\left( Y - \beta_1^\top Z\right) - \tau\left( Y - \beta_2^\top Z\right) + \tau Z^\top (\beta_1 - \beta_2) \\
    &= \beta_1^\top Z - Y = \left|Y - \beta_1^\top Z\right|.
\end{align*}

Now we study the remainder term in the quadratic approximation, which we write as  
\begin{align*}
    R_n(\beta_1,\beta_2) &= M_{n,\mathcal{T}}(\beta_1) - M_{n,\mathcal{T}}(\beta_2) - S_{n,\mathcal{T}}(\beta_2)(\beta_1 - \beta_2) - \frac{1}{2}(\beta_1 - \beta_2)^\top A_\mathcal{T}(\beta_1 - \beta_2) \\
    &= P_{n,\mathcal{T}}\left\{\xi_{\tau,\beta_1} - \xi_{\tau,\beta_2}\right\} - P_{n,\mathcal{T}}s_{\tau,\beta_2}(\beta_1 - \beta_2) - \frac{1}{2}(\beta_1 - \beta_2)^\top A_\mathcal{T}(\beta_1 - \beta_2) \\
    &= \left(P_{n,\mathcal{T}} - P_\mathcal{T}\right)\left\{\xi_{\tau,\beta_1} - \xi_{\tau,\beta_2}\right\} - \left(P_{n,\mathcal{T}} - P_\mathcal{T}\right)s_{\tau,\beta_2}(\beta_1 - \beta_2)  \\
    &\quad + P_\mathcal{T}\left\{\xi_{\tau,\beta_1} - \xi_{\tau,\beta_2} \right\} - P_\mathcal{T}s_{\tau,\beta_2}(\beta_1 - \beta_2) - \frac{1}{2}(\beta_1 - \beta_2)^\top A_\mathcal{T}(\beta_1 - \beta_2),
\end{align*}
for any $\beta_1,\beta_2$. Therefore, 
\begin{align*}
    &\quad P\left(\sup_{\|\theta_1 - \theta_2\|_2 \leq \delta, \ \|\theta_2 - \theta_\mathcal{T}\|_2 \leq \delta_{n_\mathcal{T}}} \frac{|R_n(\theta_1, \theta_2)|}{\|\theta_1 - \theta_2\|_2^2 + n_\mathcal{T}^{-1}} > \epsilon \right) \\
    &\leq P\left(\sup_{\|\beta_1 - \beta_2\|_2 \leq \delta, \ \|\beta_2 - \beta_\mathcal{T}\|_2 \leq \delta_{n_\mathcal{T}}} \frac{|\left(P_{n,\mathcal{T}} - P_\mathcal{T}\right)\left\{\xi_{\tau,\beta_1} - \xi_{\tau,\beta_2}\right\} - \left(P_{n,\mathcal{T}} - P_\mathcal{T}\right)s_{\tau,\beta_2}(\beta_1 - \beta_2)|}{\|\beta_1 - \beta_2\|_2^2 + n_\mathcal{T}^{-1}} > \frac{\epsilon}{2} \right) \\
    &\quad + P\left(\sup_{\|\beta_1 - \beta_2\|_2 \leq \delta, \ \|\beta_2 - \beta_\mathcal{T}\|_2 \leq \delta_{n_\mathcal{T}}} \frac{\left|P_\mathcal{T}\left\{\xi_{\tau,\beta_1} - \xi_{\tau,\beta_2} \right\} - P_\mathcal{T}s_{\tau,\beta_2}(\beta_1 - \beta_2) - \frac{1}{2}(\beta_1 - \beta_2)^\top A_\mathcal{T}(\beta_1 - \beta_2)\right|}{\|\beta_1 - \beta_2\|_2^2 + n_\mathcal{T}^{-1}} > \frac{\epsilon}{2} \right) \\
    &\leq \underbrace{P\left(\frac{|\left(P_{n,\mathcal{T}} - P_\mathcal{T}\right)\left\{\xi_{\tau,\beta_{1n}} - \xi_{\tau,\beta_{2n}}\right\} - \left(P_{n,\mathcal{T}} - P_\mathcal{T}\right)s_{\tau,\beta_{2n}}(\beta_{1n} - \beta_{2n})|}{\|\beta_{1n} - \beta_{2n}\|_2^2 + n_\mathcal{T}^{-1}} > \frac{\epsilon}{4} \right)}_{\textnormal{prob. 1}} \\
    &\quad + \underbrace{P\left(\sup_{\|\beta_1 - \beta_2\|_2 \leq \delta, \ \|\beta_2 - \beta_\mathcal{T}\|_2 \leq \delta_{n_\mathcal{T}}} \frac{P_\mathcal{T}\left\{\xi_{\tau,\beta_1} - \xi_{\tau,\beta_2} \right\} - P_\mathcal{T}s_{\tau,\beta_2}(\beta_1 - \beta_2) - \frac{1}{2}(\beta_1 - \beta_2)^\top A_\mathcal{T}(\beta_1 - \beta_2)}{\|\beta_1 - \beta_2\|_2^2 + n_\mathcal{T}^{-1}} > \frac{\epsilon}{2} \right)}_{\textnormal{prob. 2}} \\
    &\quad + \underbrace{P\left(\sup_{\delta_{n_\calT} < \|\beta_1 - \beta_2\|_2 \leq \delta, \ \|\beta_2 - \beta_\mathcal{T}\|_2 \leq \delta_{n_\mathcal{T}}} \frac{|\left(P_{n,\mathcal{T}} - P_\mathcal{T}\right)\left\{\xi_{\tau,\beta_1} - \xi_{\tau,\beta_2}\right\} - \left(P_{n,\mathcal{T}} - P_\mathcal{T}\right)s_{\tau,\beta_2}(\beta_1 - \beta_2)|}{\|\beta_1 - \beta_2\|_2^2 + n_\mathcal{T}^{-1}} > \frac{\epsilon}{2} \right)}_{\textnormal{prob. 3}},
\end{align*}
where $(\beta_{1n},\beta_{2n})$ is such that
\begin{align*}
    &\quad \frac{|\left(P_{n,\mathcal{T}} - P_\mathcal{T}\right)\left\{\xi_{\tau,\beta_{1n}} - \xi_{\tau,\beta_{2n}}\right\} - \left(P_{n,\mathcal{T}} - P_\mathcal{T}\right)s_{\tau,\beta_{2n}}(\beta_{1n} - \beta_{2n})|}{\|\beta_{1n} - \beta_{2n}\|_2^2 + n_\mathcal{T}^{-1}} \\
    &\geq \sup_{\|\beta_1 - \beta_2\|_2 \leq \delta_{n_\mathcal{T}}, \ \|\beta_2 - \beta_\mathcal{T}\|_2 \leq \delta_{n_\mathcal{T}}} \frac{|\left(P_{n,\mathcal{T}} - P_\mathcal{T}\right)\left\{\xi_{\tau,\beta_1} - \xi_{\tau,\beta_2}\right\} - \left(P_{n,\mathcal{T}} - P_\mathcal{T}\right)s_{\tau,\beta_2}(\beta_1 - \beta_2)|}{\|\beta_1 - \beta_2\|_2^2 + n_\mathcal{T}^{-1}} - \frac{\epsilon}{4},
\end{align*}
which exists by the definition of supremum and is random.

Now, we bound \textnormal{prob. 1} first. To this end, we define a function $g_{\beta_1,\beta_2}$ indexed by $(\beta_1,\beta_2)$ such that
\begin{equation*}
    g_{\beta_1,\beta_2}: (z,y) \mapsto \xi_{\tau,\beta_1}(z,y) - \xi_{\tau,\beta_2}(z,y) - (\beta_1 - \beta_2)^\top s_{\tau,\beta_2}(z,y).
\end{equation*}
Note that based on our simplification of this difference, we can equivalently write $g_{\beta_1,\beta_2}(z,y)$ as
\begin{equation*}
    g_{\beta_1,\beta_2}(z,y) = |y-z^\top \beta_1| \ind{\min\{z^\top \beta_1,z^\top \beta_2\} \leq y < \max\{z^\top \beta_1,z^\top \beta_2\}}.
\end{equation*}
Furthermore, define a function class $\mathcal{G}$ as 
\begin{equation*}
    \mathcal{G} = \{\|\beta_1 -\beta_2\|_2^{-1} g_{\beta_1,\beta_2}: 0<\|\beta_1 - \beta_2\|\leq \delta, \|\beta_2 - \beta_\mathcal{T}\|\leq \delta\}
\end{equation*}
for some small $\delta$. We will argue that the function class $\mathcal{G}$ is a VC-class, which is equivalent to showing the subgraphs of functions in $\mathcal{G}$ is VC. 
\begin{align*}
    \textnormal{subgraph}_\mathcal{G} 
    &= \{\{(y,z,t) \in \mathbb{R}\times\mathbb{R}^d\times\mathbb{R}:\|\beta_1 -\beta_2\|_2^{-1} g_{\beta_1,\beta_2}<t \} : 0<\|\beta_1 - \beta_2\|\leq \delta, \|\beta_2 - \beta_\mathcal{T}\|\leq \delta\} \\
    &= \{\{(y,z,t) \in \mathbb{R}\times\mathbb{R}^d\times\mathbb{R}: g_{\beta_1,\beta_2}<t\|\beta_1 -\beta_2\|_2 \} : 0<\|\beta_1 - \beta_2\|\leq \delta, \|\beta_2 - \beta_\mathcal{T}\|\leq \delta\}.
\end{align*}
This is in turn equivalent to showing that the class of indicator functions $\tilde{\mathcal{G}}$ defined below is a VC-class.
\begin{equation*}
    \tilde{\mathcal{G}} = \left\{ (z,y,t) \mapsto \ind{g_{\beta_1,\beta_2}(z,y)<t\|\beta_1 - \beta_2\|_2} :  0<\|\beta_1 - \beta_2\|\leq \delta, \|\beta_2 - \beta_\mathcal{T}\|\leq \delta\right\}.
\end{equation*}
The functions in $\tilde{\mathcal{G}}$ is indexed by finite-dimensional parameter $(\beta_1,\beta_2)$ and can be computed with finite arithmetic operations and comparisons, which implies that $\tilde{\mathcal{G}}$ is indeed a VC-class and hence implies that $\mathcal{G}$ is a VC-class. A similar argument was made in \citet[][Section 11]{pollard1990empirical} to show that a certain process is manageable \citep{pollard1990empirical}. Next, we note that
\begin{align*}
    \frac{g_{\beta_1,\beta_2}(z,y)}{\|\beta_1-\beta_2\|_2} &= \frac{|y-z^\top \beta_1| \ind{\min\{z^\top \beta_1,z^\top \beta_2\} \leq y < \max\{z^\top \beta_1,z^\top \beta_2\}}}{\|\beta_1 - \beta_2\|_2} \\
    &\leq \frac{|z^\top\beta_2-z^\top \beta_1| \ind{\min\{z^\top \beta_1,z^\top \beta_2\} \leq y < \max\{z^\top \beta_1,z^\top \beta_2\}}}{\|\beta_1 - \beta_2\|_2} \\
    &\leq \frac{\|z\|_2\|\beta_1-\beta_2\|_2}{\|\beta_1 - \beta_2\|_2} = \|z\|_2.
\end{align*}
Hence, functions in $\mathcal{G}$ has an envelope function $(z,y)\mapsto \|z\|_2$ which is square-integrable provided that $Z$ has finite second moment. Then by Lemma~19.15 in \citet{van2000asymptotic}, the class $\mathcal{G}$ is a Donsker class. Furthermore, under the assumption that $p_{Y|Z=z}(y) \leq h(z)$, we have
\begin{align*}
    \left\|\frac{g_{\beta_{1n},\beta_{2n}}}{\|\beta_{1n}-\beta_{2n}\|_2}\right\|_{L^2(P_\mathcal{T})}^2 
    &\leq \int \|z\|_2^2 P_{Y|Z=z}\left(\min\{z^\top \beta_{1n},z^\top \beta_{2n}\} \leq Y < \max\{z^\top \beta_{1n},z^\top \beta_{2n}\}\right)dP_\mathcal{T}(z) \\
    &\leq \int \|z\|_2^3\|\beta_{1n} - \beta_{2n}\|_2h(z)dP_\mathcal{T}(z) = o_P(1),
\end{align*}
since $\|\beta_{1n} - \beta_{2n}\|_2 \leq \delta_{n_\mathcal{T}} = o(1)$ provided that the integral $\int \|z\|_2^3h(z)dP_\mathcal{T}(z)$ is finite. By Lemma~19.24 in \citet{van2000asymptotic},
\begin{equation*}
    \left(P_{n,\mathcal{T}} - P_\mathcal{T}\right)\frac{g_{\beta_{1n},\beta_{2n}}}{\|\beta_{1n} - \beta_{2n}\|_2} = o_P(n_{\mathcal{T}}^{-1/2}),
\end{equation*}
and hence
\begin{equation*}
    \left(P_{n,\mathcal{T}} - P_\mathcal{T}\right)\left\{\xi_{\tau,\beta_{1n}} - \xi_{\tau,\beta_{2n}}\right\} - \left(P_{n,\mathcal{T}} - P_\mathcal{T}\right)s_{\tau,\beta_{2n}}(\beta_{1n} - \beta_{2n}) = o_P(\|\beta_{1n}-\beta_{2n}\|_2n_{\mathcal{T}}^{-1/2}) = o_P(\|\beta_{1n}-\beta_{2n}\|_2^2 + n_\mathcal{T}^{-1}).
\end{equation*}
This implies that prob. 1 converges to 0 as $n_\mathcal{T}$ approaches infinity.

Next, we study prob. 3. We note that
\begin{align*}
    &\quad P\left(\sup_{\delta_{n_\calT} < \|\beta_1 - \beta_2\|_2 \leq \delta, \ \|\beta_2 - \beta_\mathcal{T}\|_2 \leq \delta_{n_\mathcal{T}}} \frac{|\left(P_{n,\mathcal{T}} - P_\mathcal{T}\right)\left\{\xi_{\tau,\beta_1} - \xi_{\tau,\beta_2}\right\} - \left(P_{n,\mathcal{T}} - P_\mathcal{T}\right)s_{\tau,\beta_2}(\beta_1 - \beta_2)|}{\|\beta_1 - \beta_2\|_2^2 + n_\mathcal{T}^{-1}} > \frac{\epsilon}{2} \right) \\
    &\leq P\left(\sup_{\delta_{n_\calT} < \|\beta_1 - \beta_2\|_2 \leq \delta, \ \|\beta_2 - \beta_\mathcal{T}\|_2 \leq \delta_{n_\mathcal{T}}} \frac{|\left(P_{n,\mathcal{T}} - P_\mathcal{T}\right)\left\{\xi_{\tau,\beta_1} - \xi_{\tau,\beta_2}\right\} - \left(P_{n,\mathcal{T}} - P_\mathcal{T}\right)s_{\tau,\beta_2}(\beta_1 - \beta_2)|}{\delta_{n_\calT}^2 + n_\mathcal{T}^{-1}} > \frac{\epsilon}{2} \right) \\
    &= P\left(\sup_{\delta_{n_\calT} < \|\beta_1 - \beta_2\|_2 \leq \delta, \ \|\beta_2 - \beta_\mathcal{T}\|_2 \leq \delta_{n_\mathcal{T}}} \frac{|\left(P_{n,\mathcal{T}} - P_\mathcal{T}\right)g_{\beta_1,\beta_2}|}{\delta_{n_\calT}^2 + n_\mathcal{T}^{-1}} > \frac{\epsilon}{2} \right) \\
    &\leq P\left( \frac{\sup_{\|\beta_1 - \beta_2\|_2 \leq \delta, \ \|\beta_2 - \beta_\mathcal{T}\|_2 \leq \delta}|\left(P_{n,\mathcal{T}} - P_\mathcal{T}\right)g_{\beta_1,\beta_2}|}{\delta_{n_\calT}^2 + n_\mathcal{T}^{-1}} > \frac{\epsilon}{2} \right)
\end{align*}
With similar argument as before, we can show that the function class
$$\{g_{\beta_1,\beta_2}: \|\beta_1 - \beta_2\|\leq \delta, \|\beta_2 - \beta_\mathcal{T}\|\leq \delta\}$$
is a Donsker class for some small $\delta$. As a result, $\sup_{\|\beta_1 - \beta_2\|_2 \leq \delta, \ \|\beta_2 - \beta_\mathcal{T}\|_2 \leq \delta}|(P_{n,\mathcal{T}} - P_\mathcal{T})g_{\beta_1,\beta_2}| = O_p(n_\calT^{-1/2})$. Hence, the probability of interest will tend to 0 if $\delta_{n_\calT}$ is chosen such that $\delta_{n_\calT} \gg n_\calT^{-1/4}$.

Finally, we study prob. 2 and note that in fact the term inside the probability is deterministic and is related to the remainder term in a second-order Taylor expansion of $P_\mathcal{T}\xi_{\tau,\beta}$. To see this, we note that
\begin{align*}
    P_\mathcal{T}\xi_{\tau,\beta} = \iint_{-\infty}^{\beta^\top z}(\tau -1) (y-\beta^\top z)p_{Y|Z=z}(y)dydP_\mathcal{T}(z) + \iint_{\beta^\top z}^\infty \tau (y-\beta^\top z)p_{Y|Z=z}(y)dydP_\mathcal{T}(z),
\end{align*}
and
\begin{align*}
    \frac{\partial}{\partial \beta} P_\mathcal{T}\xi_{\tau,\beta} &= -\iint_{-\infty}^{\beta^\top z} (\tau-1)zp_{Y|Z=z}(y)dydP_\mathcal{T}(z) - \iint_{\beta^\top z}^{\infty} \tau zp_{Y|Z=z}(y)dydP_\mathcal{T}(z) \\
    &= \int (1-\tau) z F_{Y|Z=z}(\beta^\top z) dP_\mathcal{T}(z) - \int \tau z (1-F_{Y|Z=z}(\beta^\top z)) dP_\mathcal{T}(z) \\
    &= \int z (F_{Y|Z=z}(\beta^\top z) - \tau )dP_\mathcal{T}(z) \\
    &= \iint z (\ind{y<\beta^\top z} - \tau )p_{Y|Z=z}(y)dydP_\mathcal{T}(z) \\
    &= P_\mathcal{T}s_{\tau,\beta},
\end{align*}
and
\begin{equation*}
    \frac{\partial^2}{\partial \beta^2}P_\mathcal{T}\xi_{\tau,\beta} = \int z p_{Y|Z=z}(\beta^\top z)z^\top dP_\mathcal{T}(z) = A_\mathcal{T}(\beta).
\end{equation*}
Therefore, we get for some $\tilde\beta$ in between $\beta_1$ and $\beta_2$,
\begin{align*}
    &\quad P_\mathcal{T}\left\{\xi_{\tau,\beta_1} - \xi_{\tau,\beta_2} \right\} - P_\mathcal{T}s_{\tau,\beta_2}(\beta_1 - \beta_2) - \frac{1}{2}(\beta_1 - \beta_2)^\top A_\mathcal{T}(\beta_1 - \beta_2) \\
    &= \frac{1}{2} (\beta_1 - \beta_2)^\top \left\{A_\mathcal{T}(\tilde\beta)- A_\mathcal{T}(\beta_\mathcal{T})\right\}(\beta_1 - \beta_2) \\
    &\leq \frac{1}{2}\|\beta_1 - \beta_2\|_2^2 \left\|A_\mathcal{T}(\tilde\beta)- A_\mathcal{T}(\beta_\mathcal{T})\right\| \\
    &\leq L \|\beta_1 - \beta_2 \|_2^2 \delta,
\end{align*}
by the Lipschitz property of the function $A_\mathcal{T}(\beta)$. This implies that
\begin{equation*}
    \sup_{\|\beta_1 - \beta_2\|_2 \leq \delta, \ \|\beta_2 - \beta_\mathcal{T}\|_2 \leq \delta_{n_\mathcal{T}}} \frac{P_\mathcal{T}\left\{\xi_{\tau,\beta_1} - \xi_{\tau,\beta_2} \right\} - P_\mathcal{T}s_{\tau,\beta_2}(\beta_1 - \beta_2) - \frac{1}{2}(\beta_1 - \beta_2)^\top A_\mathcal{T}(\beta_1 - \beta_2)}{\|\beta_1 - \beta_2\|_2^2 + n_\mathcal{T}^{-1}} \leq L\delta,
\end{equation*}
which is smaller than $\epsilon/2$ with a sufficiently small choice of $\delta$.

\textbf{Verification of Condition~\ref{cond:target perturb quad}.} Recall that we have defined $R_n^\dag(\theta_1,\theta_2) = M_{n,\mathcal{T}}^\dag(\theta_1) - M_{n,\mathcal{T}}^\dag(\theta_2) - S_{n,\mathcal{T}}^\dag(\theta_2)(\theta_1 - \theta_2) - \frac{1}{2}(\theta_1 - \theta_2)^\top A_\mathcal{T}(\theta_1 - \theta_2)$. In this example, we have
\begin{align*}
    M_{n,\mathcal{T}}^\dag(\beta) &= n_\mathcal{T}^{-1}\sum_{i=1}^{n_\mathcal{T}}W_{\mathcal{T},i}\xi_{\tau,\beta}(Z_{\mathcal{T},i},Y_{\mathcal{T},i}) \\
    S_{n,\mathcal{T}}^\dag(\beta) &= n_\mathcal{T}^{-1}\sum_{i=1}^{n_\mathcal{T}}W_{\mathcal{T},i}s_{\tau,\beta}(Z_{\mathcal{T},i},Y_{\mathcal{T},i}).
\end{align*}
We verify Condition~\ref{cond:target perturb quad} when the weight $W$ corresponds to nonparametric bootstrap and therefore follows a multinomial distribution. First, we note that $\EE_W[R_n^\dag(\theta_1,\theta_2)]  = R_n(\theta_1,\theta_2)$ as $W_{\mathcal{T},i}$ has mean 1. In light of Condition~\ref{cond:target quad}, it suffices to show that
\begin{equation*}
    \lim_{n_\mathcal{T} \rightarrow \infty} P\left(\sup_{\|\theta_1 - \theta_2\|_2 \leq \delta_{n_\mathcal{T}}, \ \|\theta_2 - \theta_\mathcal{T}\|_2 \leq \delta_{n_\mathcal{T}}} \frac{\textnormal{Var}_W[\{R_n^\dagger(\theta_1, \theta_2)\}]}{\{\|\theta_1 - \theta_2\|_2^2 + n_\mathcal{T}^{-1}\}^2} > \epsilon \right) = 0.
\end{equation*}

To start, we note that
\begin{equation*}
    M_{n,\mathcal{T}}^\dag(\theta_1) - M_{n,\mathcal{T}}^\dag(\theta_2) - S_{n,\mathcal{T}}^\dag(\theta_2)(\theta_1 - \theta_2) = n_\mathcal{T}^{-1}\sum_{i=1}^{n_\mathcal{T}} W_{\mathcal{T},i}g_{\beta_1,\beta_2}(Z_{\mathcal{T},i},Y_{\mathcal{T},i}),
\end{equation*}
and therefore
\begin{align*}
    \textnormal{Var}_W[\{R_n^\dagger(\theta_1, \theta_2)\}] &= \textnormal{Var}_W\left[n_\mathcal{T}^{-1}\sum_{i=1}^{n_\mathcal{T}} W_{\mathcal{T},i}g_{\beta_{1},\beta_{2}}(Z_{\mathcal{T},i},Y_{\mathcal{T},i})\right] \\
    &= n_\mathcal{T}^{-2}\sum_{i=1}^{n_\mathcal{T}}\textnormal{Var}(W_{\mathcal{T},i})g_{\beta_{1},\beta_{2}}^2(Z_{\mathcal{T},i},Y_{\mathcal{T},i}) \\
    &\quad + n_\mathcal{T}^{-2}\sum_{i\neq j} \textnormal{Cov}(W_{\mathcal{T},i},W_{\mathcal{T},j})g_{\beta_{1},\beta_{2}}(Z_{\mathcal{T},i},Y_{\mathcal{T},i})g_{\beta_{1},\beta_{2}}(Z_{\mathcal{T},j},Y_{\mathcal{T},j}) \\
    &\leq n_\mathcal{T}^{-2}\sum_{i=1}^{n_\mathcal{T}}g_{\beta_{1},\beta_{2}}^2(Z_{\mathcal{T},i},Y_{\mathcal{T},i}),
\end{align*}
where the last line follows from the fact that $g_{\beta_1,\beta_2}(z,y)$ is nonnegative for any $(z,y)$ and $\textnormal{Var}(W_{\mathcal{T},i}) \leq 1$ and $\textnormal{Cov}(W_{\mathcal{T},i},W_{\mathcal{T},j}) < 0$ for the multinomial distribution. (The weights used in \citet{jin2001simple} is easier to analyze as the covariance terms vanish due to independence of $W_{\calT,i}$ and $W_{\calT,j}$, and we are left with the same variance terms.) To continue, we note that based on our simplification, we have
\begin{align*}
n_\mathcal{T}^{-2}\sum_{i=1}^{n_\mathcal{T}} g_{\beta_{1},\beta_{2}}^2(Z_{\mathcal{T},i},Y_{\mathcal{T},i}) &= n_\mathcal{T}^{-2}\sum_{i=1}^{n_\mathcal{T}} |Y_{\mathcal{T},i}-Z_{\mathcal{T},i}^\top \beta_1|^2 \ind{\min\{Z_{\mathcal{T},i}^\top \beta_1,Z_{\mathcal{T},i}^\top \beta_2\} \leq Y_{\mathcal{T},i} < \max\{Z_{\mathcal{T},i}^\top \beta_1,Z_{\mathcal{T},i}^\top \beta_2\}} \\
&\leq n_\mathcal{T}^{-2} \sum_{i=1}^{n_\mathcal{T}} \|Z_{\mathcal{T},i}\|_2^2\|\beta_1-\beta_2\|_2^2 \ind{\min\{Z_{\mathcal{T},i}^\top \beta_1,Z_{\mathcal{T},i}^\top \beta_2\} \leq Y_{\mathcal{T},i} < \max\{Z_{\mathcal{T},i}^\top \beta_1,Z_{\mathcal{T},i}^\top \beta_2\}},
\end{align*}
and therefore
\begin{align*}
    \frac{\textnormal{Var}_W[\{R_n^\dagger(\beta_1, \beta_2)\}]}{\{\|\beta_1 - \beta_2\|_2^2 + n_\mathcal{T}^{-1}\}^2} &\leq \frac{n_\mathcal{T}^{-2} \sum_{i=1}^{n_\mathcal{T}} \|Z_{\mathcal{T},i}\|_2^2\|\beta_1-\beta_2\|_2^2 \ind{\min\{Z_{\mathcal{T},i}^\top \beta_1,Z_{\mathcal{T},i}^\top \beta_2\} \leq Y_{\mathcal{T},i} < \max\{Z_{\mathcal{T},i}^\top \beta_1,Z_{\mathcal{T},i}^\top \beta_2\}}}{\{\|\beta_1 - \beta_2\|_2^2 + n_\mathcal{T}^{-1}\}^2} \\
    &\leq n_\mathcal{T}^{-1} \sum_{i=1}^{n_\mathcal{T}} \|Z_{\mathcal{T},i}\|_2^2 \ind{\min\{Z_{\mathcal{T},i}^\top \beta_1,Z_{\mathcal{T},i}^\top \beta_2\} \leq Y_{\mathcal{T},i} < \max\{Z_{\mathcal{T},i}^\top \beta_1,Z_{\mathcal{T},i}^\top \beta_2\}} \\
    &= P_{n,\mathcal{T}} h_{\beta_1,\beta_2},
\end{align*}
where $h_{\beta_1,\beta_2}: (z,y) \mapsto \|z\|_2^2 \ind{\min\{z^\top \beta_1,z^\top \beta_2\} \leq y < \max\{z^\top \beta_1,z^\top \beta_2\}}$. Hence, it suffices to show that
\begin{equation*}
    \lim_{n_\mathcal{T} \rightarrow \infty} P\left(\sup_{\|\theta_1 - \theta_2\|_2 \leq \delta_{n_\mathcal{T}}, \ \|\theta_2 - \theta_\mathcal{T}\|_2 \leq \delta_{n_\mathcal{T}}} P_{n,\mathcal{T}} h_{\beta_1,\beta_2} > \epsilon \right) = 0.
\end{equation*}

To establish the above display, we note that
\begin{align*}
    &\quad P\left(\sup_{\|\theta_1 - \theta_2\|_2 \leq \delta_{n_\mathcal{T}}, \ \|\theta_2 - \theta_\mathcal{T}\|_2 \leq \delta_{n_\mathcal{T}}} P_{n,\mathcal{T}} h_{\beta_1,\beta_2} > \epsilon \right) \\
    &\leq P\left(\sup_{\|\theta_1 - \theta_2\|_2 \leq \delta_{n_\mathcal{T}}, \ \|\theta_2 - \theta_\mathcal{T}\|_2 \leq \delta_{n_\mathcal{T}}} \left(P_{n,\mathcal{T}} - P_\mathcal{T}\right) h_{\beta_1,\beta_2} > \frac{\epsilon}{2} \right) + P\left(\sup_{\|\theta_1 - \theta_2\|_2 \leq \delta_{n_\mathcal{T}}, \ \|\theta_2 - \theta_\mathcal{T}\|_2 \leq \delta_{n_\mathcal{T}}} P_{\mathcal{T}} h_{\beta_1,\beta_2} > \frac{\epsilon}{2} \right) \\
    &\leq P\left(\left(P_{n,\mathcal{T}} - P_\mathcal{T}\right) h_{\beta_{1n},\beta_{2n}} > \frac{\epsilon}{4} \right) + P\left(\sup_{\|\theta_1 - \theta_2\|_2 \leq \delta_{n_\mathcal{T}}, \ \|\theta_2 - \theta_\mathcal{T}\|_2 \leq \delta_{n_\mathcal{T}}} P_{\mathcal{T}} h_{\beta_1,\beta_2} > \frac{\epsilon}{2} \right),
\end{align*}
where $(\beta_{1n},\beta_{2n})$ is such that $\left(P_{n,\mathcal{T}} - P_\mathcal{T}\right) h_{\beta_{1n},\beta_{2n}} \geq \sup_{\|\theta_1 - \theta_2\|_2 \leq \delta_{n_\mathcal{T}}, \ \|\theta_2 - \theta_\mathcal{T}\|_2 \leq \delta_{n_\mathcal{T}}} \left(P_{n,\mathcal{T}} - P_\mathcal{T}\right) h_{\beta_1,\beta_2} - \epsilon/4$. We will show that both probabilities in the last line of the above display converge to 0. That is, both $\sup_{\|\theta_1 - \theta_2\|_2 \leq \delta_{n_\mathcal{T}}, \ \|\theta_2 - \theta_\mathcal{T}\|_2 \leq \delta_{n_\mathcal{T}}} P_{\mathcal{T}} h_{\beta_1,\beta_2}$ and $\left(P_{n,\mathcal{T}} - P_\mathcal{T}\right) h_{\beta_{1n},\beta_{2n}}$ are $o_P(1)$.
\begin{align*}
    \|h_{\beta_{1n},\beta_{2n}}\|_{L^2(P_\mathcal{T})}^2 &= \int \|z\|_2^4 P_{Y|Z=z}\left(\min\{z^\top \beta_{1n},z^\top \beta_{2n}\} \leq Y < \max\{z^\top \beta_{1n},z^\top \beta_{2n}\}\right)dP_\mathcal{T}(z) \\
    &\leq \int \|z\|_2^5\|\beta_{1n} - \beta_{2n}\|_2h(z)dP_\mathcal{T}(z) = o_P(1),
\end{align*}
and $h_{\beta_{1n},\beta_{2n}}$ belongs to a fixed Donsker class. By Lemma~19.24 in \citet{van2000asymptotic}, $(P_{n,\mathcal{T}} - P_\mathcal{T}) h_{\beta_{1n},\beta_{2n}} = o_P(n_\mathcal{T}^{-1/2})$. 
\begin{align*}
    P_{\mathcal{T}} h_{\beta_1,\beta_2} &= \int \|z\|_2^2 P_{Y|Z=z}\left(\min\{z^\top \beta_1,z^\top \beta_2\} \leq Y < \max\{z^\top \beta_1,z^\top \beta_2\}\right)dP_\mathcal{T}(z) \\
    &\leq \int \|z\|_2^3\|\beta_1 - \beta_2\|_2h(z)dP_\mathcal{T}(z) = o_P(1),
\end{align*}
uniformly in $(\beta_1,\beta_2)$ as $\|\beta_1 - \beta_2\|_2 \leq \delta_{n_\mathcal{T}}$, and $\delta_{n_\mathcal{T}}$ converges to 0. 

\textbf{Verification of Conditions~\ref{cond:source quad} and \ref{cond:source perturb quad}.} We can follow the same argument to establish these conditions. Note that in the argument above, we did not use the fact that $P_\mathcal{T}\xi_{\tau,\beta}$ is optimized at $\beta_\mathcal{T}$, and we have assumed that (a) the second-order Taylor expansion of $P\xi_{\tau,\beta}$ exists around $\beta_\mathcal{T}$ in all sites, and (b) the conditional density of $Y$ is upper bounded in all sites. Therefore, all the argument above still holds if we replace all subscript $\mathcal{T}$ (target site) to subscript $k$ ($k$-th source site). 

\textbf{Verification of Condition~\ref{cond:continuity of A}.} Continuity is implied by the Lipschitz condition. For $k \in \mathcal{K}$, $\beta_k = \beta_\mathcal{T}$ and positive-definiteness follows from the expression of $A$. 

\textbf{Verification of Condition~\ref{cond:empirical process}.} This condition is equivalent to $(P_{n,k} - P_k)(s_{\tau,\widehat\beta_\mathcal{T}} - s_{\tau,\beta_\mathcal{T}}) = o_P(n_k^{-1/2})$, where $s_{\tau,\beta}(z,y) = z(\ind{y<\beta^\top z} - \tau).$ First, we note that both $s_{\tau,\widehat\beta_\mathcal{T}}$ and $s_{\tau,\beta_\mathcal{T}}$ are in a fixed Donsker class $\{s_{\tau,\beta}, \|\beta - \beta_\mathcal{T}\|_2 \leq \delta\}$. By Lemma~19.24, this condition holds as
\begin{align*}
    \|s_{\tau,\widehat\beta_\mathcal{T}} - s_{\tau,\beta_\mathcal{T}}\|_{L^2(P_k)}^2 &= \int \|z\|_2^2 P_{k,Y|Z=z}\left(\min\{z^\top \beta_\mathcal{T},z^\top \widehat\beta_\mathcal{T}\} \leq Y < \max\{z^\top \beta_\mathcal{T},z^\top \widehat\beta_\mathcal{T}\}\right)dP_k(z) \\
    &\leq \int \|z\|_2^3h(z)\|\widehat\beta_\mathcal{T} - \beta_\mathcal{T}\|_2dP_k(z) = o_P(1).
\end{align*}

\end{proof}

\begin{proof}[Proof of Lemma~\ref{lemma: quad auc maximization}]
On a high level, verifying Conditions~\ref{cond:target quad} and \ref{cond:source quad} involves the application of Theorem~\ref{thm: sherman} where we first construct a degenerate U-process. To this end, we first introduce a shorthand notation for U-processes. Let $f(\cdot,\cdot;\theta)$ be a generic function taking in two arguments indexed by $\theta$, and let $U_{n,k} f(\cdot,\cdot;\theta)$ denote the U-statistic 
$$\frac{1}{n_k(n_k-1)}\sum_{i \neq j} f(X_{k,i},X_{k,j}; \theta)$$ 
for $k \in \{\mathcal{T}\} \cup \{1,\ldots,K\}$. With this definition, we can write $M_{n,k}(\theta)$ succinctly as $U_{n,k} h(\cdot,\cdot;\theta)$ and $S_{n,k}(\theta)$ as $U_{n,k} s_k(\cdot,\cdot;\theta) = 2P_{n,k}\nabla_1 \zeta_k(\cdot;\theta) - P_k\nabla_1 \zeta_k(\cdot;\theta)$.

\textbf{Verification of Condition~\ref{cond:target quad}.} We first decompose the remainder term as follows
\begin{align*}
    R_n(\theta_1,\theta_2) &= M_{n,\mathcal{T}}(\theta_1) - M_{n,\mathcal{T}}(\theta_2) - S_{n,\mathcal{T}}(\theta_2)(\theta_1 - \theta_2) - \frac{1}{2}(\theta_1 - \theta_2)^\top A_\mathcal{T}(\theta_1 - \theta_2) \\
    &= U_{n,\calT}\left\{h(\cdot,\cdot;\theta_1) - h(\cdot,\cdot;\theta_2)\right\} - 2P_{n,\calT}\nabla_1 \zeta_\calT(\cdot;\theta_2)(\theta_1 - \theta_2) + P_\calT \nabla_1 \zeta_\calT(\cdot;\theta_2) (\theta_1 - \theta_2) \\
    &\quad - \frac{1}{2}(\theta_1 - \theta_2)^\top P_\calT \nabla_2\zeta_\calT(\cdot;\theta_\calT)(\theta_1 - \theta_2) \\
    &= U_{n,\calT}\left\{h(\cdot,\cdot;\theta_1) - h(\cdot,\cdot;\theta_2)\right\} - 2P_{n,\calT}\left\{\zeta_\calT(\cdot;\theta_1) - \zeta_\calT(\cdot;\theta_2)\right\} + P_\calT \left\{\zeta_\calT(\cdot;\theta_1) - \zeta_\calT(\cdot;\theta_2)\right\} \\
    &\quad + 2P_{n,\calT}\left\{\zeta_\calT(\cdot;\theta_1) - \zeta_\calT(\cdot;\theta_2) - \nabla_1 \zeta_\calT(\cdot;\theta_2)(\theta_1 - \theta_2)\right\}  \\
    &\quad - P_\calT \left\{\zeta_\calT(\cdot;\theta_1) - \zeta_\calT(\cdot;\theta_2) - \nabla_1 \zeta_\calT(\cdot;\theta_2) (\theta_1 - \theta_2)\right\} - \frac{1}{2}(\theta_1 - \theta_2)^\top P_\calT \nabla_2\zeta_\calT(\cdot;\theta_\calT)(\theta_1 - \theta_2) \\
    &= \underbrace{U_{n,\calT}\left\{h(\cdot,\cdot;\theta_1) - h(\cdot,\cdot;\theta_2)\right\} - 2P_{n,\calT}\left\{\zeta_\calT(\cdot;\theta_1) - \zeta_\calT(\cdot;\theta_2)\right\} + P_\calT \left\{\zeta_\calT(\cdot;\theta_1) - \zeta_\calT(\cdot;\theta_2)\right\}}_{\textnormal{term 1}} \\
    &\quad + \underbrace{2P_{n,\calT}\left\{\zeta_\calT(\cdot;\theta_1) - \zeta_\calT(\cdot;\theta_2) - \nabla_1 \zeta_\calT(\cdot;\theta_2)(\theta_1 - \theta_2) - \frac{1}{2}(\theta_1 - \theta_2)^\top \nabla_2\zeta_\calT(\cdot;\theta_\calT)(\theta_1 - \theta_2)\right\}}_{\textnormal{term 2}} \\
    &\quad + \underbrace{(\theta_1 - \theta_2)^\top \left\{P_{n,\calT} - P_\calT \right\} \nabla_2\zeta_\calT(\cdot;\theta_\calT) (\theta_1 - \theta_2)}_{\textnormal{term 3}} \\
    &\quad - \underbrace{P_\calT \left\{\zeta_\calT(\cdot;\theta_1) - \zeta_\calT(\cdot;\theta_2) - \nabla_1 \zeta_\calT(\cdot;\theta_2) (\theta_1 - \theta_2) - \frac{1}{2}(\theta_1 - \theta_2)^\top \nabla_2\zeta_\calT(\cdot;\theta_\calT)(\theta_1 - \theta_2)\right\}}_{\textnormal{term 4}}. \\
\end{align*}
To verify Condition~\ref{cond:target quad}, it suffices to show that for $j=1,2,3,4$,
\begin{equation}\label{eq: claim for each term}
    \lim_{n_\mathcal{T} \rightarrow \infty} P\left(\sup_{ \|\theta_1 - \theta_2\|_2 \leq \delta, \ \|\theta_2 - \theta_\mathcal{T}\|_2 \leq \delta_{n_\mathcal{T}} } \frac{|\textnormal{term } j|}{\|\theta_1 - \theta_2\|_2^2 + n_\mathcal{T}^{-1}} > \epsilon \right) = 0,
\end{equation}
and we now analyze each term separately.

We start with term 3.
\begin{align*}
    &\sup_{ \|\theta_1 - \theta_2\|_2 \leq \delta, \ \|\theta_2 - \theta_\mathcal{T}\|_2 \leq \delta_{n_\mathcal{T}} } \frac{|\textnormal{term } 3|}{\|\theta_1 - \theta_2\|_2^2 + n_\mathcal{T}^{-1}} \\
    \leq & \sup_{ \|\theta_1 - \theta_2\|_2 \leq \delta, \ \|\theta_2 - \theta_\mathcal{T}\|_2 \leq \delta_{n_\mathcal{T}} } \frac{ \left\|\left\{P_{n,\calT} - P_\calT \right\} \nabla_2\zeta_\calT(\cdot;\theta_\calT)\right\| \|\theta_1 - \theta_2\|_2^2 }{\|\theta_1 - \theta_2\|_2^2 } \\
    \leq & \left\|\left\{P_{n,\calT} - P_\calT \right\} \nabla_2\zeta_\calT(\cdot;\theta_\calT)\right\| = o_P(1)
\end{align*}
due to the law of large number and continuous mapping theorem. Hence \eqref{eq: claim for each term} holds for term 3. 

Term 2 and term 4 can be analyzed in very similar way. Differentiability of $\zeta_\calT$ enables the following Taylor expansion for given $x$
\begin{equation*}
    \zeta_\calT(x;\theta_1) - \zeta_\calT(x;\theta_2) = \nabla_1 \zeta_\calT(x;\theta_2) (\theta_1 - \theta_2) + \frac{1}{2}(\theta_1 - \theta_2)^\top \nabla_2 \zeta_\calT(x;\bar\theta) (\theta_1 - \theta_2),
\end{equation*}
for some $\bar\theta$ in between $\theta_1$ and $\theta_2$. Therefore,
\begin{align*}
    &\quad \left|\zeta_\calT(x;\theta_1) - \zeta_\calT(x;\theta_2) - \nabla_1 \zeta_\calT(x;\theta_2) (\theta_1 - \theta_2) - \frac{1}{2}(\theta_1 - \theta_2)^\top \nabla_2\zeta_\calT(x;\theta_\calT)(\theta_1 - \theta_2)\right| \\
    &= \frac{1}{2}\left|(\theta_1 - \theta_2)^\top \left\{ \nabla_2\zeta_\calT(x;\bar\theta) - \nabla_2\zeta_\calT(x;\theta_\calT) \right\} (\theta_1 - \theta_2) \right| \\
    &\leq \frac{1}{2} \|\theta_1 - \theta_2\|_2^2 \left\| \nabla_2\zeta_\calT(x;\bar\theta) - \nabla_2\zeta_\calT(x;\theta_\calT) \right\| \\
    &\leq L(x)\|\theta_1 - \theta_2\|_2^2 \delta,
\end{align*}
where the last line follows from the Lipschitz property of $\nabla_2\zeta_\calT$ under our assumption. Thus,
\begin{align*}
    \sup_{ \|\theta_1 - \theta_2\|_2 \leq \delta, \ \|\theta_2 - \theta_\mathcal{T}\|_2 \leq \delta_{n_\mathcal{T}} } \frac{|\textnormal{term } 4|}{\|\theta_1 - \theta_2\|_2^2 + n_\mathcal{T}^{-1}} &\leq \sup_{ \|\theta_1 - \theta_2\|_2 \leq \delta, \ \|\theta_2 - \theta_\mathcal{T}\|_2 \leq \delta_{n_\mathcal{T}} } \delta P_\calT L = \delta P_\calT L,
\end{align*}
and
\begin{align*}
    \sup_{ \|\theta_1 - \theta_2\|_2 \leq \delta, \ \|\theta_2 - \theta_\mathcal{T}\|_2 \leq \delta_{n_\mathcal{T}} } \frac{|\textnormal{term } 2|}{\|\theta_1 - \theta_2\|_2^2 + n_\mathcal{T}^{-1}} &\leq \sup_{ \|\theta_1 - \theta_2\|_2 \leq \delta, \ \|\theta_2 - \theta_\mathcal{T}\|_2 \leq \delta_{n_\mathcal{T}} } 2 \delta P_{n,\calT} L  = 2 \delta P_{n,\calT} L.
\end{align*}
Therefore, both term 2 and term 4 satisfy \eqref{eq: claim for each term} with a sufficiently small choice of $\delta$, due to the law of large numbers. 

Finally, we turn to term 1 and first unpack the notation. 
\begin{align*}
    \textnormal{term 1} &= U_{n,\calT}\left\{h(\cdot,\cdot;\theta_1) - h(\cdot,\cdot;\theta_2)\right\} - 2P_{n,\calT}\left\{\zeta_\calT(\cdot;\theta_1) - \zeta_\calT(\cdot;\theta_2)\right\} + P_\calT \left\{\zeta_\calT(\cdot;\theta_1) - \zeta_\calT(\cdot;\theta_2)\right\} \\
    &= \frac{1}{n_\calT(n_\calT-1)}\sum_{i\neq j}\left\{h(X_{\calT,i},X_{\calT,j};\theta_1) - h(X_{\calT,i},X_{\calT,j};\theta_2)\right\} - \frac{1}{n_\calT}\sum_i \left\{\zeta_\calT(X_{\calT,i};\theta_1) - \zeta_\calT(X_{\calT,i};\theta_2)\right\} \\
    &\quad - \frac{1}{n_\calT}\sum_j \left\{\zeta_\calT(X_{\calT,j};\theta_1) - \zeta_\calT(X_{\calT,j};\theta_2)\right\} + \EE_\calT\left[\zeta_\calT(X_{\calT};\theta_1) - \zeta_\calT(X_{\calT};\theta_2)\right] \\
    &= \frac{1}{n_\calT(n_\calT-1)}\sum_{i\neq j}\Big\{h(X_{\calT,i},X_{\calT,j};\theta_1) - h(X_{\calT,i},X_{\calT,j};\theta_2) - \left\{\zeta_\calT(X_{\calT,i};\theta_1) - \zeta_\calT(X_{\calT,i};\theta_2)\right\} \\
    &\quad - \left\{\zeta_\calT(X_{\calT,j};\theta_1) - \zeta_\calT(X_{\calT,j};\theta_2)\right\} + \EE_\calT\left[\zeta_\calT(X_{\calT};\theta_1) - \zeta_\calT(X_{\calT};\theta_2)\right] \Big\}
\end{align*}
Hence, term 1 is in fact a U-process indexed by $(\theta_1,\theta_2)$ with kernel function
\begin{multline*}
    \kappa_{\theta_1,\theta_2}: (x_1,x_2) \mapsto h(x_1,x_2;\theta_1) - h(x_1,x_2;\theta_2) - \left\{\zeta_\calT(x_1;\theta_1) - \zeta_\calT(x_1;\theta_2)\right\} - \left\{\zeta_\calT(x_2;\theta_1) - \zeta_\calT(x_2;\theta_2)\right\} \\ + \EE_\calT\left[\zeta_\calT(X_{\calT};\theta_1) - \zeta_\calT(X_{\calT};\theta_2)\right].
\end{multline*}
This U-process is in fact degenerate, as by construction, $\zeta_\calT(\cdot;\theta) = P_\calT h(\cdot,x;\theta) = P_\calT h(x,\cdot;\theta)$. Furthermore, when $(\theta_1,\theta_2) = (\theta_\calT,\theta_\calT)$, the kernel function is constant 0. The Euclidean property of the function class $\{\kappa_{\theta_1,\theta_2}: \|\theta_1-\theta_2\| \leq \delta, \|\theta_2 - \theta_\calT\| \leq \delta \}$ can be established given that it contains linear combination of indicator functions (see Section 5 in \citet{sherman1993limiting} for more details.) Finally, to apply Theorem~\ref{thm: sherman}, we need to establish that norm of $\kappa_{\theta_1,\theta_2}$ converges to 0 as $(\theta_1,\theta_2) \rightarrow (\theta_\calT, \theta_\calT)$. 
\begin{equation*}
\lim_{(\theta_1,\theta_2) \rightarrow(\theta_\calT,\theta_\calT)} \int \kappa^2_{\theta_1,\theta_2}(x_1,x_2) dP_\calT(x_1)dP_\calT(x_2) = \int \lim_{(\theta_1,\theta_2) \rightarrow(\theta_\calT,\theta_\calT)}\kappa^2_{\theta_1,\theta_2}(x_1,x_2) dP_\calT(x_1)dP_\calT(x_2) = 0,
\end{equation*}
where we apply the dominated convergence theorem and note that (1) the pointwise limit of $\kappa^2_{\theta_1,\theta_2}(x_1,x_2)$ is 0 for almost every $(x_1,x_2)$ and (2) $\kappa^2$ is upper bounded by some absolute constant. Theorem~\ref{thm: sherman} then implies that term 1 is $o_P(n_\calT^{-1})$ uniformly over a shrinking neighborhood of $(\theta_\calT,\theta_\calT)$, that is,
\begin{align*}
    \sup_{ \|\theta_1 - \theta_2\|_2 \leq \delta_{n_\mathcal{T}}, \ \|\theta_2 - \theta_\mathcal{T}\|_2 \leq \delta_{n_\mathcal{T}} } \frac{|\textnormal{term } 1|}{\|\theta_1 - \theta_2\|_2^2 + n_\mathcal{T}^{-1}} &= o_P(1).
\end{align*}
Next, we note that with $\delta_{n_\calT} \gg n_\calT^{-1/2}$,
\begin{align*}
    \sup_{ \delta_{n_\mathcal{T}} < \|\theta_1 - \theta_2\|_2 \leq \delta, \ \|\theta_2 - \theta_\mathcal{T}\|_2 \leq \delta_{n_\mathcal{T}} } \frac{|\textnormal{term } 1|}{\|\theta_1 - \theta_2\|_2^2 + n_\mathcal{T}^{-1}} &\leq \frac{\sup_{\|\theta_1 - \theta_2\|_2 \leq \delta, \ \|\theta_2 - \theta_\mathcal{T}\|_2 \leq \delta }|\textnormal{term } 1|}{\delta_{n_\calT}^2 + n_\mathcal{T}^{-1}}= o_P(1).
\end{align*}
This is due to the uniform convergence result on degenerate U-processes in, for example, \citet{arcones1993limit} (see Theorem 5.6 therein), as $\kappa$ is uniformly bounded, degenerate, and belongs to a Donsker class. Thus, \eqref{eq: claim for each term} is satisfied for term 1.

\textbf{Verification of Condition~\ref{cond:target perturb quad}.} Let $W_i$'s correspond to the weights in nonparametric bootstrap. First, we give a general calculation of the variance of a perturbed degree-2 U-statistics conditional on the observed data. To this end, let $\xi(\cdot,\cdot)$ denote a generic symmetric kernel function and write $\xi_{ij}$ as a shorthand notation for $\xi(x_i,x_j)$, which is deterministic given the observed data. We consider the quantity
$$ M=\frac{2}{n(n-1)}\sum_{1\le i<j \le n}W_iW_j\xi_{ij}$$
Then we have
$$ E_W(M)= \frac{2}{n(n-1)}\sum_{1\le i<j \le n} (1-n^{-1})\xi_{ij},$$
\begin{align*}
    \Var_W(M) &= \frac{4}{n^2(n-1)^2} E_W \left[\sum_{1\le i<j \le n}(W_iW_j-1+n^{-1})\xi_{ij}\sum_{1\le l<k\le n}(W_lW_k-1+n^{-1})\xi_{lk} \right] \\
    &= \frac{4}{n^2(n-1)^2} \sum_{i<j, l<k,  |\{i,j\}\cap \{l, k\}|=1}E\left[(W_1W_2-1+n^{-1})(W_1W_3-1+n^{-1})\right] \xi_{ij}\xi_{lk} \\
    &\quad + \frac{4}{n^2(n-1)^2} \sum_{i<j, l<k,  |\{i,j\}\cap \{l, k\}|=2}E\left[(W_1W_2-1+n^{-1})(W_1W_2-1+n^{-1})\right] \xi_{ij}\xi_{lk} \\
    &\quad + \frac{4}{n^2(n-1)^2} \sum_{i<j, l<k,  \{i,j\}\cap \{l, k\}=\emptyset}E\left[(W_1W_2-1+n^{-1})(W_3W_4-1+n^{-1})\right] \xi_{ij}\xi_{lk} \\
    &\doteq  \frac{4(n^2-6n+6)}{n^5(n-1)}\sum_{i<j, l<k,  |\{i,j\}\cap \{l, k\}|=1}   \xi_{ij}\xi_{lk} -\frac{8(2n-3)}{n^5(n-1)} \sum_{i<j, l<k,  \{i,j\}\cap \{l, k\}=\emptyset}  \xi_{ij}\xi_{lk} \\
    &\doteq \frac{4(n-2)}{n^4(n-1)} \sum_{i<j, l<k, |\{i,j\}\cap \{l, k\}|=1}   \xi_{ij}\xi_{lk} -\frac{(2n-2)(2n-3)}{n^3} \left(\frac{2}{n(n-1)}\sum_{i<j}  \xi_{ij} \times \frac{2}{n(n-1)} \sum_{l<k}\xi_{lk}\right)^2 \\
    &\doteq \frac{4(n-2)^2}{n^3} \times \left[\frac{1}{n(n-1)(n-2)}\sum_{i<j, l<k, |\{i,j\}\cap\{l,k\}|=1}\xi_{ij}\xi_{lk}  - \frac{2n-3}{2n-2} \left(\frac{2}{n(n-1)}\sum_{i<j}  \xi_{ij} \right)^2\right] \\
    &\doteq \frac{4}{n} \times \left[\frac{1}{n(n-1)(n-2)}\sum_{i<j, l<k, |\{i,j\}\cap\{l,k\}|=1}\xi_{ij}\xi_{lk}  -  \left(\frac{2}{n(n-1)}\sum_{i<j}  \xi_{ij} \right)^2\right],
\end{align*}
where we use $\doteq$ to mean equality up to some asymptotically negligible terms.


Recall that we have
\begin{align*}
    M^\dag_{n,\mathcal{T}}(\theta) &= \frac{1}{n_\calT(n_\calT-1)} \sum_{i\neq j} W_{\mathcal{T},i} W_{\mathcal{T},j}h\left(X_{\calT,i}, X_{\calT,j};\theta\right) \\
    S^\dag_{n,\calT}(\theta) &= \frac{1}{n_\calT(n_\calT-1)} \sum_{i\neq j} W_{\mathcal{T},i} W_{\mathcal{T},j} s_\calT \left(X_{\calT,i}, X_{\calT,j};\theta\right),
\end{align*}
and consequently, the remainder term takes the form of $R_n^\dag(\theta_1,\theta_2) = M_{n,\mathcal{T}}^\dag(\theta_1) - M_{n,\mathcal{T}}^\dag(\theta_2) - S_{n,\mathcal{T}}^\dag(\theta_2)(\theta_1 - \theta_2) - \frac{1}{2}(\theta_1 - \theta_2)^\top A_\mathcal{T}(\theta_1 - \theta_2)$.
\begin{multline*}
    R^\dag_n(\theta_1,\theta_2) = \frac{1}{n_\calT(n_\calT - 1)}\sum_{i \neq j} W_{\calT,i} W_{\calT,j} \Big\{h\left(X_{\calT,i}, X_{\calT,j};\theta_1\right) - h\left(X_{\calT,i}, X_{\calT,j};\theta_2\right) \\
    - s_\calT \left(X_{\calT,i}, X_{\calT,j};\theta_2\right)(\theta_1 - \theta_2)\Big\} - \frac{1}{2}(\theta_1 - \theta_2)^\top A_\mathcal{T}(\theta_1 - \theta_2).
\end{multline*}
Similar to the proof of Lemma~\ref{lemma: quad quantile regression}, given Condition~\ref{cond:target quad}, it suffices to focus on the variance of $R^\dag_n(\theta_1,\theta_2)$ condition on the observed data. Given the general calculation above, it suffices to upper bound the following quantity
\begin{align*}
    &\frac{4}{n_\calT^2(n_\calT-1)^2}\sum_{i=1}^{n_\calT}\sum_{j\neq i}\sum_{k \neq i} \Big[ \left\{h\left(X_{\calT,i}, X_{\calT,j};\theta_1\right) - h\left(X_{\calT,i}, X_{\calT,j};\theta_2\right)
    - s_\calT \left(X_{\calT,i}, X_{\calT,j};\theta_2\right)(\theta_1 - \theta_2)\right\} \\
    &\quad \quad \times \left\{h\left(X_{\calT,i}, X_{\calT,k};\theta_1\right) - h\left(X_{\calT,i}, X_{\calT,k};\theta_2\right)
    - s_\calT \left(X_{\calT,i}, X_{\calT,k};\theta_2\right)(\theta_1 - \theta_2)\right\} \Big],
\end{align*}
and we see that this quantity involves a degree-3 U-statistics which we now analyze.

First, we decompose the following function
\begin{align*}
    &\quad h\left(x_1,x_2;\theta_1\right) - h\left(x_1,x_2;\theta_2\right)
    - s_\calT \left(x_1,x_2;\theta_2\right)(\theta_1 - \theta_2) \\
    &= h\left(x_1,x_2;\theta_1\right) - h\left(x_1,x_2;\theta_2\right) - \nabla_1\zeta (x_1;\theta_2)(\theta_1-\theta_2) - \nabla_1\zeta (x_2;\theta_2)(\theta_1-\theta_2) + P_\calT\nabla_1\zeta (\cdot;\theta_2)(\theta_1-\theta_2) \\
    &= h\left(x_1,x_2;\theta_1\right) - h\left(x_1,x_2;\theta_2\right) - \left\{\zeta(x_1;\theta_1) - \zeta(x_1;\theta_2)\right\} - \left\{\zeta(x_2;\theta_1) - \zeta(x_2;\theta_2)\right\} + P_\calT\left\{\zeta(\cdot;\theta_1) - \zeta(\cdot;\theta_2)\right\} \\
    &\quad + \zeta(x_1;\theta_1) - \zeta(x_1;\theta_2) - \nabla_1\zeta (x_1;\theta_2)(\theta_1-\theta_2) \\
    &\quad + \zeta(x_2;\theta_1) - \zeta(x_2;\theta_2) - \nabla_1\zeta (x_2;\theta_2)(\theta_1-\theta_2) \\
    &\quad - P_\calT\left\{\zeta(\cdot;\theta_1) - \zeta(\cdot;\theta_2) - \nabla_1\zeta (\cdot;\theta_2)(\theta_1-\theta_2)\right\} \\
    &= \kappa(x_1,x_2;\theta_1,\theta_2) + r(x_1;\theta_1,\theta_2) + r(x_2;\theta_1,\theta_2) - P_\calT r(\cdot;\theta_1,\theta_2),
\end{align*}
where we define the following functions for notational convenience
\begin{align*}
    \kappa(x_1,x_2;\theta_1,\theta_2) &= h\left(x_1,x_2;\theta_1\right) - h\left(x_1,x_2;\theta_2\right) - \left\{\zeta(x_1;\theta_1) - \zeta(x_1;\theta_2)\right\} - \left\{\zeta(x_2;\theta_1) - \zeta(x_2;\theta_2)\right\} \\
    &\quad + P_\calT\left\{\zeta(\cdot;\theta_1) - \zeta(\cdot;\theta_2)\right\} \\
    r(x;\theta_1,\theta_2) &= \zeta(x;\theta_1) - \zeta(x;\theta_2) - \nabla_1\zeta (x;\theta_2)(\theta_1-\theta_2).
\end{align*}
This allows us to rewrite the quantity of interest as 
\begin{align*}
    &\frac{4}{n_\calT^2(n_\calT-1)^2}\sum_{i=1}^{n_\calT}\sum_{j\neq i}\sum_{k \neq i} \Big[ \left\{\kappa(X_{\calT,i},X_{\calT,j};\theta_1,\theta_2) + r(X_{\calT,i};\theta_1,\theta_2) + r(X_{\calT,j};\theta_1,\theta_2) - P_\calT r(\cdot;\theta_1,\theta_2)\right\} \\
    &\quad \quad \times \left\{\kappa(X_{\calT,i},X_{\calT,k};\theta_1,\theta_2) + r(X_{\calT,i};\theta_1,\theta_2) + r(X_{\calT,k};\theta_1,\theta_2) - P_\calT r(\cdot;\theta_1,\theta_2)\right\} \Big].
\end{align*}
There are 6 types of terms involved, 3 ``square terms" and 3 ``cross terms", and we now bound each of them separately in a total of 6 steps.
\\
\\
\noindent{\textbf{Step I.}}
\begin{align*}
    \textnormal{term (1)} &= \frac{4}{n_\calT^2(n_\calT-1)^2}\sum_{i=1}^{n_\calT}\sum_{j\neq i}\sum_{k \neq i} \left\{r(X_{\calT,j};\theta_1,\theta_2) - P_\calT r(\cdot;\theta_1,\theta_2)\right\}\left\{ r(X_{\calT,k};\theta_1,\theta_2) - P_\calT r(\cdot;\theta_1,\theta_2)\right\},
\end{align*}
which is asymptotically equivalent to
\begin{align*}
    \textnormal{term (1)}^* &= \frac{4}{n_\calT^4}\sum_{i=1}^{n_\calT}\sum_{j=1}^{n_\calT}\sum_{k=1}^{n_\calT} \left\{r(X_{\calT,j};\theta_1,\theta_2) - P_\calT r(\cdot;\theta_1,\theta_2)\right\}\left\{ r(X_{\calT,k};\theta_1,\theta_2) - P_\calT r(\cdot;\theta_1,\theta_2)\right\} \\
    &= \frac{4}{n_\calT}\left[\frac{1}{n_\calT}\sum_{j=1}^{n_\calT}\left\{r(X_{\calT,j};\theta_1,\theta_2) - P_\calT r(\cdot;\theta_1,\theta_2)\right\}\right] \left[\frac{1}{n_\calT}\sum_{k=1}^{n_\calT}\left\{ r(X_{\calT,k};\theta_1,\theta_2) - P_\calT r(\cdot;\theta_1,\theta_2)\right\}\right] \\
    &= \frac{4}{n_\calT} \left\{\left(P_{n,\calT} - P_\calT\right)r(\cdot;\theta_1,\theta_2)\right\}^2
\end{align*}
\begin{align*}
    \left(P_{n,\calT} - P_\calT\right)r(\cdot;\theta_1,\theta_2) &= (\theta_1 - \theta_2)^\top \left\{\left(P_{n,\calT} - P_\calT\right)\nabla_2\zeta(\cdot;\bar\theta(\cdot))\right\}(\theta_1 - \theta_2) \\
    &= (\theta_1 - \theta_2)^\top \left\{\left(P_{n,\calT} - P_\calT\right)\nabla_2\zeta(\cdot;\theta_\calT)\right\}(\theta_1 - \theta_2) \\
    &\quad + \left(P_{n,\calT} - P_\calT\right) \left\{ (\theta_1 - \theta_2)^\top \left(\nabla_2\zeta(\cdot;\bar\theta(\cdot)) - \nabla_2\zeta(\cdot;\theta_\calT)\right)(\theta_1 - \theta_2) \right\} \\
    &\leq \|\theta_1-\theta_2\|_2^2 \left\|\left(P_{n,\calT} - P_\calT\right)\nabla_2\zeta(\cdot;\theta_\calT)\right\| + P_{n,\calT}\|\theta_1 - \theta_2\|_2^2 L(\cdot)\delta_{n_\calT} + P_\calT\|\theta_1 - \theta_2\|_2^2 L(\cdot)\delta_{n_\calT},
\end{align*}
where the last line follows from the Lipschitz property of $\nabla_2\zeta$. Therefore,
\begin{align*}
    \sup \frac{\textnormal{term (1)}}{\{\|\theta_1 - \theta_2\|_2^2 + n_\mathcal{T}^{-1}\}^2} \leq \sup \frac{C}{\|\theta_1 - \theta_2\|_2^2 + n_\mathcal{T}^{-1}} \Big\{\|\theta_1-\theta_2\|_2^4 \left\|\left(P_{n,\calT} - P_\calT\right)\nabla_2\zeta(\cdot;\theta_\calT)\right\|^2 \\
    + \|\theta_1 - \theta_2\|_2^4 \left(P_{n,\calT}L\right)^2 \delta_{n_\calT}^2 + \|\theta_1 - \theta_2\|_2^4 \left(P_\calT L\right)^2\delta_{n_\calT}^2 \Big\} = o_P(1).
\end{align*}
\noindent{\textbf{Step II.}}
\begin{align*}
    \textnormal{term (2)} &= \frac{4}{n_\calT^2(n_\calT-1)^2}\sum_{i=1}^{n_\calT}\sum_{j\neq i}\sum_{k\neq i} r(X_{\calT,i};\theta_1,\theta_2) \left\{ r(X_{\calT,k};\theta_1,\theta_2) - P_\calT r(\cdot;\theta_1,\theta_2)\right\},
\end{align*}
which is asymptotically equivalent to
\begin{align*}
    \textnormal{term (2)}^* &= \frac{4}{n_\calT^4}\sum_{i=1}^{n_\calT}\sum_{j=1}^{n_\calT}\sum_{k=1}^{n_\calT} r(X_{\calT,i};\theta_1,\theta_2) \left\{ r(X_{\calT,k};\theta_1,\theta_2) - P_\calT r(\cdot;\theta_1,\theta_2)\right\} \\
    &= \frac{4}{n_\calT} \left[\frac{1}{n_\calT}\sum_{i=1}^{n_\calT} r(X_{\calT,i};\theta_1,\theta_2)\right] \left[\frac{1}{n_\calT}\sum_{k=1}^{n_\calT} \left\{ r(X_{\calT,k};\theta_1,\theta_2) - P_\calT r(\cdot;\theta_1,\theta_2)\right\}\right],
\end{align*}
and
\begin{align*}
    \frac{1}{n_\calT}\sum_{i=1}^{n_\calT} r(X_{\calT,i};\theta_1,\theta_2) &\leq \|\theta_1 - \theta_2\|_2^2 \left\{ \|P_{n,\calT}\nabla_2\zeta(\cdot;\theta_\calT)\| + P_{n,\calT}L \delta_{n_\calT}\right\}.
\end{align*}
Therefore,
\begin{multline*}
    \sup \frac{\textnormal{term (2)}}{\{\|\theta_1 - \theta_2\|_2^2 + n_\mathcal{T}^{-1}\}^2} \leq \sup \frac{C}{\|\theta_1 - \theta_2\|_2^2 + n_\mathcal{T}^{-1}}\|\theta_1 -\theta_2\|_2^4 \left\{ \|P_{n,\calT}\nabla_2\zeta(\cdot;\theta_\calT)\| + P_{n,\calT}L \delta_{n_\calT}\right\} \\
    \times \left\{\left\|\left(P_{n,\calT} - P_\calT\right)\nabla_2\zeta(\cdot;\theta_\calT)\right\| + P_{n,\calT}L\delta_{n_\calT} + P_\calT L\delta_{n_\calT}\right\} = o_P(1).
\end{multline*}
\noindent{\textbf{Step III.}}
\begin{align*}
    \textnormal{term (3)} &= \frac{4}{n_\calT^2(n_\calT-1)^2}\sum_{i=1}^{n_\calT}\sum_{j\neq i}\sum_{k\neq i} \kappa(X_{\calT,i},X_{\calT,j};\theta_1,\theta_2) \left\{ r(X_{\calT,k};\theta_1,\theta_2) - P_\calT r(\cdot;\theta_1,\theta_2)\right\},
\end{align*}
which is asymptotically equivalent to 
\begin{align*}
    \textnormal{term (3)}^* &= \frac{4}{n_\calT} \left[\frac{1}{n_\calT^2}\sum_{i=1}^{n_\calT}\sum_{j=1}^{n_\calT} \kappa(X_{\calT,i},X_{\calT,j};\theta_1,\theta_2)\right] \left[\frac{1}{n_\calT}\sum_{k=1}^{n_\calT}\left\{ r(X_{\calT,k};\theta_1,\theta_2) - P_\calT r(\cdot;\theta_1,\theta_2)\right\}\right].
\end{align*}
By definition, $\kappa(\cdot,\cdot;\theta_1,\theta_2)$ is $P_\calT$-degenerate for any $(\theta_1, \theta_2)$, and thus Theorem~\ref{thm: sherman} implies that $\sum_{i=1}^{n_\calT}\sum_{j=1}^{n_\calT} \kappa(X_{\calT,i},X_{\calT,j};\theta_1,\theta_2)/n_\calT^2$ is $o_P(n_\calT^{-1})$ uniformly over a neighborhood of $(\theta_\calT,\theta_\calT)$ (see the verification of Condition~\ref{cond:target quad} for more details.) Note that we have already established an upper bound for the term in the second square brackets. Therefore,
\begin{equation*}
    \sup \frac{\textnormal{term (3)}}{\{\|\theta_1 - \theta_2\|_2^2 + n_\mathcal{T}^{-1}\}^2} = o_P(1).
\end{equation*}
\noindent{\textbf{Step IV.}}
\begin{align*}
    \textnormal{term (4)} &= \frac{4}{n_\calT^2(n_\calT-1)^2}\sum_{i=1}^{n_\calT}\sum_{j\neq i}\sum_{k\neq i} r(X_{\calT,i};\theta_1,\theta_2)^2 = \frac{4}{n_\calT^2}\sum_{i=1}^{n_\calT} r(X_{\calT,i};\theta_1,\theta_2)^2,
\end{align*}
and we again note that
\begin{align*}
    r(x;\theta_1,\theta_2) &= (\theta_1 - \theta_2)^\top \nabla_2 \zeta(x,\theta_\calT)(\theta_1 - \theta_2) + (\theta_1 - \theta_2)^\top \left(\nabla_2 \zeta(x,\bar\theta(x)) - \nabla_2 \zeta(x,\theta_\calT)\right)(\theta_1 - \theta_2) \\
    &\leq \|\theta_1 - \theta_2\|_2^2 \left\{ \|\nabla_2\zeta(x,\theta_\calT)\| + L(x)\delta_{n_\calT}\right\},
\end{align*}
and thus
\begin{equation*}
    r(x;\theta_1,\theta_2)^2 \leq 2\|\theta_1 - \theta_2\|_2^4 \left\{ \|\nabla_2\zeta(x,\theta_\calT)\|^2 + L(x)^2\delta_{n_\calT}^2\right\}.
\end{equation*}
Therefore,
\begin{align*}
    \sup \frac{\textnormal{term (4)}}{\{\|\theta_1 - \theta_2\|_2^2 + n_\mathcal{T}^{-1}\}^2} &\leq \sup \frac{C}{\|\theta_1 - \theta_2\|_2^2 + n_\mathcal{T}^{-1}} \|\theta_1 - \theta_2\|_2^4 \left\{ P_{n,\calT}\|\nabla_2\zeta(\cdot,\theta_\calT)\|^2 + P_{n,\calT}L^2\delta_{n_\calT}^2\right\} = o_P(1).
\end{align*}
\noindent{\textbf{Step V.}}
\begin{align*}
    \textnormal{term (5)} &= \frac{4}{n_\calT^2(n_\calT-1)^2}\sum_{i=1}^{n_\calT}\sum_{j\neq i}\sum_{k\neq i} \kappa(X_{\calT,i},X_{\calT,j};\theta_1,\theta_2) r(X_{\calT,i};\theta_1,\theta_2) \\
    &= \frac{4}{n_\calT^2(n_\calT-1)}\sum_{(i,j): j\neq i} \kappa(X_{\calT,i},X_{\calT,j};\theta_1,\theta_2) r(X_{\calT,i};\theta_1,\theta_2) \\
    &= \underbrace{\frac{4}{n_\calT^2(n_\calT-1)}\sum_{(i,j): j\neq i} \left\{\kappa(X_{\calT,i},X_{\calT,j};\theta_1,\theta_2) r(X_{\calT,i};\theta_1,\theta_2) - P_\calT \kappa(\cdot,X_{\calT,j};\theta_1,\theta_2) r(\cdot;\theta_1,\theta_2)\right\}}_{\textnormal{term (5.1)}} \\
    &\quad + \underbrace{\frac{4}{n_\calT^2}\sum_{j=1}^{n_\calT} P_\calT \kappa(\cdot,X_{\calT,j};\theta_1,\theta_2) r(\cdot;\theta_1,\theta_2)}_{\textnormal{term (5.2)}}.
\end{align*}
For term (5.1), we note that the kernel function
\begin{equation}\label{eq:kernal term 5.1}
   \kappa(x_1,x_2;\theta_1,\theta_2) r(x_1;\theta_1,\theta_2) - P_\calT \kappa(\cdot,x_2;\theta_1,\theta_2) r(\cdot;\theta_1,\theta_2) 
\end{equation}
is $P_\calT$ degenerate. Indeed, we note that the degeneracy of $\kappa$ implies that when holding $x_1$ fixed and integrating out $x_2$ we get 0. We have shown that the class of functions $\kappa(\cdot,\cdot;\theta_1,\theta_2)$ for $(\theta_1,\theta_2)$ in a neighborhood of $(\theta_\calT,\theta_\calT)$ is Donsker and uniformly bounded. Moreover, by assumption, $r(\cdot;\theta_1,\theta_2): x \mapsto \zeta(x;\theta_1) - \zeta(x;\theta_2) - \nabla_1\zeta (x;\theta_2)(\theta_1-\theta_2)$ is Donsker with an integrable envelope function. Thus, the class of functions containing all kernel functions of the form \eqref{eq:kernal term 5.1} when varying $(\theta_1,\theta_2)$ over a neighborhood of $(\theta_\calT,\theta_\calT)$ is Donsker. To apply Theorem \ref{thm: sherman}, it suffices to show that the norm of the function in \eqref{eq:kernal term 5.1} converges to 0. As $\kappa$ is uniformly bounded, we have
\begin{align*}
    &\quad \iint \kappa(x_1,x_2;\theta_1,\theta_2)^2 r(x_1;\theta_1,\theta_2)^2 dP_\calT(x_1)dP_\calT(x_2) \\
    &\leq C\int r(x;\theta_1,\theta_2)^2 dP_\calT(x) \\
    &\leq 2C\|\theta_1 - \theta_2\|_2^4 \left\{ P_\calT\|\nabla_2\zeta(x,\theta_\calT)\|^2 + P_\calT L^2\delta_{n_\calT}^2\right\} \rightarrow 0,
\end{align*}
and
\begin{align*}
    P_\calT \kappa(\cdot,x_2;\theta_1,\theta_2) r(\cdot;\theta_1,\theta_2)  &\leq C P_\calT |r(\cdot;\theta_1,\theta_2)| \leq C\int \|\theta_1 - \theta_2\|_2^2 \left\{ \|\nabla_2\zeta(x,\theta_\calT)\| + L(x)\delta_{n_\calT}\right\} dP_\calT(x) \\
    &= C\|\theta_1 - \theta_2\|_2^2\left\{ P_\calT \|\nabla_2\zeta(\cdot,\theta_\calT)\| + P_\calT L\delta_{n_\calT}\right\} \rightarrow 0.
\end{align*}
Thus, Theorem~\ref{thm: sherman} implies that term (5.1) is $o_P(n_\calT^{-2})$ uniformly over $(\theta_1,\theta_2)$.

Now we turn to term (5.2). Applying Cauchy-Schwarz inequality and Jensen's inequality, we have
\begin{align*}
    \textnormal{term (5.2)} &= \frac{4}{n_\calT^2}\sum_{j=1}^{n_\calT} P_\calT \kappa(\cdot,X_{\calT,j};\theta_1,\theta_2) r(\cdot;\theta_1,\theta_2) \\
    &\leq \frac{4}{n_\calT^2}\sum_{j=1}^{n_\calT} \left\{P_\calT r^2(\cdot;\theta_1,\theta_2)\right\}^{1/2} \left\{P_\calT \kappa^2(\cdot,X_{\calT,j};\theta_1,\theta_2)\right\}^{1/2} \\
    &= \frac{4}{n_\calT}\left\{P_\calT r^2(\cdot;\theta_1,\theta_2)\right\}^{1/2} \frac{1}{n_\calT}\sum_{j=1}^{n_\calT}  \left\{P_\calT \kappa^2(\cdot,X_{\calT,j};\theta_1,\theta_2)\right\}^{1/2} \\
    &\leq \frac{4}{n_\calT}\left\{P_\calT r^2(\cdot;\theta_1,\theta_2)\right\}^{1/2} \frac{1}{n_\calT}\sum_{j=1}^{n_\calT}  \left\{P_\calT \kappa^2(\cdot,X_{\calT,j};\theta_1,\theta_2)\right\}^{1/2} \\
    &\leq \frac{4}{n_\calT}\left\{P_\calT r^2(\cdot;\theta_1,\theta_2)\right\}^{1/2} \left\{ \frac{1}{n_\calT}\sum_{j=1}^{n_\calT}  P_\calT \kappa^2(\cdot,X_{\calT,j};\theta_1,\theta_2)\right\}^{1/2} \\
    &\leq \frac{4\sqrt{2}}{n_\calT}\|\theta_1 - \theta_2\|_2^2(P_\calT \|\nabla_2\zeta(x,\theta_\calT)\|^2 + P_\calT L^2\delta_{n_\calT}^2)^{1/2}\left\{ \frac{1}{n_\calT}\sum_{j=1}^{n_\calT}  P_\calT \kappa^2(\cdot,X_{\calT,j};\theta_1,\theta_2)\right\}^{1/2}.
\end{align*}
Therefore, to show that term (5.2) is $o_P(\{\|\theta_1-\theta_2\|_2^2+n_\calT^{-1}\}^2)$ uniformly, it suffices to show that the following quantity is $o_P(1)$
\begin{align*}
    &\quad \sup_{\|\theta_1-\theta_2\|_2\leq \delta_{n_\calT},\|\theta_2 -\theta_\calT\|_2\leq \delta_{n_\calT}} \frac{1}{n_\calT}\sum_{j=1}^{n_\calT}  P_\calT \kappa^2(\cdot,X_{\calT,j};\theta_1,\theta_2) \\
    &\leq \sup_{\|\theta_1-\theta_2\|_2\leq \delta_{n_\calT},\|\theta_2 -\theta_\calT\|_2\leq \delta_{n_\calT}} \left\{ \frac{1}{n_\calT}\sum_{j=1}^{n_\calT}  P_\calT \kappa^2(\cdot,X_{\calT,j};\theta_1,\theta_2) - P_\calT^2 \kappa^2(\cdot,\cdot;\theta_1,\theta_2)\right\} \\
    &\quad + \sup_{\|\theta_1-\theta_2\|_2\leq \delta_{n_\calT},\|\theta_2 -\theta_\calT\|_2\leq \delta_{n_\calT}}P_\calT^2 \kappa^2(\cdot,\cdot;\theta_1,\theta_2).
\end{align*}
Donsker property of $\kappa$ implies that the term in the second line of the above display is $o_P(1)$. As for the term in the last line, we note that for any $n_\calT$, we can find $(\theta_{1n},\theta_{2n})$ (converging to $(\theta_\calT,\theta_\calT)$ as $n_\calT \rightarrow \infty$) such that $2P_\calT^2 \kappa^2(\cdot,\cdot;\theta_{1n},\theta_{2n}) \geq \sup_{\|\theta_1-\theta_2\|_2\leq \delta_{n_\calT},\|\theta_2 -\theta_\calT\|_2\leq \delta_{n_\calT}}P_\calT^2 \kappa^2(\cdot,\cdot;\theta_1,\theta_2).$ By dominated convergence theorem, this term is $o(1)$.
\\
\\
\noindent{\textbf{Step VI.}}
\begin{align*}
    \textnormal{term (6)} &= \frac{4}{n_\calT^2(n_\calT-1)^2}\sum_{i=1}^{n_\calT}\sum_{j\neq i}\sum_{k\neq i} \kappa(X_{\calT,i},X_{\calT,j};\theta_1,\theta_2)\kappa(X_{\calT,i},X_{\calT,k};\theta_1,\theta_2),
\end{align*}
which is asymptotically equivalent to
\begin{align*}
    \textnormal{term (6)}^* &= \frac{4}{n_\calT^4}\sum_{i=1}^{n_\calT}\sum_{j=1}^{n_\calT}\sum_{k=1}^{n_\calT} \kappa(X_{\calT,i},X_{\calT,j};\theta_1,\theta_2)\kappa(X_{\calT,i},X_{\calT,k};\theta_1,\theta_2) \\
    &= \underbrace{\frac{4}{n_\calT^4}\sum_{i=1}^{n_\calT}\sum_{j=1}^{n_\calT}\sum_{k=1}^{n_\calT} \left\{\kappa(X_{\calT,i},X_{\calT,j};\theta_1,\theta_2)\kappa(X_{\calT,i},X_{\calT,k};\theta_1,\theta_2) - P_\calT \kappa(\cdot,X_{\calT,j};\theta_1,\theta_2)\kappa(\cdot,X_{\calT,k};\theta_1,\theta_2)\right\}}_{\textnormal{term (6.1)}} \\
    &\quad + \underbrace{\frac{4}{n_\calT^3}\sum_{j=1}^{n_\calT}\sum_{k=1}^{n_\calT} P_\calT \kappa(\cdot,X_{\calT,j};\theta_1,\theta_2)\kappa(\cdot,X_{\calT,k};\theta_1,\theta_2)}_{\textnormal{term (6.2)}}.
\end{align*}
We will show that term (6) is $o_P(n_\calT^{-2})$ uniformly.

We study term (6.2) first, which involves a degree-2 U-statistic with kernel function $\kappa_2(\cdot,\cdot;\theta_1,\theta_2): (x_1,x_2) \mapsto P_\calT \kappa(\cdot,x_1;\theta_1,\theta_2)\kappa(\cdot,x_2;\theta_1,\theta_2)$. The function $\kappa_2(\cdot,\cdot;\theta_1,\theta_2)$ is $P_\calT$-degenerate given the degeneracy of $\kappa$. The Euclidean or Donsker property of $\{\kappa_2(\cdot,\cdot;\theta_1,\theta_2): \|\theta_1-\theta_2\|\leq \delta, \|\theta_2-\theta_\calT\|\leq\delta\}$ follows from the fact that multiplication preserves such property and $\kappa(\cdot,\cdot;\theta_1,\theta_2)$ involves linear combinations of indicator functions. Finally, for the norm convergence, we have
\begin{align*}
    \|\kappa_2(\cdot,\cdot;\theta_1,\theta_2)\|_{L^2(P_\calT^2)}^2 &= \iint \left\{\int \kappa(x,x_1;\theta_1,\theta_2)\kappa(x,x_2;\theta_1,\theta_2)dP_\calT(x)\right\}^2dP_\calT(x_1)dP_\calT(x_2) \\
    &\leq \iint \left\{\int \kappa^2(x,x_1;\theta_1,\theta_2)dP_\calT(x)\int \kappa^2(x,x_2;\theta_1,\theta_2)dP_\calT(x)\right\}dP_\calT(x_1)dP_\calT(x_2) \\
    &\leq C \iint \kappa^2(x,x_1;\theta_1,\theta_2)dP_\calT(x)dP_\calT(x_1) \rightarrow 0
\end{align*}
as $(\theta_1,\theta_2) \rightarrow (\theta_\calT,\theta_\calT)$ as we have shown when establishing Condition~\ref{cond:target quad}. Also note that $\kappa_2(\cdot,\cdot;\theta_\calT,\theta_\calT)$ is constant 0. Theorem~\ref{thm: sherman} implies that term (6.2) is $o_P(n_\calT^{-2})$ uniformly over $\{(\theta_1,\theta_2):\|\theta_1-\theta_2\|\leq \delta_{n_\calT},\|\theta_2-\theta_\calT\|\leq \delta_{n_\calT}\}$. 

We study term (6.1) next, which involves a degree-3 U-statistic with kernel function $\kappa_3(\cdot,\cdot,\cdot;\theta_1,\theta_2): (x,x_1,x_2) \mapsto \kappa(x,x_1;\theta_1,\theta_2)\kappa(x,x_2;\theta_1,\theta_2) - P_\calT \kappa(\cdot,x_1;\theta_1,\theta_2)\kappa(\cdot,x_2;\theta_1,\theta_2)$. Note that this kernel is again $P_\calT$-degenerate due to the degeneracy of $\kappa$ and $\kappa_3(\cdot,\cdot,\cdot;\theta_\calT,\theta_\calT)$ is constant 0. To show the norm of $\kappa_3$ converges to 0, it suffices to show that the norm of $\kappa(x,x_1;\theta_1,\theta_2)\kappa(x,x_2;\theta_1,\theta_2)$ converges to 0 and apply triangle inequality, and because of the uniform boundedness of $\kappa$ it suffices to show that the norm of $\kappa(x,x_1;\theta_1,\theta_2)$ converges to 0 (as we have shown before.) Thus, Theorem~\ref{thm: sherman stronger} implies that term (6.1) is $o_P(n_\calT^{-5/2})$ uniformly over $\{(\theta_1,\theta_2):\|\theta_1-\theta_2\|\leq \delta_{n_\calT},\|\theta_2-\theta_\calT\|\leq \delta_{n_\calT}\}$.
\\
\\
\noindent \textbf{Combining all the results}, we get
\begin{equation*}
    \sup_{\|\theta_1 - \theta_2\|_2 \leq \delta_{n_\mathcal{T}}, \ \|\theta_2 - \theta_\mathcal{T}\|_2 \leq \delta_{n_\mathcal{T}}} \frac{\textnormal{Var}_W[\{R_n^\dagger(\theta_1, \theta_2)\}]}{\{\|\theta_1 - \theta_2\|_2^2 + n_\mathcal{T}^{-1}\}^2} = o_P(1).
\end{equation*}

\textbf{Verification of Conditions~\ref{cond:source quad} and \ref{cond:source perturb quad}} follows the same argument as above if we replace subscript $\calT$ (target site) with $k$ ($k$-th source site.)

\textbf{Verification of Condition~\ref{cond:continuity of A}.} As the operator norm is a convex function, we apply Jensen's inequality for given $k \in \{1,\ldots,K\}$,
\begin{align*}
    \|A_k(\theta_1) - A_k(\theta_2)\| &= \|P_k\nabla_2\zeta_k(\cdot;\theta_1) - P_k\nabla_2\zeta_k(\cdot;\theta_2)\| = \|P_k (\nabla_2\zeta_k(\cdot;\theta_1) - \nabla_2\zeta_k(\cdot;\theta_2))\| \\
    &\leq P_k \|\nabla_2\zeta_k(\cdot;\theta_1) - \nabla_2\zeta_k(\cdot;\theta_2)\| \\
    &\leq P_kL\|\theta_1 - \theta_2\|,
\end{align*}
where the last inequality follows from the Lipschitz condition in assumption (b). This shows that $A_k(\theta)$ is in fact Lipschitz-continuous in a neighborhood of $\theta_\calT$ which implies continuity. Positive-definiteness follows directly from the definition of $A_\calT$ and $A_k$ and assumption (c). 

\textbf{Verification of Condition~\ref{cond:empirical process}.} Recall that $s_k(x_1,x_2;\theta) = \nabla_1 \zeta_k(x_1;\theta) + \nabla_1\zeta_k(x_2;\theta) - P_k\nabla_1\zeta_k(\cdot;\theta)$, and $S_{n,k} = 2\sum_{i=1}^{n_k} \nabla_1\zeta_k(X_{k,i};\theta)/n_k - P_k\nabla_1\zeta_k(\cdot;\theta)$. Therefore,
\begin{align*}
    &\quad S_{n,k}(\widehat\theta_\mathcal{T}) - S_{n,k}(\theta_\mathcal{T}) - P_ks_k(\widehat\theta_\mathcal{T}) + P_ks_k(\theta_\mathcal{T}) \\
    &= 2P_{n,k}\nabla_1\zeta_k(\cdot;\widehat\theta_\calT) - P_k\nabla_1\zeta_k(\cdot;\widehat\theta_\calT) - \left\{2P_{n,k}\nabla_1\zeta_k(\cdot;\theta_\calT) - P_k\nabla_1\zeta_k(\cdot;\theta_\calT)\right\} \\
    &\quad - \left\{P_k\nabla_1\zeta_k(\cdot;\widehat\theta_\calT) - P_k\nabla_1\zeta_k(\cdot;\theta_\calT)\right\} \\
    &= 2\left(P_{n,k} - P_k\right)\left\{\nabla_1\zeta_k(\cdot;\widehat\theta_\calT) - \nabla_1\zeta_k(\cdot;\theta_\calT)\right\}
\end{align*}
Assumption (d) implies that $\nabla_1\zeta_k(\cdot;\widehat\theta_\calT)$ belongs to a Donsker class. Furthermore, we have
\begin{align*}
    &~~~~\left\|\nabla_1\zeta_k(\cdot;\widehat\theta_\calT) - \nabla_1\zeta_k(\cdot;\theta_\calT)\right\|_{L^2(P_k)}^2 \\
    &= \int \left\|\nabla_1\zeta_k(x;\widehat\theta_\calT) - \nabla_1\zeta_k(x;\theta_\calT)\right\|^2 dP_k(x) \\
    &\leq \int \left\|\nabla_2\zeta_k(x;\bar\theta_\calT(x))\right\|^2\|\widehat\theta_\calT - \theta_\calT\|_2^2 dP_k(x) \\
    &\leq 2\int \left\{\left\|\nabla_2\zeta_k(x;\bar\theta_\calT(x)) - \nabla_2\zeta_k(x;\theta_\calT)\right\|^2 + \left\|\nabla_2\zeta_k(x;\theta_\calT)\right\|^2\right\}\|\widehat\theta_\calT - \theta_\calT\|_2^2 dP_k(x) \\
    &\leq 2\int \left\{L^2(x)\|\widehat\theta_\calT - \theta_\calT\|_2^2 + \left\|\nabla_2\zeta_k(x;\theta_\calT)\right\|^2\right\}\|\widehat\theta_\calT - \theta_\calT\|_2^2 dP_k(x) \\
    &= 2\|\widehat\theta_\calT - \theta_\calT\|_2^4P_kL^2 + 2\|\widehat\theta_\calT - \theta_\calT\|_2^2 P_k\left\|\nabla_2\zeta_k(\cdot;\theta_\calT)\right\|^2 \xrightarrow[]{p} 0
\end{align*}
as $\widehat\theta_\calT \xrightarrow[]{p} \theta_\calT$, where $\bar\theta_\calT(x)$ is some value in between $\widehat\theta_\calT$ and $\theta_\calT$ that can depend on the value $x$. In the above display, the second line follows from a Taylor expansion, the fourth line follows from the Lipschitz condition in assumption (b) with square-integrable $L$. Hence, Lemma 19.24 in \citet{van2000asymptotic} implies that Condition~\ref{cond:empirical process} is satisfied.

\end{proof}

\end{document}